\documentclass[acmsmall,screen,nonacm]{acmart}
\usepackage{url}
\usepackage{xspace}
\usepackage{booktabs}
\usepackage{makecell}
\usepackage{graphicx}
\usepackage{colortbl}
\usepackage{multirow}
\usepackage{subfigure}
\usepackage{tablefootnote}
\usepackage{color, xcolor}
\usepackage{amsthm,amsmath,amsfonts}
\usepackage{pifont}
\usepackage{enumitem}

\newcommand{\papernum}{189\xspace}
\newcommand{\llmnum}{78\xspace}
\newcommand{\bugnum}{20\xspace}
\newcommand{\plnum}{24\xspace}
\newcommand{\datanum}{111\xspace}

\usepackage[most]{tcolorbox}
\newcommand{\summary}[2]{
\begin{center}
\begin{tcolorbox}[title={Summary of Results for RQ#1}, breakable]
{#2}
\end{tcolorbox}
\end{center}
}

\colorlet{BLUE}{blue} 
\colorlet{ORANGE}{orange}
\colorlet{BLACK}{black}
\usepackage[normalem]{ulem}
\newcommand{\revise}[1]{{\color{black}{#1}}}
\newcommand{\delete}[1]{}

\newcommand{\revisenew}[1]{{\color{black}{#1}}}
\newcommand{\deletenew}[1]{}

\AtBeginDocument{
  \providecommand\BibTeX{{
    \normalfont B\kern-0.5em{\scshape i\kern-0.25em b}\kern-0.8em\TeX}}}

\setcopyright{acmcopyright}
\copyrightyear{2024}
\acmYear{2024}
\acmDOI{1}

\acmJournal{TOSEM}
\acmVolume{1}
\acmNumber{1}
\acmArticle{1}
\acmMonth{1}

\begin{document}

\title{A Systematic Literature Review on Large Language Models for Automated Program Repair}

\author{Quanjun Zhang}
\email{quanjunzhang@njust.edu.cn}
\orcid{0000-0002-2495-3805}
\affiliation{
\institution{School of Computer Science and Engineering, Nanjing University of Science and Technology}
\city{Nanjing}\country{China}\postcode{210094}
}

\author{Chunrong Fang}
\email{fangchunrong@nju.edu.cn}
\orcid{0000-0002-9930-7111}

\author{Yang Xie} 
\email{serialxy@outlook.com}
\orcid{0009-0000-8032-8468}

\author{Yuxiang Ma} 
\email{502022320009@smail.nju.edu.cn}
\orcid{0009-0007-6841-8410}

\affiliation{
\institution{State Key Laboratory for Novel Software Technology, Nanjing University}
\city{Nanjing}\country{China}\postcode{210093}
}

\author{Weisong Sun}
\email{weisong.sun@ntu.edu.sg}
\orcid{0000-0001-9236-8264}
\affiliation{
  \institution{School of Computer Science and Engineering, Nanyang Technological University}
  \state{50 Nanyang Avenue}
  \country{Singapore}
  \postcode{639798}
}

\author{Yun Yang}
\email{yyang@swin.edu.au}
\orcid{0000-0002-7868-5471}
\affiliation{
\institution{Department of Computing Technologies, Swinburne University of Technology}
\city{Melbourne}\country{Australia}\postcode{3122}
}

\author{Zhenyu Chen}
\email{zychen@nju.edu.cn}
\orcid{0000-0002-9592-7022}
\affiliation{
\institution{State Key Laboratory for Novel Software Technology, Nanjing University}
\city{Nanjing}\country{China}\postcode{210093}
}

\begin{abstract}
Automated Program Repair (APR) attempts to patch software bugs and reduce manual debugging efforts.
Very recently, with the advances in Large Language Models (LLMs), a rapidly increasing number of APR techniques have been proposed, significantly facilitating software development and maintenance and demonstrating remarkable performance.
However, due to ongoing explorations in the LLM-based APR field, it is challenging for researchers to understand the current achievements, challenges, and potential opportunities.
This work provides the first systematic literature review to summarize the applications of LLMs in APR between 2020 and \delete{2024}\revise{2025}.
We analyze \papernum relevant papers from LLMs, APR and their integration perspectives.
First, we categorize existing popular LLMs that are applied to support APR and outline \delete{three types}\revise{four types} of utilization strategies for their deployment.
Besides, we detail some specific repair scenarios that benefit from LLMs, e.g., semantic bugs and security vulnerabilities.
Furthermore, we discuss several critical aspects of integrating LLMs into APR research, e.g., input forms and open science.
Finally, we highlight a set of challenges remaining to be investigated and the potential guidelines for future research.
Overall, our paper provides a systematic overview of the research landscape to the APR community, helping researchers gain a comprehensive understanding of achievements and promote future research.
Our artifacts are publicly available at the GitHub repository: \url{https://github.com/iSEngLab/AwesomeLLM4APR}.

\end{abstract}



\maketitle

\section{Introduction}
Software bugs are recognized as inevitable and destructive~\cite{natella2016assessing,planning2002economic,liblit2005scalable}, posing safety issues for users worldwide and costing billions of dollars in financial losses annually~\cite{boulder2013university,theguardian2013,mozilla2025,google2025}.
It is non-trivial and time-consuming for developers to fix detected software bugs manually~\cite{britton2013reversible,weiss2007long,tian2013drone,anvik2005coping}.
\textbf{Automated Program Repair (APR)} plays a crucial role in software development and maintenance, aiming to fix software bugs \delete{without}\revise{with minimal} human intervention.
Following the foundational work GenProg~\cite{weimer2009automatically,le2012genprog} in 2009, APR has been extensively investigated over the past decades~\cite{monperrus2018automatic,gazzola2019automatic}, and researchers have proposed a variety of APR techniques, including heuristic-based~\cite{le2012genprog,martinez2016astor,yuan2018arja,jiang2018shaping}, constraint-based~\cite{xuan2016nopol,martinez2018ultra,xiong2017precise,durieux2016dynamoth}, and pattern-based~\cite{liu2019tbar,koyuncu2020fixminer,liu2019avatar} ones.
Recently, inspired by the advances of \textbf{Deep Learning (DL)}, an increasing number of learning-based APR techniques have been proposed that utilize neural network models to automatically learn bug-fixing patterns~\cite{tufano2019empirical,li2020dlfix,chen2019sequencer,lutellier2020coconut,jiang2021cure,li2022dear,zhu2021syntax,ye2022selfapr,ye2022neural,zhu2023tare,ye2024iter}.
Thanks to the powerful ability of DL models to learn hidden repair patterns from massive code corpora, learning-based APR has achieved remarkable performance in the last couple of years~\cite{zhang2023survey}, attracting considerable attention from both academia and industry~\cite{jin2023inferfix,kim2022empirical,joshi2023repair}.

Very recently, \textbf{Large Language Models (LLMs)} have been successfully applied to a broad range of source code-related tasks~\cite{zhang2023llmsurvey,wang2024software,zhang2023unifying}, such as code generation~\cite{li2023skcoder,wang2023two,zhu2024hot,wang2023natural}, code summarization~\cite{su2024distilled,wang2023one,sun2023automatic}, and test generation~\cite{schafer2024empirical,dinella2022toga,nashid2023retrieval,hu2023identify,alagarsamy2023a3test,zhang2025exploring,zhang2025improvingtosem,zhang2025improving}.
Benefiting from massive model parameters and vast training data, LLMs have demonstrated impressive performance and fundamentally revolutionized the research paradigm in the \textbf{Software Engineering (SE)} community.
In the domain of APR, beginning with pioneering studies, e.g., TFix~\cite{berabi2021tfix}, CIRCLE~\cite{yuan2022circle} and AlphaRepair~\cite{xia2022less}, the community has witnessed an explosion of repair studies utilizing LLMs, already achieving considerable advantages and further indicating significant potential for future research.
However, the integration of LLMs within APR is a considerably complex undertaking, making it difficult for interested researchers to understand existing work.
For example, existing LLM-based APR studies encompass different research perspectives (e.g., empirical~\cite{xia2023automated}, technical~\cite{xia2022less} and benchmark studies~\cite{zhang2023critical}), repair phases (e.g., patch generation~\cite{zhang2023gamma} and correctness assessment~\cite{zhang2024appt}), repair scenarios (e.g., static warnings~\cite{jin2023inferfix} and syntax errors~\cite{joshi2023repair}), mode architectures (e.g., encoder-only~\cite{zhang2023pre} and decoder-only~\cite{mashhadi2021applying}) and model utilization paradigms (e.g., fine-tuning~\cite{yuan2022circle}, few-shot~\cite{nashid2023retrieval} and zero-shot~\cite{zhang2023gamma}).
Despite ongoing explorations in the field, the literature currently lacks a detailed and systematic review of the applications of LLMs in APR, making it challenging for researchers to understand the multitudinous design choices of existing work and conduct follow-up research.

\textbf{{This Paper.}}
To bridge this gap, our work provides the first systematic literature review on the deployment of rapidly emerging LLM-based APR studies.
Based on this, the community can gain a comprehensive understanding of the strengths, weaknesses, and gaps in existing LLM-based APR techniques.
We discuss what LLMs are widely adopted in state-of-the-art APR research and how they are integrated into the repair workflow.
We collect \papernum{} relevant papers and perform a systematic analysis from LLMs, APR, and integration perspectives.
From our analysis, we reveal the current challenges and point out possible future directions for LLM-based APR research.
Overall, this work offers a thorough overview of the ongoing progress within the LLM-based APR community, aiding researchers in navigating this burgeoning field and advancing toward innovative practices.

\textbf{{Contributions.}}
To sum up, this work makes the following contributions:
\begin{itemize}
    \item \textit{Survey Methodology.}
    We conduct the first systematic literature review with \papernum{} high-quality APR papers that utilize recent LLMs to address repair challenges from 2020 to \delete{April 2024}\revise{September 2025}.

    \item \textit{Trend Analysis.}
    We perform a detailed analysis of selected APR studies in terms of publication trends, distribution of publication venues, and types of contributions.

    \item \textit{LLMs Perspective.}
    We summarize \llmnum{} LLMs utilized to support program repair and provide a summary of the typical usage and trends of different LLM categories in the APR domain.

    \item \textit{APR Perspective.}
    We describe common repair scenarios that LLMs are applied to, encompassing \bugnum{} bug types, such as security vulnerabilities and programming problems.

    \item \textit{Integration Perspective.}
    We discuss some key factors, including datasets, input representations and open science, that impact the performance of integrating LLMs into APR.

    \item \textit{{Challenges and Opportunities.}}
    We summarize some crucial challenges of applying LLMs in the APR field, and pinpoint some potential guidelines for future LLM-based APR research.
    
\end{itemize}

\textbf{{Paper Organization.}}
Section~\ref{sec:background} introduces some basic concepts about APR and LLMs, and Section~\ref{sec:related_work} discuss the 
Then, according to the contributions listed above, Section~\ref{sec:methodology} lists our research questions (RQs) and the research methodology to collect papers related to our work.
Section~\ref{sec:rq1} investigates the trend and distribution of LLM-based APR studies.
Section~\ref{sec:rq2} summarizes LLMs that are used by existing APR studies.
Section~\ref{sec:rq3_repair_scenarios} illustrates the primary repair scenarios that LLMs are applied to and provides a brief description of each work.
Section~\ref{sec:rq4} discusses some crucial factors during the integration of LLMs and APR, including datasets, input representation, patch correctness, and open science.
Section~\ref{sec:challenges} discusses some challenges and practical guidelines.
Section~\ref{sec:conclusion} draws the conclusions.

\section{Background\delete{ and Related Work}}
\label{sec:background}
\revise{In this section, we introduce the core concepts relevant to this work, including APR and LLMs.
Section~\ref{sec:apr} outlines the principles and evolution of APR, and Section~\ref{sec:llm} summarizes recent advances in LLM classifications and representative models.}

\subsection{Automated Program Repair}
\label{sec:apr}
\subsubsection{Problem Description}

\delete{APR, one of the most challenging activities in the SE field, aims to fix software bugs without human intervention and is regarded as a fundamental aspect of software automation.}
\delete{For example, in}\revise{In} the software development and maintenance process, developers usually attempt to implement a designed functionality according to requirement specifications and write test suites to validate its correctness.
If all test cases pass, the functionality is deemed correctly implemented; otherwise, developers must analyze the symptoms of failures and make necessary modifications to the buggy code snippets to ensure all test cases pass.
Formally, given a buggy program $P$ and its corresponding specification $S$ that $P$ fails to satisfy, APR is defined to find a minimal transformation $P'$ of $P$ that satisfies $S$.
In practice, test cases are usually utilized as the specification, and in such a typical test-driven repair scenario, APR aims to find a minimal program variant that passes the available test suite, \delete{which contains at least one test case making the original buggy program $P$ does not satisfy.}\revise{which contains at least one test case that the original buggy program $P$ fails to satisfy.}

\begin{figure}[htbp]
\centering
    \includegraphics[width=0.7\linewidth]{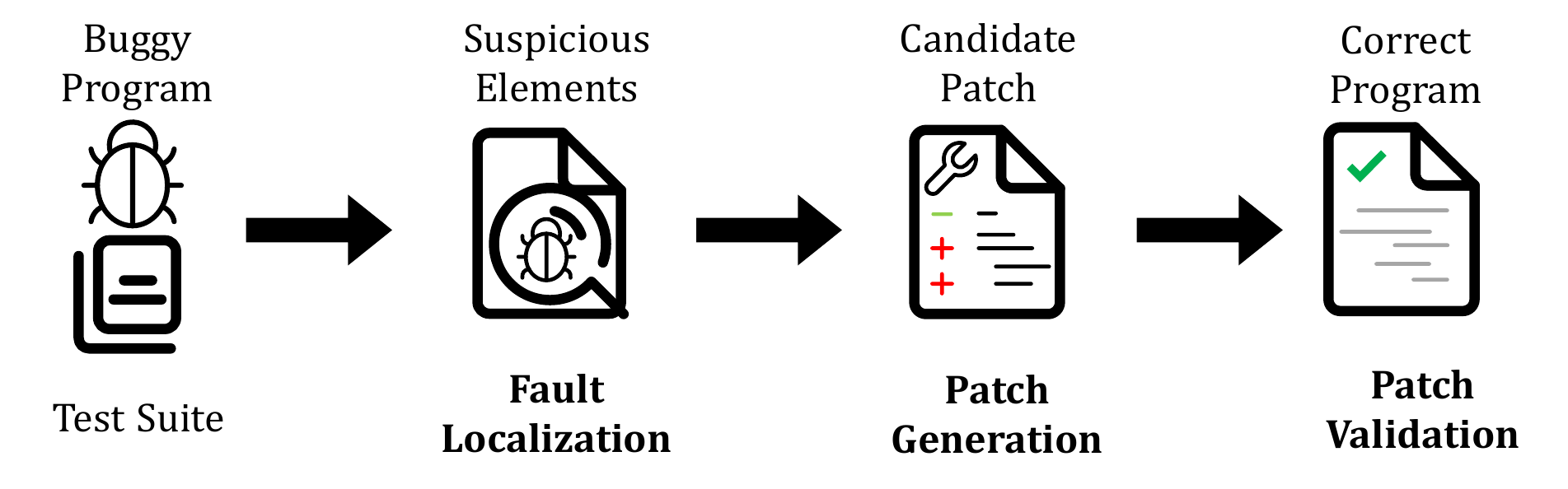}
    \caption{The basic pipeline of automated program repair.}
    \label{fig:apr}
\end{figure}

\subsubsection{Repair Workflow}
Fig.~\ref{fig:apr} illustrates the typical \textit{generate-and-validate} workflow of APR, which is usually composed of three parts.
Specifically, for a detected bug\footnote{In practice, APR is generally triggered after a bug has been detected through testing techniques, such as static analysis and fuzzing tools.}, 
(1) the \textbf{fault localization phase} identifies suspicious code elements that need to be fixed based on off-the-shelf localization techniques~\cite{wong2016survey};
(2) the \textbf{patch generation phase} generates program variants (i.e., candidate patches) via transformation rules~\cite{zhang2023survey};
(3) the \textbf{patch validation phase} utilizes available test suites as the oracle to identify correct patches by dynamic execution~\cite{zhang2024appt}.
It is important to note that a candidate patch that successfully passes the available test suite is termed a plausible patch. 
However, if such a plausible patch fails to generalize to additional test cases, it is considered an overfitting patch. 
Conversely, a plausible patch that remains effective across broader test cases is recognized as a correct patch, one that is semantically in alignment with patches written manually by developers.

\subsubsection{Repair Techniques}

The literature has introduced a multitude of APR techniques to generate correct patches from various perspectives.
Such APR techniques can be categorized into four categories:
(1) \textbf{heuristic-based APR} utilizes genetic programming to explore the search space of the correct patch, such as GenProg~\cite{le2012genprog}, Astor~\cite{martinez2016astor}, ARJA~\cite{yuan2018arja}, and SimFix~\cite{jiang2018shaping};
(2) \textbf{constraint-based APR} usually focuses on the condition synthesis problem by treating program repair as a constraint-solving task, such as Nopol~\cite{xuan2016nopol},  Cardumen~\cite{martinez2018ultra}, ACS~\cite{xiong2017precise}, and Dynamoth~\cite{durieux2016dynamoth};
(3) \textbf{pattern-based APR} utilizes pre-defined repair templates that are usually hand-crafted by experts to transform buggy code snippets into correct ones, such as TBar~\cite{liu2019tbar}, FixMiner~\cite{koyuncu2020fixminer} and Avatar~\cite{liu2019avatar};
(4) \textbf{learning-based APR} treats patch generation as a neural machine translation task with the advance of DL models, such as \delete{Tufano~et al.~\cite{tufano2019empirical}, SequenceR~\cite{chen2019sequencer}, CoCoNut~\cite{lutellier2020coconut}, DLFix~\cite{li2020dlfix}, CURE~\cite{jiang2021cure}, Recoder~\cite{zhu2021syntax}, DEAR~\cite{li2022dear}, SelfAPR~\cite{ye2022selfapr}, RewardAPR~\cite{ye2022neural}, and Tare~\cite{zhu2023tare}}\revise{CURE~\cite{jiang2021cure}, Recoder~\cite{zhu2021syntax}, DEAR~\cite{li2022dear} and SelfAPR~\cite{ye2022selfapr}}.

Among the four types of APR techniques, learning-based APR has achieved remarkable performance by learning hidden bug-fixing patterns automatically from extensive source code databases~\cite{zhang2023survey}.
Recently, inspired by the success of LLMs in NLP and SE tasks, researchers have increasingly been utilizing advanced LLMs to address software bugs~\cite{xia2023automated,xia2022less,zhang2023gamma,zhang2023pre}.
Compared with learning-based APR, such LLM-based APR techniques have demonstrated significantly better performance and have received growing attention in the community, which is the focus of our work.

\subsection{Large Language Models}
\label{sec:llm}
LLMs refer to advanced Artificial Intelligence (AI) models that undergo extensive pre-training on large text corpora to enhance their capabilities of natural language understanding, generation, and interpretation in a human-like manner~\cite{zhao2023survey}.
In recent years, their training scope has been extended beyond natural languages to include programming languages, enabling LLMs to reason about and generate source code effectively. 
The literature has seen a variety of LLMs supporting NLP and SE research~\cite{shang2025large}, which can be categorized into three main categories based on their model architectures:
(1) \textbf{Encoder-only LLMs}, such as CodeBERT~\cite{feng2020codebert}, GraphCodeBERT~\cite{guo2020graphcodebert}, train the encoder part of the Transformer to generate a fixed-dimensional bidirectional representation with Masked Language Modeling (MLM) and Next Sentence Prediction (NSP).
MLM aims to predict the original tokens that have been randomly masked out, and NSP predicts whether two given sentences actually follow each other in a text.
(2) \textbf{Decoder-only LLMs}, such as CodeGPT~\cite{lu2021codexglue}, train the decoder part of the Transformer to support auto-regressive tasks with Causal Language Modeling (CLM), which aims to predict new tokens in a sequence based on previous tokens.
(3) \textbf{Encoder-decoder LLMs}, such as CodeT5~\cite{wang2021codet5}, train both encoder and decoder parts of the Transformer to support sequence-to-sequence generation tasks with denoising objectives.
We will summarize existing LLMs and how they are leveraged to support program repair in Section~\ref{sec:rq2_llms}.

\section{\revise{Related Work}}
\label{sec:related_work}
In this survey, we systematically collect, summarize, and analyze APR studies empowered with advanced LLMs.
Although our primary focus is the integration of LLMs with APR, given its growing prominence, there exists some related studies that explore similar aspects of either LLMs or APR.
We clarify the fundamental differences between our work and prior work below.

The first group of related surveys attempts to analyze the applications of LLM more generally in SE.
Zhang~et al.~\cite{zhang2023llmsurvey} provide a detailed summarization of the LLM-based SE research from two intuitive perspectives: LLMs and their applications in SE.
They introduce representative work across 30 LLMs, 15 pre-training tasks, 16 downstream tasks, and 43 specific code-related tasks.
Besides, Wang~et al.~\cite{wang2024software} provide a review of LLMs in software testing; Fan~et al.~\cite{fan2023large} discuss the achievements and challenges of LLMs in SE;
Hou~et al.~\cite{hou2023large} conduct a systematic literature review on LLM4SE.
However, these works target the whole software engineering/testing workflow rather than our specific program repair task.
They only provide a bird's-eye view of a limited number of LLM-based APR papers, while our work provides an in-depth analysis from various perspectives.
Their work indicates that program repair is the most popular scenario in which LLMs are applied within software engineering.
Given the complexity and significance of program repair, we are motivated to conduct a more in-depth review of APR achievements involving LLMs.

The second group of related surveys attempts to review the achievements of APR.
Gazzola~et al.~\cite{gazzola2019automatic} present a survey to organize repair studies, and Monperrus~et al.~\cite{monperrus2018automatic} present a bibliography of behavioral and state repair studies.
Both works collect publications up to 2017, hence there is no overlap with our research as the first LLM-based APR work is published in 2020.
\revise{For completeness, we note that an earlier undergraduate thesis~\cite{MacNeil2023SystematicReview} also reviews LLMs for program repair; however, our study differs substantially in scope, methodological rigor, and analytical depth.}
The most highly related work to this paper is Zhang~et al.~\cite{zhang2023survey}, which provides a systematic survey to summarize the state-of-the-art in the learning-based APR community.
However, they mainly focus on DL models and devote merely a single section to discussing LLM-based APR studies (only five papers), failing to include the surge of works that have emerged during the last two years.
In contrast, our work solely focuses on the technology of LLM applications in the APR community until \delete{April 2024}\revise{September 2025}, involving the trend of utilized LLMs, optimization techniques, repair scenarios, and input forms.

\section{Survey Methodology}
\label{sec:methodology}

\delete{
In this section, guided by the principles outlined by Petersen~et al.~\cite{petersen2015guidelines} and Kitchenham~et al.~\cite{kitchenham2007guidelines}, we present details of our systematic literature review methodology.
}
\revise{
In this section, guided primarily by the SEGRESS guidelines~\cite{kitchenham2023segress}, we present details of our systematic literature review methodology.
}

\subsection{Research Questions}
We attempt to provide a comprehensive overview of recent LLMs' applications to APR by summarizing the relevant studies and further providing guidelines on the follow-up research.
\revise{Although the scope of this study is relatively broad, we adopt an SLR approach ~\cite{petersen2015guidelines,kitchenham2007guidelines} rather than a systematic mapping study, as our goal is to synthesize detailed evidence, identify methodological patterns, and derive analytical insights rather than merely classify or map existing research topics.}
To achieve this, this systematic literature review answers the following Research Questions (RQs):

\begin{itemize}
    \item \textbf{RQ1 \revise{(Trend Analysis)}: What is the trend of APR studies that utilize LLMs?}
    \item \textbf{RQ2 \revise{(LLM Perspective)}: Which popular LLMs have been applied to support APR?}
    \item \textbf{RQ3 \revise{(APR Perspective)}: What repair scenarios have been facilitated by LLMs?}
    \item \textbf{RQ4 \revise{(Integration Perspective)}: What key factors contribute to the integration of LLMs for APR?}
\end{itemize}

\subsection{Search Strategy}
\begin{figure}[t]
\centering
    \includegraphics[width=0.9\linewidth]{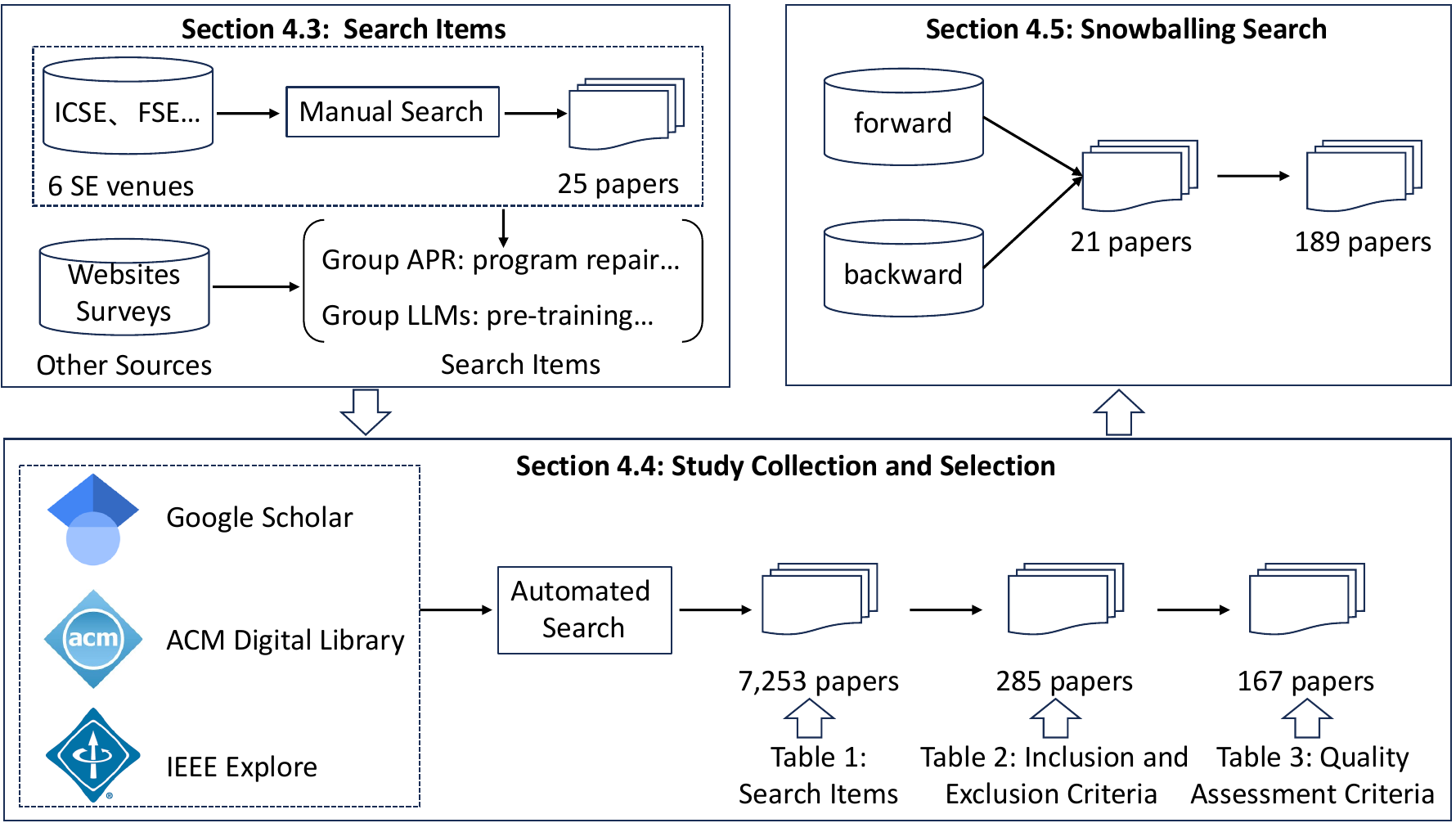}
    \caption{\revisenew{Process of paper collection and selection.}}
    \label{fig:workflow}
\end{figure}

We search and identify primary papers following the ``Quasi-Gold Standard'' (QGS)~\cite{zhang2011identifying} method, which is a common practice to construct a set of known studies for refining search strings by combining manual and automated search processes.
\revisenew{Fig.~\ref{fig:workflow} illustrates the general workflow of paper collection and section.}
We first conduct a manual search to identify a set of relevant studies and derive a search string from them, as detailed in Section~\ref{sec:search_items}.
We then utilize the search string to perform an automated search and employ a series of relatively strict filtering steps to obtain the most relevant studies, as detailed in Section~\ref{sec:study_colection}.
We finally use a snowballing search to supplement the search results further, as detailed in Section~\ref{sec:snowballing}.
Given the abundance of relevant papers from different research communities, such as SE and AI, this search strategy allows us to capture the most pertinent papers while achieving greater efficiency than a purely manual search with a relatively rigorous process.
Particularly, we undertake the following five phases to search for and identify relevant studies.

\begin{enumerate}
    \item \textbf{Search Sources}.
    This phase selects high-quality publication venues for the initial manual search and well-known digital databases for the subsequent automated search.
    
    \item \textbf{QGS Establishment}.
    This phase inspects all papers identified in the manual search and filters them by inclusion/exclusion criteria.
    
    \item \textbf{Search Items.}
    This phase defines search items with domain knowledge and establishes QGS.

    \item \textbf{Paper Collection and Selection}.
    This phase conducts an automated search using the above search items and filters the collected papers by inclusion/exclusion criteria.
    
    \item \textbf{Snowballing Search}.
     This phase conducts a snowballing search to complement the final search results.
\end{enumerate}

\subsection{Search Items}
\label{sec:search_items}
To perform the manual search and establish QGS, we select four top-tier SE conferences (ICSE, ESEC/FSE, ASE, ISSTA) and two journals (TOSEM and TSE) and search for relevant papers involving both program repair and LLMs.
We manually inspect and identify 25 papers from 2020 to 2024 that meet our criteria, which then form the basis for constructing QGS.
Following prior surveys~\cite{wang2022machine,yang2022survey,zhang2023survey}, we divide the search items used for searching papers automatically into two groups: (1) an APR-related group containing some commonly-used keywords related to program repair;
and (2) an LLM-related group containing some popular keywords related to large language models or pre-trained models.
Besides, as the APR community benefits from well-maintained forums and real-time updated works, we further refine our search items by concluding some frequent keywords from three sources: a community-driven website\footnote{\url{http://program-repair.org/bibliography.html}}, a living review of APR by Monperrus~\cite{monperrus2020living} and the most recent learning-based APR survey~\cite{zhang2023survey}.
Finally, we identify a search string that includes several LLM-related and APR-related keywords frequently appearing in program repair studies that use LLMs.
The complete set of search keywords is shown in Table~\ref{tab:search_items}.

\begin{table}[t]

\begin{center}
\caption{Search items used for retrieving LLM-based APR papers.}
\label{tab:search_string}
    \setlength{\fboxrule}{0.5pt}
    \fbox{
    \parbox{0.95\textwidth}{%
    \textit{
    (``program repair'' OR ``software repair'' OR ``automatic repair'' OR ``code repair'' OR ``bug repair'' OR ``bug fix'' OR ``code fix'' OR ``automatic fix'' OR ``patch generation'' OR ``patch correctness'' OR ``patch validation'' OR ``fix generation'' OR ``code transformation'' OR ``code edit'' OR ``fix error'')
    \textbf{AND} (``LLM(s)'' OR ``Large Language Model(s)'' OR ``Pre-trained'' OR ``Pretrained'' OR ``Pre-training'' OR ``Pretraining'' OR ``PLM(s)'' OR  ``(Code)BERT'' OR ``(Code)T5'' OR ``(Code)GPT'' OR ``Codex'' OR ``ChatGPT'' OR ``(Code)Llama'' OR ``GPT-*'' OR ``neural'' OR ``machine'' OR ``deep'' OR ``learning'' OR ``transformer/transformers'' OR ``model/models'' OR ``transfer'' OR ``supervised'')
    }
    }%
    }
\end{center}
\end{table}

It is noteworthy that we also include the list of keywords related to machine/deep learning from Zhang~et al.~\cite{zhang2023survey}, so as to avoid missing any papers related to our work as much as possible.

\subsection{Study Collection and Selection}
\label{sec:study_colection}

\begin{table}[t]
\caption{Inclusion criteria and exclusion criteria.}
\footnotesize
\resizebox{0.9\linewidth}{!}{
\begin{tabular}{lp{12cm}}
\toprule
\multicolumn{2}{l}{\textbf{Inclusion criteria}} \\
\midrule
\ding{172} & The paper utilizes an LLM in its framework. \\
\ding{173} & The paper addresses the task of program repair. \\
\ding{174} & The paper is accessible with full text. \\
\ding{175} & The paper is only written in English. \\
\midrule

\multicolumn{2}{l}{\textbf{Exclusion criteria}} \\ 
\midrule

\ding{182} & The paper is fewer than seven pages. \\
\ding{183} & The paper is an old conference version extended to journals with the same authors. \\
\ding{184} & The paper uses repair methods to contribute to LLMs. \\
\ding{185} & The paper is published as an SLR, review, or survey only mentioning the development of LLMs and APR. \\
\ding{186} & The paper is a duplicate publication, with the preprint and the published version presenting the same content under distinct titles.\\
\ding{187} & The paper is published in a workshop or a doctoral symposium. \\
\ding{188} & The paper is a grey publication, e.g., a technical report or thesis. \\
\ding{189} & The paper falls into the category of short papers, tool demonstrations, and editorials. \\

\bottomrule
\end{tabular}
}
\label{tab:criteria}
\end{table}

\begin{table}[t]
\caption{Checklist of Quality Assessment Criteria (QAC) for LLM-based APR studies.}
\resizebox{0.8\linewidth}{!}{
\begin{tabular}{rl}
\toprule
\textbf{ID} & \textbf{Quality Assessment Criteria} \\ 
\midrule
QAC1 & Is the LLM adopted as the research subject rather than only as a baseline method? \\
QAC2 & Is the impact of the paper on the APR community clearly stated? \\
QAC3 & Is the contribution of the paper clearly stated? \\
QAC4 & Is the paper published in a high-repute venue? \\
QAC5 & Is the relevant artifact made open source, e.g., datasets? \\
QAC6 & Is there a clear motivation for the paper? \\
QAC7 & Is the implementation of the proposed approach described clearly? \\
QAC8 & Is the experimental setup described in detail, e.g., hyper-parameters and environments? \\
QAC9 & Are the findings clearly supported by experimental results? \\
QAC10 & Are the key contributions and limitations of the study discussed? \\
\midrule 
\end{tabular}
}
\label{tab:qac}
\end{table}

\revise{
To perform the automated search, we collect potentially relevant papers by primarily searching the Google Scholar repository, ACM Digital Library, and IEEE Explorer Digital Library at the end of September 2025.
Based on our search strategy, we obtain a total of 7,253 papers.
Once we gather the studies based on our automated search strategy, we conduct a filtering phase followed by a quality assessment phase to eliminate papers that do not align with the objectives of this review.
}

\subsubsection{Inclusion and Exclusion Criteria}
\revise{The search conducted on the selected databases is intentionally broad to ensure the inclusion of as many potentially relevant studies as possible.
However, this inclusiveness inevitably resulted in collecting papers outside the scope of our survey.
Thus, to refine the selection, we defined a set of specific inclusion and exclusion criteria and applied them to all collected papers, ensuring that every retained study aligned with our research scope.
Table~\ref{tab:criteria} presents the designed inclusion and exclusion criteria.
For example, following Exclusion criterion~\ding{183}, we filter out papers with fewer than seven pages~\cite{zhang2023survey}.
Following Exclusion criterion~\ding{182}, we retain only the extended journal version~\cite{huang2023empirical} when a study expanded a previously published conference paper.
Following Exclusion criterion~\ding{186}, we identify 21 papers whose published versions have different titles compared to their preprint.
In such cases, to avoid duplication, we retain only the
published versions.
It is worth noting that, given the positioning of this study, we do not adopt a multivocal review\cite{garousi2019guidelines} and thus exclude broader grey literature (e.g., technical blogs and theses). 
However, considering the rapid evolution of this field, we incorporate a limited number of \textit{arXiv} papers that have passed our rigorous quality assessment, following recent LLM4SE surveys to ensure the timeliness of our review.
During this process, two authors independently inspect the titles, abstracts, keywords, and venues of all remaining papers according to our inclusion/exclusion criteria in Table~\ref{tab:criteria}, reducing the number of candidates to 285.
Any disagreements between two authors are resolved through discussion, and if a consensus cannot be reached, the third author makes the final decision.
}

\subsubsection{Quality Assessment}

To ensure the inclusion of high-quality papers, we conduct a full-text inspection of the remaining papers to assess their relevance and quality within the LLM-based APR field.
\revisenew{This step is adopted to mitigate the impact of substantial quality variations observed in the collected studies, particularly among very recent and non-peer-reviewed papers, and is consistent with common practices in most recent LLM-based SE surveys~\cite{zhang2023llmsurvey,zhang2025large}.}
Following Wang~et al.~\cite{wang2022machine}, we design ten quality assessment questions, where each paper is independently rated by two authors on a three-tier scale:``yes'' (1 point), ``partial'' (0.5 points), or ``no'' (0 points).
Papers scoring below the threshold of 8 points are excluded from further analysis, resulting in 167 papers being retained.
Any disagreement is resolved through discussion, and when necessary, by consultation with the third author.
The designed quality assessment criteria (QAC) are presented in Table~\ref{tab:qac}.
It is worth noting that, for QAC4, given the emerging nature of this research area and the rapid evolution of LLMs, many recent works have not yet completed the peer-review process.
Therefore, we also consider papers from \textit{arXiv} that satisfy our quality assessment criteria.
Before inclusion, we carefully check whether each arXiv paper had been subsequently published by comparing its title, authorship, and content.
If a published version is found, only the peer-reviewed version is retained; otherwise, the arXiv version is included and explicitly marked as such in our dataset.
This strategy aligns with the common practice in recent LLM-based SE surveys, and ensures our work remains both comprehensive and up to date while maintaining a high standard of quality.

\subsection{Snowballing Search}
\label{sec:snowballing}
To ensure maximal comprehensiveness of the collected papers, we additionally employ the snowballing search ~\cite{watson2022systematic} to manually incorporate any potential papers that are overlooked during our previous search process.
The snowballing search is a common practice to examine the references (i.e., backward snowballing) or citations (i.e., forward snowballing) within a list of papers to discover additional relevant works.
Specifically, we scrutinize every reference and citation within the collected papers to assess their relevance to our research objectives. 
This meticulous manual analysis of all potential papers through the entire study selection process ultimately leads to the inclusion of \papernum{} papers in our survey.
\revisenew{Through this process, we identified 21 additional studies, leading to the inclusion of \papernum{} final papers in our survey}.
All included papers and their bibliographic information are publicly available in our replication repository for transparency and reproducibility~\cite{myurl}.

\section{RQ1: What is the trend of APR studies that utilize LLMs?}
\label{sec:rq1}

\subsection{What are the trends of publication over time?}

 \begin{figure}[t]
 \centering
     \subfigure[\revise{Number of publications per year.}]
     {
         \includegraphics[width=0.45\linewidth]{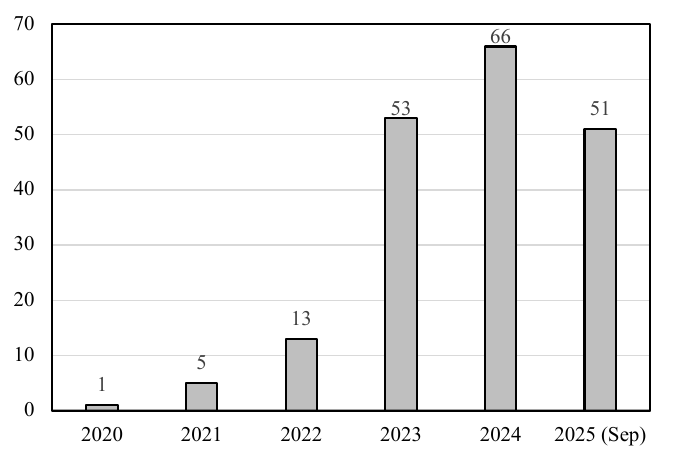}
         \label{fig:trend_time1}
     }
     \subfigure[\revise{Cumulative number of publications per year.}]
     {
       \includegraphics[width=0.45\linewidth]{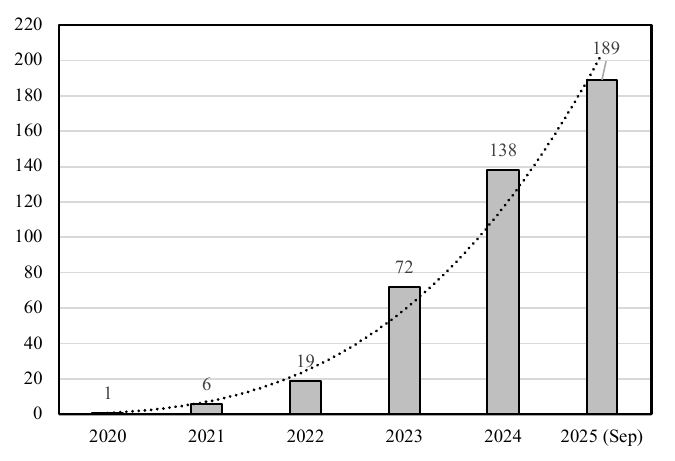}
       \label{fig:trend_time2}
     }
     \caption{\revise{Publication trends of LLM-based APR studies.}}
     \label{fig:trend_time}
 \end{figure}

\revise{We analyze the publication trends of APR studies empowered with LLMs between 2020 and September 2025.
Although the Transformer architecture~\cite{vaswani2017attention} was proposed in 2017, and the groundbreaking LLM, BERT~\cite{devlin2019bert}, was introduced in 2018, we do not find any studies using LLMs to fix bugs before 2020.
Fig.~\ref{fig:trend_time1} illustrates the number of relevant studies, and we find that the number of publications from 2020 to 2024 shows a significant increase, with the number reaching 66 papers in 2024.
It should be noted that we collect papers as of September 2025; thus the number of relevant studies (i.e., 51 papers) in 2025 does not fully reflect the overall trend of LLM-based APR for the entire year.
We also calculate the cumulative number of publications as shown in Fig.~\ref{fig:trend_time2}.
To predict the publication trend in the last six years, we fit the cumulative number of publications as a power function, and find that the slope of the curve fitting the distribution increases substantially between 2020 and 2025.
Particularly, the coefficient of determination attains the peak value ($R^2 = 0.98833$)\footnote{The publication trend is fitted using a power function $y = a \times x^b$, where a=0.8451 and b=3.0876. It is worth noting that this trend analysis offers only an indicative projection. 
Given the rapidly evolving nature of LLM-based research, future publication trends may vary as the field continues to advance.}, estimating that there may be over 210 relevant LLM-based APR publications by the end of 2025.
Overall, utilizing LLMs to address bug-fixing has become a prevalent trend since 2020, and a large number of studies will adopt LLMs to address the challenges of APR in the future.}

\begin{table}[t]
  \centering
  \caption{\revise{Publication venues with LLM-based APR studies.}}
  \resizebox{\linewidth}{!}{
        \begin{tabular}{llll}
    \toprule
    \revise{\textbf{Acronym}} & \revise{\textbf{Venues}} & \revise{\textbf{Papers}} & \revise{\textbf{Proportion}} \\
    \midrule
    \revise{ICSE} & \revise{International Conference on Software Engineering} & \revise{26} & \revise{12.94\%} \\
    \revise{arxiv} & \revise{N.A.} & \revise{25} & \revise{12.44\%} \\
    \revise{FSE/ESEC} & \revise{European Software Engineering Conference and International Symposium on Foundations of Software Engineering} & \revise{17} & \revise{8.46\%} \\
    \revise{ASE} & \revise{International Conference on Automated Software Engineering} & \revise{13} & \revise{6.47\%} \\
    \revise{TOSEM} & \revise{Transactions on Software Engineering Methodology} & \revise{12} & \revise{5.97\%} \\
    \revise{ISSTA} & \revise{International Symposium on Software Testing and Analysis} & \revise{11} & \revise{5.47\%} \\
    \revise{ICLR} & \revise{International Conference on earning Representations} & \revise{11} & \revise{5.47\%} \\
    \revise{TSE} & \revise{Transactions on Software Engineering} & \revise{9} & \revise{4.48\%} \\
    \revise{NeurIPS} & \revise{Neural Information Processing Systems} & \revise{7} & \revise{3.48\%} \\
    \revise{ACL} & \revise{The Association for Computational Linguistics} & \revise{6} & \revise{2.99\%} \\
    \revise{APR} & \revise{International Automated Program Repair Workshop} & \revise{6} & \revise{2.99\%} \\
    \revise{AAAI} & \revise{AAAI Conference on Artificial Intelligence} & \revise{3} & \revise{1.49\%} \\
    \revise{ICSME} & \revise{The International Conference on Software Maintenance and Evolution} & \revise{3} & \revise{1.49\%} \\
    \revise{MSR} & \revise{International Conference on Mining Software Repositories} & \revise{3} & \revise{1.49\%} \\
    \revise{USENIX Security} & \revise{USENIX Security Symposium} & \revise{3} & \revise{1.49\%} \\
    \revise{EMSE} & \revise{Empirical Software Engineering} & \revise{2} & \revise{1.00\%} \\
    \revise{ICST} & \revise{IEEE International Conference on Software Testing, Verification and Validation} & \revise{2} & \revise{1.00\%} \\
    \revise{OOPSLA} & \revise{Conference on Object-Oriented Programming Systems, Languages,and Applications} & \revise{2} & \revise{1.00\%} \\
    \revise{AUSE} & \revise{Automated Software Engineering} & \revise{1} & \revise{0.50\%} \\
    \revise{COMPSAC} & \revise{International Computer Software and Applications Conference} & \revise{1} & \revise{0.50\%} \\
    \revise{DAC} & \revise{Design Automation Conference} & \revise{1} & \revise{0.50\%} \\
    \revise{JSS} & \revise{Journal of Systems and Software} & \revise{1} & \revise{0.50\%} \\
    \revise{NAACL} & \revise{North American Chapter of the Association for Computational Linguistics} & \revise{1} & \revise{0.50\%} \\
    \revise{PLDI} & \revise{ACM SIGPLAN Conference on Programming Language Design\textbackslash{}\& Implementation} & \revise{1} & \revise{0.50\%} \\
    \revise{PMLR} & \revise{Proceedings of Machine Learning Research} & \revise{1} & \revise{0.50\%} \\
    \revise{S\&P} & \revise{IEEE Symposium on Security and Privacy} & \revise{1} & \revise{0.50\%} \\
    \revise{SLE} & \revise{ACM SIGPLAN International Conference on Software Language Engineering} & \revise{1} & \revise{0.50\%} \\
    \revise{TDSC} & \revise{IEEE Transactions on Dependable and Secure Computing} & \revise{1} & \revise{0.50\%} \\
    \revise{TIFS} & \revise{IEEE Transactions on Information Forensics and Security} & \revise{1} & \revise{0.50\%} \\
    \revise{TLT} & \revise{IEEE Transactions on Learning Technologies} & \revise{1} & \revise{0.50\%} \\
    \revise{TODAES} & \revise{ACM Transactions on Design Automation of Electronic Systems} & \revise{1} & \revise{0.50\%} \\
    \revise{WWW} & \revise{Proceedings of the ACM Web Conference} & \revise{1} & \revise{0.50\%} \\
    \revise{Others} & \revise{N.A.} & \revise{14} & \revise{6.97\%} \\
    \midrule
    \revise{Total} & \revise{N.A.} & \revise{189} & \revise{94.03\%} \\
    \bottomrule
    \end{tabular}%
    }
  \label{tab:venues}%
\end{table}%

\subsection{What is the distribution of the publication venues?}

\delete{We analyze \papernum{} studies published in various publication venues.
Table~\ref{tab:venues} lists the number of relevant papers published in each publication venue.
We find that 56.69\% of papers are published in peer-reviewed venues, and ICSE and ESEC/FSE are the most popular venues that contain 13 papers, followed by APR (six papers), ICLR (five papers), and ASE (four papers).
Besides, the top popular venues are top-tier conferences and symposiums, and only four journals have at least one relevant publication, i.e., TOSEM (four papers), TSE (three papers), TDSC (one paper), and TIFS (one paper).
This finding indicates a preference for presenting the latest work at conferences due to the timeliness of conference proceedings. 
These collected papers are published in different research fields, including SE (e.g., TOSEM and ICSE), AI (e.g., NeurIPS, AAAI and ICLR), NLP (e.g., ACL), Programming Language (e.g., OOPSLA and PLDI) and Security (e.g., S\&P, TDSC and TIFS). 
We also find that the remaining 43.31\% of papers are non-peer-reviewed and published on arXiv.
This phenomenon can likely be attributed to the rapid emergence of relevant studies in a short development time.
Considering the varying quality levels of such non-peer-reviewed papers, we conducted a rigorous evaluation to ensure the inclusion of high-quality relevant papers in this work.}
\revise{We analyze \papernum{} studies published in various publication venues.
Table~\ref{tab:venues} lists the number of relevant papers published in each publication venue.
We find that 87.56\% of papers are published in peer-reviewed venues, and ICSE is the most popular venues that contains 26 papers, followed by ESEC/FSE (17 papers), ASE (13 papers), TOSEM (12 papers), and ISSTA (11 papers).
Besides, most top popular venues are top-tier conferences and symposiums, and only three journals have more than one relevant publication, i.e., TOSEM (12 papers), TSE (9 papers), EMSE (2 papers).
This finding indicates a preference for presenting the latest work at conferences due to the timeliness of conference proceedings. 
Although LLM-based APR research primarily appears in top-tier SE venues (e.g., ICSE, FSE, ASE, TOSEM, ISSTA), the field has also gained attention from the broader AI (e.g., NeurIPS, AAAI, and ICLR), Programming Language (e.g., OOPSLA and PLDI), and Security (e.g., S\&P, TDSC, and TIFS) communities. 
This interdisciplinary diffusion reflects the growing influence and popularity of LLM-based techniques in program repair.
We also find that the remaining 12.44\% of papers are non-peer-reviewed and published on arXiv.
This phenomenon can likely be attributed to the rapid emergence of relevant studies in a short development time.
Considering the varying quality levels of such non-peer-reviewed papers, we conducted a rigorous evaluation to ensure the inclusion of high-quality relevant papers in this work.}

\begin{figure}[t]
\centering
    \includegraphics[width=0.5\linewidth]{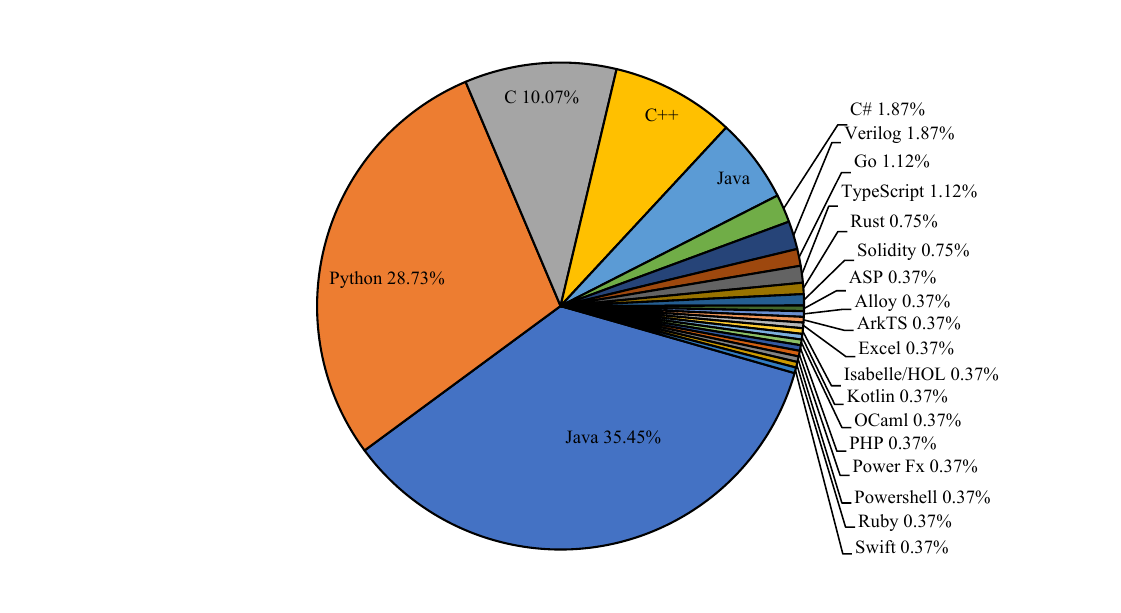}
    \caption{\revise{Distribution of LLM-based studies across different programming languages.}}
    \label{fig:pls}
\end{figure}

\subsection{What is the distribution of the programming languages?}

\revise{Fig.~\ref{fig:pls} presents the distribution of different programming languages targeted by existing LLM-based APR techniques.
We find that Java is the most widely utilized programming language, accounting for 35.45\% of all collected papers, followed by Python (28.73\%) and C (10.07\%).
The possible reason lies in the availability of mature datasets for these languages, which serve as recognized benchmarks for evaluating repair techniques, such as Defects4J~\cite{just2014defects4j},
QuixBugs~\cite{lin2017quixbugs}, BFP~\cite{tufano2019empirical}, CVEfixes~\cite{bhandari2021cvefixes}, Big-Vul~\cite{fan2020ac}, and TFix~\cite{berabi2021tfix}.

We also find that LLM-based APR encompasses a broader range of programming languages compared to traditional APR.
For example, our collected papers involve \plnum{} different programming languages in total, including both general-purpose languages (e.g., Java, Python, C, C++, JavaScript, Rust) and domain-specific languages (e.g., Verilog for hardware description and Solidity for smart contracts), whereas learning-based APR techniques~\cite{zhang2023survey} are typically limited to only five common languages (i.e., Java, JavaScript, Python, C, and C++).
Importantly, we notice that some rare languages, previously overlooked in the APR community, are now being addressed.
This broader language adaptability of LLM-based APR might stem from the inherent capabilities of LLMs to encapsulate general programming knowledge that can be transferred across multiple languages by fine-tuning~\cite{yuan2022circle}.
In particular, recent LLMs are much larger and are trained on more diverse datasets, allowing them to incorporate knowledge from many different languages~\cite{zhang2024lpr}.
Besides, LLMs' robust natural language understanding capabilities facilitate the few-shot or zero-shot repair settings with limited learning samples, which is a significant advantage over DL models~\cite{zhang2023survey} that typically require extensive repair corpora for training. 
Consequently, LLMs can efficiently handle lesser-known programming languages, like Verilog~\cite{ahmad2024hardware} and Rust~\cite{deligiannis2023fixing}, which are often underrepresented in previous APR research~\cite{monperrus2020living}.
This phenomenon highlights the promising prospects and scalability of APR brought about by recent LLMs.
}

\subsection{What are the types of publication contributions?}

\begin{table}[htbp]
  \centering
  \footnotesize
  \caption{\revise{Categories of four main contributions in collected studies.}}
    \begin{tabular}{lm{7cm}l}
    \toprule
    Category & Description & \multicolumn{1}{l}{Papers} \\
    \midrule
    New Technique or Methodology & The study proposes a novel approach to address specific challenges in the APR community, empowered by LLMs. & 118 \\
    Empirical Study & The study explores the performance of LLMs in fixing software bugs with a quantitative and qualitative analysis. & 56 \\
    Benchmark & The paper introduces a new benchmark specifically designed to evaluate the fix capabilities of LLMs for various bug types & 15 \\
    Human Study & The study conducts a survey to explore practitioners’ perspectives or experiences of LLMs when fixing software bugs. & 3 \\
    \bottomrule
    \end{tabular}%
  \label{tab:contribution}%
\end{table}%

We categorize collected papers according to their main contributions into four categories: new technique or methodology, empirical study, benchmark, and human study, as illustrated in Table~\ref{tab:contribution}.
\revisenew{It is worth noting that, although most studies are assigned to a single category, a small number of studies make equally substantial contributions to multiple categories.}
We find that 118 relevant papers are published, with the aim of proposing a novel repair approach or framework with LLMs to address various issues in the APR community. 
Besides, 56 papers concentrate on conducting empirical studies to explore the actual benefits of LLMs in fixing various bugs, such as the potential of fine-tuning LLMs in vulnerability repair~\cite{zhang2023pre,wu2023empirical}.
We further notice 15 relevant studies constructing new benchmarks to evaluate the performance of LLMs and three papers~\cite{wu2023empirical,meem2024exploring} administering a survey to offer insights into how practitioners or developers think about and employ LLMs to fix software bugs in practice.

\summary{1}{
(1) LLMs have shown a booming trend in fixing software bugs, with \papernum{} papers between 2020 and 2025.
(2) The number of conference papers employing LLMs for APR significantly exceeds that of journal papers, with ICSE and TOSEM being the most popular conference and journal venues, respectively.
(3) LLM-based APR papers are published in different research fields, including SE, AI, and Security.
(4) There are \plnum{} programming languages that LLM-based APR has been applied to, with Java, Python, C, and C++ being the most frequently targeted.
(5) LLMs have been applied to some underrepresented programming languages, such as Verilog and Rust.
(6) The vast majority of collected studies primarily focus on introducing new techniques and conducting empirical research, while two papers perform user studies to understand practitioners' attitudes and experiences regarding leveraging various LLMs for solving bug-fixing tasks.
}

\section{RQ2: Which popular LLMs have been applied to support APR?}
\label{sec:rq2}

\subsection{What is the distribution of involved LLMs?}
\label{sec:rq2_llms}
\begin{table}[htbp]
  \centering
  \caption{\revise{Distribution of existing LLMs utilized in APR}}
  \resizebox{0.98\linewidth}{!}{
    \begin{tabular}{lrrrrrrr||lrrrrrrr}
    \toprule
    \textbf{Model} & \textbf{2020} & \textbf{2021} & \textbf{2022} & \textbf{2023} & \textbf{2024} & \textbf{2025} & \multicolumn{1}{l||}{\textbf{Total}} & \textbf{Model} & \textbf{2020} & \textbf{2021} & \textbf{2022} & \textbf{2023} & \textbf{2024} & \textbf{2025} & \multicolumn{1}{l}{\textbf{Total}} \\
    \midrule
    GPT-3.5 & 0     & 0     & 1     & 24    & 32    & 15    & 72    & LLaMA & 0     & 0     & 0     & 1     & 2     & 0     & 3 \\
    GPT-4 & 0     & 0     & 0     & 9     & 27    & 15    & 51    & PaLM  & 0     & 0     & 0     & 2     & 1     & 0     & 3 \\
    CodeT5 & 0     & 0     & 1     & 16    & 12    & 2     & 31    & RoBERTa & 0     & 0     & 1     & 1     & 0     & 1     & 3 \\
    CodeLLaMA & 0     & 0     & 0     & 4     & 14    & 10    & 28    & T5    & 0     & 1     & 1     & 1     & 0     & 0     & 3 \\
    Codex & 0     & 0     & 2     & 12    & 6     & 2     & 22    & TFix  & 0     & 0     & 1     & 1     & 1     & 0     & 3 \\
    GPT-4o & 0     & 0     & 0     & 0     & 3     & 12    & 15    & BART  & 0     & 1     & 1     & 0     & 0     & 0     & 2 \\
    StarCoder & 0     & 0     & 0     & 0     & 12    & 3     & 15    & Bard  & 0     & 0     & 0     & 0     & 2     & 0     & 2 \\
    CodeBERT & 0     & 1     & 2     & 7     & 2     & 1     & 13    & Claude & 0     & 0     & 0     & 1     & 1     & 0     & 2 \\
    CodeGen & 0     & 0     & 0     & 3     & 7     & 2     & 12    & CodeGen-2 & 0     & 0     & 0     & 1     & 1     & 0     & 2 \\
    InCoder & 0     & 0     & 0     & 7     & 4     & 0     & 11    & CodeQwen1.5 & 0     & 0     & 0     & 0     & 1     & 1     & 2 \\
    PLBART & 0     & 0     & 0     & 7     & 3     & 0     & 10    & DeepSeekCoder-V2 & 0     & 0     & 0     & 0     & 1     & 1     & 2 \\
    GPT-4o-mini & 0     & 0     & 0     & 0     & 0     & 9     & 9     & Deepseek-v2.5 & 0     & 0     & 0     & 0     & 0     & 2     & 2 \\
    DeepseekCoder & 0     & 0     & 0     & 0     & 4     & 4     & 8     & Falcon & 0     & 0     & 0     & 1     & 1     & 0     & 2 \\
    Claude-3.5 & 0     & 0     & 0     & 0     & 1     & 6     & 7     & GPT-J & 0     & 0     & 0     & 2     & 0     & 0     & 2 \\
    GraphCodeBERT & 0     & 0     & 0     & 4     & 2     & 1     & 7     & GPT-Neo & 0     & 0     & 0     & 2     & 0     & 0     & 2 \\
    Llama-2 & 0     & 0     & 0     & 2     & 4     & 1     & 7     & Gemini-1.5 & 0     & 0     & 0     & 0     & 1     & 1     & 2 \\
    PolyCoder & 0     & 0     & 1     & 3     & 2     & 0     & 6     & Gemma & 0     & 0     & 0     & 0     & 0     & 2     & 2 \\
    UniXcoder & 0     & 0     & 0     & 4     & 0     & 1     & 5     & LLaMA-3.1 & 0     & 0     & 0     & 0     & 0     & 2     & 2 \\
    BERT  & 1     & 1     & 1     & 1     & 1     & 0     & 5     & LLaMA-3.3 & 0     & 0     & 0     & 0     & 0     & 2     & 2 \\
    CodeGPT & 0     & 0     & 2     & 1     & 1     & 0     & 4     & Phi 3 & 0     & 0     & 0     & 0     & 0     & 2     & 2 \\
    LLaMA-3 & 0     & 0     & 0     & 0     & 1     & 3     & 4     & Qwen-2.5 & 0     & 0     & 0     & 0     & 0     & 2     & 2 \\
    Mistral & 0     & 0     & 0     & 0     & 1     & 3     & 4     & Qwen2.5-Coder & 0     & 0     & 0     & 0     & 0     & 2     & 2 \\
    Claude-3 & 0     & 0     & 0     & 0     & 1     & 2     & 3     & SantaCoder & 0     & 0     & 0     & 0     & 2     & 0     & 2 \\
    CodeGemma & 0     & 0     & 0     & 0     & 1     & 2     & 3     & StarCoder2 & 0     & 0     & 0     & 0     & 1     & 1     & 2 \\
    CodeT5+ & 0     & 0     & 0     & 1     & 2     & 0     & 3     & WizardCoder & 0     & 0     & 0     & 0     & 1     & 1     & 2 \\
    \bottomrule
    \end{tabular}%
    }
  \label{tab:llms}%
\end{table}%

Plenty of LLMs, with millions to billions (even more) of parameters, have been proposed and adapted to automatically fix software bugs.
\delete{We analyze all collected papers and summarize 46 relevant LLMs that are already utilized in existing APR studies.
Fig.~\ref{fig:llms} illustrates the variety of different LLMs and lists the number of times these models have been utilized by prior APR studies.
We only include those LLMs with more than one time in collected papers due to page limitations.
As can be seen from Fig.~\ref{fig:llms}, ChatGPT (37)\footnote{In this paper, we follow prior LLM4SE studies~\cite{wang2024software,hou2023large} and refer to ChatGPT as the GPT-3.5-Turbo model released in November 2022, which served as the basis for most early LLM-based APR studies.} and GPT-4 (25) from OpenAI are the two most popular LLMs, followed by CodeT5 (23) and Codex (21).
We categorize these LLMs into three groups according to their architectures: encoder-only, encoder-decoder, and decoder-only LLMs.}
\revise{We analyze all collected papers and identify \llmnum{} distinct LLMs that are already utilized in existing APR studies.
Table~\ref{tab:llms} illustrates the variety of different LLMs and presents their usage counts across different years from 2020 to 2025.
We only include the top 50 most frequently used models due to page limitations.
It can be seen, GPT-3.5 (72)\footnote{We note that several prior LLM4SE studies~\cite{wang2024software,hou2023large} use the term ChatGPT to refer to the underlying GPT-3.5-Turbo model released in November 2022.
Since ChatGPT is technically a service interface rather than a specific model, to avoid ambiguity, we unify the terminology in this paper and refer to it simply as GPT-3.5.}
and GPT-4 (51) from OpenAI are the two most popular LLMs, followed by CodeT5 (31) and CodeLLaMA (28).
We categorize these LLMs into three groups according to their architectures: encoder-only, encoder-decoder, and decoder-only LLMs.}

\subsubsection{Encoder-only LLMs}
Encoder-only LLMs denote a category of LLMs that only utilize the encoder stack of the Transformer architecture.
Typically, these models are pre-trained on a massive corpus using a Masked Language Modeling (MLM) task, which is used to learn to predict the identity of masked words based on their context.
In the APR community, prominent models like CodeBERT (11)~\cite{feng2020codebert} and GraphCodeBERT (6)~\cite{guo2020graphcodebert} have been investigated in semantic bugs~\cite{xia2022less, mashhadi2021applying,hu2022fix,zirak2023improving} and security vulnerabilities~\cite{zhang2023pre}.
As such LLMs only contain an encoder component that is capable of generating context-aware representations for inputs, they are particularly suited for code understanding tasks, such as code search, while not directly applicable to code generation tasks, such as program repair.
Thus, in the APR community, as Zhang~et al.~\cite{zhang2023pre} mention, researchers may need to integrate a new decoder that initializes from scratch to the pre-trained encoder to construct an encoder-decoder architecture for patch generation.
We also notice such encoder-only LLMs are utilized to identify patch correctness~\cite{tian2020evaluating,tian2022change,tian2023best,le2023invalidator,zhou2024leveraging,zhang2024appt}, as discussed in Section~\ref{sec:apca}.
    
\subsubsection{Enocoder-decoder LLMs}
Encoder-decoder LLMs denote a category of LLMs that utilize both the encoder and decoder stacks of the Transformer architecture, thus inherently suitable for transforming one sequence into another.
Particularly, the encoder takes one sequence as input and encodes it into a fixed-size hidden state, which effectively captures the semantics and meaning of the input sequence.
Then the decoder processes the hidden state and produces the corresponding output sequence using attention mechanisms to refer back to parts of the input sequence as needed.
Thanks to the encoder-decoder architecture, such LLMs are particularly suited for code generation tasks in a sequence-to-sequence learning setting, such as program repair.
In the APR community, a mass of encoder-decoder LLMs have been widely adopted, such as CodeT5 (23)~\cite{wang2021codet5}, PLBART (10)~\cite{ahmad2021unified_plbart}, UniXcoder (4)~\cite{guo2022unixcoder} and T5 (3)~\cite{raffel2020exploring}.
Similar to traditional learning-based APR~\cite{zhang2023survey}, such studies usually treat APR as a \delete{natural}\revise{neural} machine translation (NMT) task by supervised sequence-to-sequence learning, such as CIRCLE~\cite{yuan2022circle}, TFix~\cite{berabi2021tfix}, VulRepair~\cite{fu2022vulrepair} and RAP-Gen~\cite{wang2023rap}.

\subsubsection{Decoder-only LLMs}
Decoder-only LLMs denote a category of LLMs that utilize only the decoder stack of the Transformer architecture.
Decoder-only LLMs are typically pre-trained using a causal language modeling (CLM) objective, learning to predict the next word in a sentence based on the preceding words.
These models are specifically designed for generating sequences of text autoregressively, i.e., producing one token at a time and using what has been generated so far as context for subsequent tokens.
In the APR community, decoder-only LLMs are the most popular and widely used group, compared with encoder-only and encoder-decoder LLMs.
\delete{Notable repair applications of decoder-only LLMs include GPT-series models (e.g., GPT-1~\cite{gpt-1}, GPT-2~\cite{radford2019language_gpt2}, GPT3~\cite{brown2020language_gpt3}, GPT-3.5~\cite{gpt3.5}, ChatGPT~\cite{chatgpt}, and GPT-4~\cite{gpt4}), some open-sourced models, (e.g., CodeGPT~\cite{lu2021codexglue}, GPT-Neo~\cite{gpt-neo}, GPT-NeoX~\cite{gpt-neox-20b}, GPT-J~\cite{gpt-j}, InCoder~\cite{fried2023incoder}, CodeGen~\cite{nijkamp2023codegen}, CodeLLaMA~\cite{roziere2023code} and StarCoder~\cite{li2023starcoder}), as well as some closed-source models (e.g., Codex~\cite{chen2021evaluating} in Fan~et al.~\cite{fan2023automated} and CEDAR~\cite{nashid2023retrieval}).}
\revise{Notable repair applications of decoder-only LLMs include a variety of open-sourced models, (e.g., GPT-1~\cite{gpt-1}, GPT-2~\cite{radford2019language_gpt2}, CodeGPT~\cite{lu2021codexglue}, GPT-Neo~\cite{gpt-neo}, GPT-NeoX~\cite{gpt-neox-20b}, GPT-J~\cite{gpt-j}, InCoder~\cite{fried2023incoder}, CodeGen~\cite{nijkamp2023codegen}, CodeLLaMA~\cite{roziere2023code} and StarCoder~\cite{li2023starcoder}), and closed-source models (e.g., GPT3~\cite{brown2020language}, GPT-3.5~\cite{gpt3.5}, \delete{ChatGPT~\cite{chatgpt}}, GPT-4~\cite{gpt4} and Codex~\cite{chen2021evaluating}), categorized according to their accessibility.}
The emergence of decoder-only LLM-based APR studies is primarily due to \delete{two}\revise{three} reasons.
The first reason is that these models can naturally perform program repair from a few examples or simple instructions without any fine-tuning.
The second reason is the recent surge in decoder-only LLMs, marked by the introduction of commercial products by leading Internet companies, such as \delete{ChatGPT}\revise{GPT-3.5} and GPT-4 by OpenAI.
\revise{The third reason is that decoder-only LLMs primarily rely on unsupervised learning during the pre-training phase, allowing them to exploit large-scale unlabeled text and code corpora for scalable and powerful language modeling, unlike encoder-based models that depend more on labeled data.}

\subsubsection{Trend Analysis}

\revise{
Table~\ref{tab:llms} reveals several clear trends in the evolution of LLM utilization for program repair from 2020 to 2025.
First, there is a rapid shift toward large-scale decoder-only models, particularly those developed by OpenAI such as GPT-3.5 and GPT-4, which together account for over 120 studies. 
This dominance highlights the strong influence of general-purpose conversational models on APR research and reflects a growing interest in zero-shot and few-shot prompting.
Second, the diversity of LLMs has expanded significantly since 2023, driven by the release of many open-source alternatives (e.g., CodeLLaMA, StarCoder, DeepSeekCoder).
This trend indicates the research community's increasing emphasis on transparency, reproducibility, and cost-effective experimentation.
Third, encoder-decoder models such as CodeT5, PLBART, and UniXcoder continue to play an important role, particularly in studies that rely on fine-tuning for task-specific adaptation. Although their overall proportion among newly proposed APR approaches has decreased in recent years, they remain foundational for sequence-to-sequence repair tasks requiring explicit supervision.
Finally, the emergence of agent-aware and multimodal LLMs (e.g., GPT-4o and Claude-3.5) during 2024–2025 represents a new phase emphasizing interactive, repository-level, and tool-integrated APR. 
These models are being actively explored for autonomous reasoning and end-to-end repair, signaling a transition toward dynamic and environment-aware repair workflows.

Overall, the trend demonstrates a gradual evolution from specialized fine-tuned encoder-decoder architectures toward general-purpose and prompt-driven LLMs, reflecting the field's continuous pursuit of scalability, flexibility, and autonomy in APR.
}

\subsection{What approaches are employed to optimize LLMs for program repair?}
\label{sec:three_ways}

\begin{table}[t]
\footnotesize
  \centering
    \caption{\revise{Breakdown of LLM utilization strategies for APR}}
    \begin{tabular}{crrp{10cm}}
    \toprule
    Approach & Year & Count & Studies \\
    \midrule
    \multirow{5}[2]{*}{\makecell{Fine-tuning\\(55 papers)}} & 2021  & 4     & \cite{berabi2021tfix}, \cite{jiang2021cure}, \cite{drain2021generating}, \cite{mashhadi2021applying} \\
          & 2022  & 8     & \cite{ahmed2022synshine}, \cite{fu2022vulrepair}, \cite{li2022dear}, \cite{garg2022deepdev}, \cite{yuan2022circle}, \cite{lajko2022towards}, \cite{hu2022fix}, \cite{kim2022empirical} \\
          & 2023  & 17    & \cite{fu2023vision}, \cite{fu2023llm4sechw}, \cite{huang2023empirical}, \cite{jin2023inferfix}, \cite{wu2023effective}, \cite{wang2023rap}, \cite{zhang2023pre}, \cite{zirak2023improving}, \cite{moon2023coffee}, \cite{xia2023plastic}, \cite{horvath2023extensive}, \cite{jiang2023impact}, \cite{koutcheme2023training}, \cite{paul2023enhancing}, \cite{zhang2023neural}, \cite{liu2023automated}, \cite{zhang2023evaluating} \\
          & 2024  & 13    & \cite{chow2024pyty}, \cite{li2024exploring}, \cite{liu2024fastfixer}, \cite{nong2024chain}, \cite{saboor2024automated}, \cite{zhang2024vuladvisor}, \cite{zhou2024large}, \cite{wang2024divide}, \cite{gharibi2024t5apr}, \cite{hossain2024deep}, \cite{lin2024one}, \cite{zhao2024repair}, \cite{prenner2023out} \\
          & 2025  & 13    & \cite{brancas2025combining}, \cite{dai2025less}, \cite{ma2025swe}, \cite{wong2024investigating}, \cite{gao2025teaching}, \cite{wei2025swe}, \cite{feng2025integrating}, \cite{luo2024fine}, \cite{machavcek2025impact}, \cite{ruiz2025art}, \cite{yang2024multi}, \cite{huang2025template}, \cite{silva2023repairllama} \\
    \midrule
    \multirow{3}[2]{*}{\makecell{Few-shot\\(24 papers)}} & 2023  & 8     & \cite{madaan2023selfrefine}, \cite{huang2023chain}, \cite{joshi2023repair}, \cite{nashid2023retrieval}, \cite{mohajer2023skipanalyzer}, \cite{fakhoury2023towards}, \cite{gao2023makes}, \cite{xia2023automated} \\
          & 2024  & 10    & \cite{fatima2024flakyfix}, \cite{guo2024codeeditorbench}, \cite{hao2024retyper}, \cite{zhao2024enhancing}, \cite{lin2024drplanner}, \cite{liu2023chatgpt}, \cite{zhang2024pydex}, \cite{xia2024automated}, \cite{xiang2024far}, \cite{yin2024thinkrepair} \\
          & 2025  & 6     & \cite{ehsani2025towards}, \cite{liu2025error}, \cite{orvalho2025counterexample}, \cite{nong2025appatch}, \cite{lin2025haprepair}, \cite{zhang2025repair} \\
    \midrule
    \multirow{4}[2]{*}{\makecell{Zero-shot\\(54 papers)}} & 2022  & 5     & \cite{xia2022less}, \cite{kolak2022patch}, \cite{prenner2022can}, \cite{ribeiro2022framing}, \cite{sobania2023analysis} \\
          & 2023  & 17    & \cite{du2023resolving}, \cite{fan2023automated}, \cite{first2023baldur}, \cite{pearce2023examining}, \cite{tol2023zeroleak}, \cite{wu2023exploring}, \cite{zhao2023right}, \cite{napoli2023evaluating}, \cite{zhang2023gamma}, \cite{ribeiro2023gpt}, \cite{koutcheme2023automated}, \cite{wei2023copiloting}, \cite{liventsev2023fully}, \cite{zhang2023steam}, \cite{tian2023chatgpt}, \cite{wuisang2023evaluation}, \cite{zhang2023critical} \\
          & 2024  & 15    & \cite{ahmad2024hardware}, \cite{alhanahnah2025empirical}, \cite{pan2024lost}, \cite{peng2024domain}, \cite{tang2024code}, \cite{tian2024debugbench}, \cite{tsai2023rtlfixer}, \cite{wadhwa2023frustrated}, \cite{yang2024cref}, \cite{chen2024large}, \cite{zheng2024opencodeinterpreter}, \cite{zhong2024ldb}, \cite{ruiz2024novel}, \cite{yang2024revisiting}, \cite{yan2023codescope} \\
          & 2025  & 17    & \cite{deligiannis2023fixing}, \cite{garg2023rapgen}, \cite{huang2025seeing}, \cite{mundler2024swt}, \cite{charalambous2023new}, \cite{xu2023guiding}, \cite{xia2025demystifying}, \cite{yuan2024designrepair}, \cite{zhang2024acfix}, \cite{steenhoek2025closing}, \cite{cao2023study}, \cite{hu2025aprmcts}, \cite{kong2024contrastrepair}, \cite{li2025hybrid}, \cite{ouyang2025knowledge}, \cite{parasaram2024fact}, \cite{kang2023explainable} \\
    \midrule
    \multirow{2}[2]{*}{\makecell{Agent\\(15 papers)}} & 2024  & 5     & \cite{arora2024masai}, \cite{chen2024coder}, \cite{yang2024swe}, \cite{zhang2024autocoderover}, \cite{liu2024marscode} \\
          & 2025  & 10    & \cite{ke2025niodebugger}, \cite{antoniades2024swe}, \cite{ouyang2024repograph}, \cite{tao2024magis}, \cite{wang2024openhands}, \cite{yu2025patchagent}, \cite{bouzenia2024repairagent}, \cite{lee2024unified}, \cite{ma2025alibaba}, \cite{ye2025adverintent} \\
    \bottomrule
    \end{tabular}%
  \label{tab:llm_ways}%
\end{table}%

\delete{LLMs typically acquire general knowledge from extensive datasets.
Thus, a fundamental research issue arises when integrating off-the-shelf LLMs with APR: \textit{how to adapt general-propose LLMs to the specific program repair task}.
Fig.~\ref{fig:three_ways} displays the prevalence of three common adaptation strategies in LLM-based APR research: fine-tuning, few-shot learning, and zero-shot learning. 
Our findings indicate that zero-shot learning, employed in 48\% of the studies, is the most popular approach, suggesting a trend towards using LLMs as-is for program repair tasks. 
Meanwhile, fine-tuning is utilized in 37\% of the cases, followed by few-shot learning at 15\%.}
\revise{LLMs typically acquire general knowledge from extensive datasets.
Thus, when integrating off-the-shelf LLMs into APR, a fundamental research issue arises: \textit{how to adapt general-purpose LLMs to the specific program repair task}.
Table~\ref{tab:llm_ways} displays the prevalence of four common adaptation strategies in LLM-based APR research: fine-tuning, few-shot prompting, zero-shot prompting, and agent. 
Among them, fine-tuning (55 papers) and zero-shot prompting (54 papers) are the two most prevalent strategies, together accounting for more than two-thirds of all studies.
Meanwhile, few-shot prompting is utilized in 24 papers, followed by agent in 15 papers.}

\subsubsection{Fine-tuning}
Fine-tuning refers to a process where LLMs are further trained on a smaller, task-specific dataset.
This is an intuitive way to allow LLMs to adjust their weights and biases through supervised learning, enabling them to perform as expected on new tasks that are similar to, but not exactly the same as, those they were initially trained on.
In the APR community, fine-tuning is widely utilized during the early emergence of LLMs with millions of parameters, such as T5 and CodeT5, as it can significantly improve performance on the target program repair task without the need to train an LLM from scratch.

We summarize the existing APR studies via fine-tuning LLMs into three stages.
First, researchers directly regard program repair as a downstream task by specific datasets, such as fine-tuning T5~\cite{berabi2021tfix}, CodeT5~\cite{fu2022vulrepair}, CodeBERT~\cite{mashhadi2021applying}, and GPT-2~\cite{lajko2022towards}.
Second, researchers utilize more advanced fine-tuning strategies for better performance.
For example, CIRCLE utilizes continual learning to repair multiple languages with a single model, and RAP-Gen~\cite{wang2023rap} utilizes retrieval-augmented generation to guide patch search space.
Recently, Zirak~et al.~\cite{zirak2023improving} empirically explore the domain shift problem in APR with two LLMs, i.e., TFix and CodeBERT, and three fine-tuning methods, i.e., Full-Fine-Tuning, Tuning-With-Light-Weight-Adapter-Layers, and Curriculum-Learning.
Third, researchers conduct empirical studies to explore the actual fix capabilities of various LLMs in different repair scenarios.
For example, Zhang~et al.~\cite{zhang2023pre} fine-tune five LLMs to repair C/C++ security vulnerabilities, Wu~et~al.~\cite{wu2023effective} involve four LLMs in Java vulnerabilities, and Jiang~et al.~\cite{jiang2023impact} consider four LLMs on Java semantic bugs.

\subsubsection{\delete{Few-shot Learning}\revise{Few-shot Prompting}}

\delete{Few-shot learning}\revise{Few-shot prompting} refers to the ability of LLMs to learn or adapt to new tasks with a very limited amount of data—often only a few examples.
This is an effective way to use examples to help LLMs understand the targeted task and generate appropriate responses without any explicit retraining or fine-tuning.
In the APR community, \delete{few-shot learning}\revise{few-shot prompting} is usually utilized and impressive in LLMs, particularly with billions of parameters, as it requires LLMs' powerful ability to generalize from very limited data.
Researchers typically provide LLMs with a small number of repair examples directly in the input prompt and require LLMs to generate correct patches. 
For example, Nashid~et al.~\cite{nashid2023retrieval} construct effective prompts by retrieving similar repair demonstrations for CodeX, and Xia~et al.~\cite{xia2023automated} provide LLMs with examples from the same buggy project to learn the coding style.

\subsubsection{Zero-shot \delete{Learning}\revise{Prompting}}
Zero-shot \delete{learning}\revise{prompting} takes the concept of few-shot \delete{learning}\revise{prompting} even further by requiring LLMs to perform program repair without any explicit examples. 
This is a recently popular way to query LLMs to perform a variety of unseen tasks, where LLMs are given a task description and must use their pre-existing knowledge and understanding to generate a response or solution.
In the APR community, zero-shot rapidly emerges following the advent of LLMs with record-breaking parameters, exemplified by ChatGPT, as it requires a powerful foundation model to perform human-like chats.

There are two typical development routes that utilize LLMs for program repair in a zero-shot \delete{learning}\revise{prompting} setting.
The first one is \textbf{cloze-style repair}, i.e., reframing program repair as a cloze-style task, and then invoking LLMs to predict partially correct code with the help of repair patterns, such as AlphaRepair~\cite{xia2022less}, GAMMA~\cite{zhang2023gamma}, FitRepair~\cite{xia2023plastic} and Repilot~\cite{wei2023copiloting}.
The second one is \textbf{conversational-based repair}, i.e., constructing complex prompts with various valuable information (e.g., buggy code, failure diagnostics, even execution feedback), and then chatting with LLMs to generate correct patches, such as Pearce~et al.~\cite{pearce2023examining}, TypeFix~\cite{peng2024domain}, RustAssistant~\cite{deligiannis2023fixing}, Zhang~et al.~\cite{zhang2023critical}, Prenner~et al.~\cite{prenner2022can}, Sobania~et al.~\cite{sobania2023analysis}, and Napoli~et al.~\cite{napoli2023evaluating}.
Such repair routes usually require LLMs capable of processing long-text prompts and human-like conversations, thus predominantly employing powerful LLMs with billion-level parameters, like ChatGPT and GPT-4.
Besides, zero-shot gets rid of training datasets, thus generalizing to various repair scenarios where gathering training data is challenging or impossible, such as hardware bugs~\cite{ahmad2024hardware}, DL programs~\cite{cao2023study} and crash bugs~\cite{du2023resolving}.

\subsubsection{Agent}
An Agent refers to an autonomous system capable of perceiving its environment, reasoning about its current state, and taking a sequence of actions to achieve specific goals.
Unlike prior prompting-based APR techniques, which hard-code the repair process into a static, pre-defined pipeline, LLM-based agents allow dynamic adaptation based on intermediate outcomes,  contextual feedback.
In the repair process, an LLM-based agent equips the underlying LLM with a set of external tools (e.g., compilers, debuggers, and test runners), enabling it to interact with and modify its environment.
This flexibility allows LLMs to iteratively plan, execute, and refine repair actions, forming a closed-loop repair workflow.
In the APR community, LLM-based agent research can be broadly divided into two directions.

The first direction focuses on enhancing the effectiveness of prompting-based APR methods on traditional benchmarks such as Defects4J and QuixBugs. 
Representative work includes RepairAgent~\cite{bouzenia2024repairagent}, which is the first autonomous LLM-based repair agent. 
RepairAgent integrates ChatGPT with a state machine to manage iterative debugging actions, enabling it to autonomously plan repair steps, select tools, and refine patches until a correct fix is achieved.
Similarly, FixAgent proposes a multi-agent framework for end-to-end unified debugging, which mimics the cognitive debugging process by structuring agent roles across three levels of complexity.
Each level activates a subset of agents specialized in tasks such as summarization, slicing, localization, and patch generation.

The second category involves repository-level repair agents, which target real-world software issues, guided by the development of new benchmarks, such as SWE-bench~\cite{jimenez2024swebench}.
Unlike traditional approaches~\cite{zhang2023survey} that often rely on predefined conditions (e.g., assuming known bug locations, failure-triggering tests, or single-function repair assumptions), these agents require end-to-end bug repair across full repositories, requiring reasoning across multiple files, dependencies, and test suites. 
Recent studies have proposed a series of SWE-oriented techniques to address the challenges of repository-level repair, including designing specialized agent-computer interfaces (e.g., SWE-Agent~\cite{yang2024swe}), encoding the whole repository via AST representations (e.g., AutoCodeRover~\cite{zhang2024autocoderover}), enabling multi-agent collaboration (e.g., MAGIS~\cite{tao2024magis}), and developing generalist and specialist agents (e.g., OpenHands~\cite{wang2024openhands}).
\revisenew{More recently, unlike most prior agent-based approaches that rely on offline-trained LLMs and fixed prompting strategies, the agent-based paradigm opens new opportunities for incorporating online learning into APR systems~\cite{mu2025experepair,xia2025live}
, enabling agents to continuously refine their behavior during the repair process in evolving software projects.}

\subsubsection{Trend analysis}
\revise{As shown in Table~\ref{tab:llm_ways}, he evolution of LLM-based program repair strategies reflects the continuous advancement of LLM capabilities and adaptation techniques.
Fine-tuning, which uses supervised learning to adapt LLMs more closely to the specifics of program repair, becomes popular with the emergence of early million-parameter-level models like CodeT5.
Few-shot prompting demonstrates LLMs' capability to generalize from a few examples, popular with the emergence of subsequent billion-parameter-level models like Codex.
Zero-shot prompting showcases LLMs' capability to tackle the program repair task without any prior direct exposure to training or examples, particularly with the emergence of recent models with tens or hundreds of billions of parameters like ChatGPT and GPT-4.
The recent introduction of agent-based APR represents a paradigm shift toward autonomous repair, where LLMs can plan, execute, and refine repair actions dynamically using tool-assisted feedback loops.
Overall, these approaches illustrate a clear trend: as LLMs become more capable, adaptation moves from explicit training toward dynamic reasoning and autonomous interaction, broadening the applicability of LLMs in APR.
}

\summary{2}{
(1) We summarize \llmnum different LLMs already utilized to fix bugs, and these LLMs can be classified into three categories based on model architectures, i.e., encoder-only, encoder-decoder, and decoder-only.
(2) Decoder-only LLMs are the most frequently utilized model architecture, and four of the top popular LLMs are decoder-only models.
(3) \delete{ChatGPT, GPT-4, CodeT5, and Codex are the most popular LLMs in existing LLM-based APR studies, utilized 37, 25, 23, and 21 times, respectively.}\revise{ChatGPT, GPT-4, CodeT5, and CodeLLaMA are the most popular LLMs in existing LLM-based APR studies, utilized 72, 52, 31, and 28 times, respectively.}
(4) We summarize \delete{three}\revise{four} typical ways of leveraging the vast knowledge encapsulated in LLMs for the specific program repair task, i.e., fine-tuning, few-shot, \delete{and zero-shot}\revise{zero-shot, and agents}.
}

\section{RQ3: What repair scenarios have been facilitated by LLMs?}
\label{sec:rq3_repair_scenarios}

\begin{table}[htbp]
\footnotesize
\renewcommand{\arraystretch}{0.8}
  \centering
  \caption{\revise{Distribution of LLM-based APR papers across different bug types}}
    \resizebox{\linewidth}{!}{
    \begin{tabular}{lllp{10cm}}
    \toprule
    \textbf{Bug Type} & \textbf{Year} & \textbf{Count} & \textbf{Studies} \\
    \midrule
    \multicolumn{1}{l}{\multirow{6}[2]{*}{Semantic Bug}} & 2020  & 1     & \cite{tian2020evaluating} \\
    \multicolumn{1}{l}{} & 2021  & 4     & \cite{jiang2021cure} \cite{tian2022change} \cite{drain2021generating} \cite{mashhadi2021applying} \\
    \multicolumn{1}{l}{} & 2022  & 9     & \cite{li2022dear} \cite{xia2022less} \cite{yuan2022circle} \cite{kolak2022patch} \cite{lajko2022towards} \cite{hu2022fix} \cite{prenner2022can} \cite{ribeiro2022framing} \cite{sobania2023analysis} \\
    \multicolumn{1}{l}{} & 2023  & 28    & \cite{huang2023empirical} \cite{hao2023enhancing} \cite{joshi2023repair} \cite{le2023invalidator} \cite{wu2023exploring} \cite{tian2022best} \cite{wang2023rap} \cite{zhao2023right} \cite{zirak2023improving} \cite{nashid2023retrieval} \cite{moon2023coffee} \cite{fakhoury2023towards} \cite{xia2023plastic} \cite{zhang2023gamma} \cite{gao2023makes} \cite{horvath2023extensive} \cite{jiang2023impact} \cite{koutcheme2023automated} \cite{koutcheme2023training} \cite{paul2023enhancing} \cite{zhang2023neural} \cite{wei2023copiloting} \cite{liu2023automated} \cite{liventsev2023fully} \cite{zhang2023steam} \cite{tian2023chatgpt} \cite{wuisang2023evaluation} \cite{xia2023automated} \\
    \multicolumn{1}{l}{} & 2024  & 18    & \cite{chen2024large} \cite{gharibi2024t5apr} \cite{hidvegi2024cigar} \cite{hossain2024deep} \cite{lin2024one} \cite{ouyang2024benchmarking} \cite{zhou2024leveraging} \cite{ruiz2024novel} \cite{shi2024code} \cite{silva2024gitbug} \cite{xia2024automated} \cite{xiang2024far} \cite{xue2024exploring} \cite{yang2024revisiting} \cite{yin2024thinkrepair} \cite{zhang2024appt} \cite{prenner2023out} \cite{yan2023codescope} \\
    \multicolumn{1}{l}{} & 2025  & 22    & \cite{bouzenia2024repairagent} \cite{cao2023study} \cite{ehsani2025bug} \cite{feng2025integrating} \cite{hu2025aprmcts} \cite{kong2024contrastrepair} \cite{kong2025demystifying} \cite{lee2024unified} \cite{li2025hybrid} \cite{lin2025haprepair} \cite{luo2024fine} \cite{machavcek2025impact} \cite{ouyang2025knowledge} \cite{parasaram2024fact} \cite{ruiz2025art} \cite{xu2025aligning} \cite{yang2024multi} \cite{ye2025adverintent} \cite{huang2025template} \cite{zhang2025repair} \cite{kang2023explainable} \cite{silva2023repairllama} \\
    \midrule
    \multicolumn{1}{l}{\multirow{3}[2]{*}{Programming Problem}} & 2023  & 4     & \cite{fan2023automated} \cite{haque2023fixeval} \cite{liu2024refining} \cite{zhang2023critical} \\
    \multicolumn{1}{l}{} & 2024  & 13    & \cite{guo2024codeeditorbench} \cite{ishizue2024improved} \cite{li2024exploring} \cite{liu2024fastfixer} \cite{omari2024investigating} \cite{tang2024code} \cite{tian2024debugbench} \cite{wan2024automated} \cite{wu2023condefects} \cite{yang2024cref} \cite{zhang2024pydex} \cite{zhao2024peer} \cite{zhao2024repair} \\
    \multicolumn{1}{l}{} & 2025  & 7     & \cite{brancas2025combining} \cite{dai2025less} \cite{orvalho2025counterexample} \cite{wong2024investigating} \cite{kong2024contrastrepair} \cite{lee2024unified} \cite{xu2025aligning} \\
    \midrule
    \multicolumn{1}{l}{\multirow{4}[2]{*}{Security Vulnerability}} & 2022  & 1     & \cite{fu2022vulrepair} \\
    \multicolumn{1}{l}{} & 2023  & 7     & \cite{fu2023vision} \cite{pearce2023examining} \cite{tol2023zeroleak} \cite{huang2023empirical} \cite{wu2023effective} \cite{wu2023exploring} \cite{zhang2023pre} \\
    \multicolumn{1}{l}{} & 2024  & 7     & \cite{kulsum2024case} \cite{le2024study} \cite{liu2023chatgpt} \cite{nong2024chain} \cite{zhang2024vuladvisor} \cite{zhou2024large} \cite{lin2024one} \\
    \multicolumn{1}{l}{} & 2025  & 6     & \cite{charalambous2023new} \cite{gao2025teaching} \cite{nong2025appatch} \cite{steenhoek2025closing} \cite{yu2025patchagent} \cite{huang2025template} \\
    \midrule
    \multicolumn{1}{l}{\multirow{2}[2]{*}{Repository-level Issue}} & 2024  & 7     & \cite{arora2024masai} \cite{chen2024coder} \cite{jimenez2024swebench} \cite{yang2024swe} \cite{zhang2024autocoderover} \cite{liu2024marscode} \cite{zhao2024enhancing} \\
    \multicolumn{1}{l}{} & 2025  & 11    & \cite{ehsani2025towards} \cite{guo2025omnigirl} \cite{ma2025swe} \cite{mundler2024swt} \cite{antoniades2024swe} \cite{ouyang2024repograph} \cite{wei2025swe} \cite{xia2025demystifying} \cite{tao2024magis} \cite{wang2024openhands} \cite{ma2025alibaba} \\
    \midrule
    \multicolumn{1}{l}{\multirow{4}[2]{*}{Static Warning}} & 2021  & 1     & \cite{berabi2021tfix} \\
    \multicolumn{1}{l}{} & 2022  & 1     & \cite{kim2022empirical} \\
    \multicolumn{1}{l}{} & 2023  & 5     & \cite{alrashedy2023can} \cite{jin2023inferfix} \cite{wang2023rap} \cite{zirak2023improving} \cite{mohajer2023skipanalyzer} \\
    \multicolumn{1}{l}{} & 2024  & 2     & \cite{wadhwa2023frustrated} \cite{wadhwa2024core} \\
    \midrule
    \multicolumn{1}{l}{\multirow{4}[2]{*}{Syntax Error}} & 2022  & 1     & \cite{ahmed2022synshine} \\
    \multicolumn{1}{l}{} & 2023  & 4     & \cite{huang2023chain} \cite{huang2023empirical} \cite{joshi2023repair} \cite{zhao2023right} \\
    \multicolumn{1}{l}{} & 2024  & 1     & \cite{ruiz2024novel} \\
    \multicolumn{1}{l}{} & 2025  & 2     & \cite{deligiannis2023fixing} \cite{lin2025haprepair} \\
    \midrule
    \multicolumn{1}{l}{\multirow{2}[2]{*}{Hardware Bug}} & 2023  & 1     & \cite{fu2023llm4sechw} \\
    \multicolumn{1}{l}{} & 2024  & 4     & \cite{ahmad2024hardware} \cite{liu2024craftrtl} \cite{tsai2023rtlfixer} \cite{yao2024hdldebugger} \\
    \midrule
    \multicolumn{1}{l}{\multirow{2}[2]{*}{Test Case}} & 2024  & 2     & \cite{fatima2024flakyfix} \cite{saboor2024automated} \\
    \multicolumn{1}{l}{} & 2025  & 2     & \cite{ke2025niodebugger} \cite{xu2023guiding} \\
    \midrule
    \multicolumn{1}{l}{\multirow{2}[2]{*}{Type Error}} & 2023  & 1     & \cite{ribeiro2023gpt} \\
    \multicolumn{1}{l}{} & 2024  & 3     & \cite{chow2024pyty} \cite{hao2024retyper} \cite{peng2024domain} \\
    \midrule
    GUI Bug & 2025  & 3     & \cite{huang2025seeing} \cite{yang2024swem} \cite{yuan2024designrepair} \\
    \midrule
    Code Review & 2024  & 2     & \cite{guo2023exploring} \cite{wang2024divide} \\
    \midrule
    \multicolumn{1}{l}{\multirow{2}[2]{*}{Performance Bug}} & 2022  & 1     & \cite{garg2022deepdev} \\
    \multicolumn{1}{l}{} & 2025  & 1     & \cite{garg2023rapgen} \\
    \midrule
    \multicolumn{1}{l}{\multirow{2}[2]{*}{Smart Contract}} & 2023  & 1     & \cite{napoli2023evaluating} \\
    \multicolumn{1}{l}{} & 2025  & 1     & \cite{zhang2024acfix} \\
    \midrule
    API Misuse & 2023  & 1     & \cite{zhang2023evaluating} \\
    \midrule
    Crash Bug & 2023  & 1     & \cite{du2023resolving} \\
    \midrule
    Error-handling Bug & 2025  & 1     & \cite{liu2025error} \\
    \midrule
    Formal Proof & 2023  & 1     & \cite{first2023baldur} \\
    \midrule
    Formal Specification & 2024  & 1     & \cite{alhanahnah2025empirical} \\
    \midrule
    Motion Planner & 2024  & 1     & \cite{lin2024drplanner} \\
    \midrule
    Translation Bug & 2024  & 1     & \cite{pan2024lost} \\
    \bottomrule
    \end{tabular}%
    }
  \label{tab:bug_types}%
\end{table}%

\delete{In this section, we conduct a comprehensive analysis of the utilization of LLMs in various repair scenarios. 
We categorize \bugnum{} existing repair scenarios addressed with LLMs, following the taxonomy from Monperrus~et al.~\cite{monperrus2020living} and Zhang~et al.~\cite{zhang2023survey}, including semantic bugs, syntax bugs, and static warnings.
Fig.~\ref{fig:bug_types} presents the distribution of repair studies across different repair scenarios that LLMs are applied to.
We find that semantic bugs triggered by test suites have attracted the highest attention from the LLM-based APR research, constituting approximately 48\% of the total research volume.
This phenomenon is also observed in traditional learning-based APR~\cite{zhang2023survey}, and is mainly driven by well-known benchmarks, such as Defects4J~\cite{just2014defects4j}.
Security vulnerabilities account for about 14\% of the research proportion, highlighting the significance of LLMs in addressing cybersecurity challenges.
The third highest number of studies is observed in the programming problems domain, constituting approximately 9\% of the total research volume.
We also find a growing interest in rare bug types that are usually ignored by prior work, such as static warnings (7\%), syntax errors (6\%), and hardware bugs (3\%).
This underscores that, thanks to LLMs' general knowledge gleaned from vast amounts of data, researchers have begun to explore repair scenarios not previously addressed in prior works.}

\revise{In this section, we conduct a comprehensive analysis of the utilization of LLMs in various repair scenarios. 
We categorize \bugnum{} existing repair scenarios addressed with LLMs, following the taxonomy from Monperrus~et al.~\cite{monperrus2020living} and Zhang~et al.~\cite{zhang2023survey}, including semantic bugs, syntax bugs, and static warnings.
Table~\ref{tab:bug_types} presents the distribution of repair studies across different repair scenarios to which LLMs are applied.
We find that semantic bugs triggered by test suites have attracted the highest attention from the LLM-based APR research, constituting approximately 42.93\% of the total research volume.
This phenomenon is also observed in traditional learning-based APR~\cite{zhang2023survey}, and is mainly driven by well-known benchmarks, such as Defects4J~\cite{just2014defects4j}.
The second highest number of studies is observed in the programming problems domain, constituting approximately 12.57\% of the total research volume.
Security vulnerabilities account for about 10.99\% of the research proportion, highlighting the significance of LLMs in addressing cybersecurity challenges.
We also find a growing interest in rare bug types that are usually ignored by prior work, such as repository-level issues (9.42\%), static warnings (4.71\%), syntax errors (4.19\%), and hardware bugs (2.62\%).
This underscores that, thanks to LLMs' general knowledge gleaned from vast amounts of data, researchers have begun to explore repair scenarios not previously addressed in prior works.}

\subsection{Semantic Bugs}

\begin{table}[t]
\footnotesize
  \centering
  \caption{\revise{A Summary of representative LLM-based technical studies on semantic bugs}}
  \resizebox{\textwidth}{!}{
    \begin{tabular}{lllllp{1cm}}
    \toprule
    {Year} & Technique & {Venue} & {Paper Type} & Model & {Language} \\
    \midrule
    2021  & CURE~\cite{jiang2021cure} & ICSE  & GPT   & Java  & \href{https://github.com/lin-tan/CURE}{URL} \\
    2021  & DeepDebug~\cite{drain2021generating} & PLDI  & BART  & Java  & N.A. \\
    2022  & DEAR~\cite{li2022dear} & ICSE  & BERT  & Java  & \href{https://github.com/AutomatedProgramRepair-2021/dear-auto-fix}{URL} \\
    2022  & AlphaRepair~\cite{xia2022less} & FSE/ESEC & CodeBERT & Java,Python & \href{https://zenodo.org/records/6819444}{URL} \\
    2022  & CIRCLE~\cite{yuan2022circle} & ISSTA & T5    & Java,C,JS,Python & \href{https://github.com/2022CIRCLE/CIRCLE}{URL} \\
    2022  & NSEdit~\cite{hu2022fix} & ICLR  & CodeBERT,CodeGPT & Java  & N.A. \\
    2023  & APRFiT~\cite{hao2023enhancing} & ICSME & GraphCodeBERT,CodeT5 & Java  & N.A. \\
    2023  & RING~\cite{joshi2023repair} & AAAI  & Codex & \makecell[l]{C,JS,Python,Excel,\\Powershell,Power Fx} & N.A. \\
    2023  & RAP-Gen~\cite{wang2023rap} & FSE/ESEC & CodeT5 & Java  & \href{https://figshare.com/s/a4e95baee01bba14bf4b}{URL} \\
    2023  & CEDAR~\cite{nashid2023retrieval} & ICSE  & Codex & JS    & \href{https://github.com/prompt-learning/cedar}{URL} \\
    2023  & FitRepair~\cite{xia2023plastic} & ASE   & CodeT5 & Java  & \href{https://zenodo.org/records/8244813}{URL} \\
    2023  & Gamma~\cite{zhang2023gamma} & ASE   & UniXcoder,CodeBERT,GPT-3.5 & Java  & \href{https://github.com/iSEngLab/GAMMA}{URL} \\
    2023  & Repilot~\cite{wei2023copiloting} & FSE/ESEC & CodeT5,InCoder & Java  & \href{https://github.com/ise-uiuc/Repilot}{URL} \\
    2023  & SarGaM~\cite{liu2023automated} & TSE   & PLBART,CoditT5,NatGen & Java  & \href{https://github.com/SarGAMTEAM/SarGAM.git}{URL} \\
    2023  & STEAM~\cite{zhang2023steam} & TOSEM & GPT-3.5 & Java  & N.A. \\
    2024  & T5APR~\cite{gharibi2024t5apr} & JSS   & CodeT5 & Java,Python & \href{https://github.com/h4iku/T5APR}{URL} \\
    2024  & CigaR~\cite{hidvegi2024cigar} & arxiv & GPT-3.5 & Java  & \href{https://github.com/ASSERT-KTH/cigar}{URL} \\
    2024  & Toggle~\cite{hossain2024deep} & FSE/ESEC & \makecell[l]{CodeT5,CodeGPT,CodeParrot,\\CodeGen,PolyCoder} & Java  & N.A. \\
    2024  & Mulpor~\cite{lin2024one} & ISSTA & CodeT5 & Java  & \href{https://zenodo.org/records/12660892}{URL} \\
    2024  & ThinkRepair~\cite{yin2024thinkrepair} & ISSTA & \makecell[l]{CodeLLaMA,StarCoder,\\DeepseekCoder,GPT-3.5} & Java,Python & \href{https://github.com/vinci-grape/ThinkRepair}{URL} \\
    2024  & ChatRepair~\cite{xia2024automated} & ISSTA & GPT-3.5 & Java, Python & N.A \\
    2025  & RepairAgent~\cite{bouzenia2024repairagent} & ICSE  & GPT-3.5 & Java  & N.A. \\
    2025  & Ehsani~et~al.~\cite{ehsani2025bug} & ASE   & GPT-4o-mini,Llama-3.3 & Python & \href{https://github.com/SOAR-Lab/llm-apr-knowledge-injection}{URL} \\
    2025  & DEVLoRe~\cite{feng2025integrating} & TOSEM & GPT-4o-mini & Java  & \href{https://github.com/XYZboom/DEVLoRe}{URL} \\
    2025  & APRMCTS~\cite{hu2025aprmcts} & ASE   & GPT-4o-mini,GPT-3.5 & Java,Python & \href{https://github.com/Tomsawyerhu/APR-MCTS}{URL} \\
    2025  & ContrastRepair~\cite{kong2024contrastrepair} & TOSEM & GPT-3.5 & Java  & N.A. \\
    2025  & FixAgent~\cite{lee2024unified} & arxiv & \makecell[l]{CodeLLaMA,Llama-2,DeepseekCoder,\\GPT-4o,Gemini-1.5,GPT-3.5,Claude-3.5} & Java,Python,C & \href{https://github.com/AcceptePapier/UniDebugger}{URL} \\
    2025  & GIANTREPAIR~\cite{li2025hybrid} & TOSEM & StarCoder,CodeLLaMA,Llama-2,GPT-3.5 & Java  & \href{https://github.com/Feng-Jay/GiantRepair}{URL} \\
    2025  & HapRepair~\cite{lin2025haprepair} & FSE/ESEC & GPT-4o & ArkTS & \href{https://github.com/SMAT-Lab/HapRepair}{URL} \\
    2025  & Ouyang~et~al.~\cite{ouyang2025knowledge} & ICSE  & \makecell[l]{GPT-3.5,GPT-4o-mini,\\DeepSeekCoder-V2,Codestral} & Python & \href{https://github.com/ShuyinOuyang/DSrepair}{URL} \\
    2025  & D4C~\cite{xu2025aligning} & ICSE  & GPT-4,Mixtral & Java,C++,Python & \href{https://github.com/CUHK-Shenzhen-SE/D4C}{URL} \\
    2025  & MORepair~\cite{yang2024multi} & TOSEM & CodeLLaMA & Java,C++ & N.A. \\
    2025  & AdverIntent-Agent~\cite{ye2025adverintent} & ISSTA & GPT-4o & Java  & \href{https://doi.org/10.5281/zenodo.15367930}{URL} \\
    2025  & NTR~\cite{huang2025template} & ICSE  & CodeT5,StarCoder,CodeLLaMA & Java  & \href{https://sites.google.com/view/neuraltemplaterepair}{URL} \\
    2025  & ReinFix~\cite{zhang2025repair} & ICSE  & GPT-3.5,GPT-4o,GPT-4 & Java  & \href{https://sites.google.com/view/repairingredients}{URL} \\
    2025  & AutoSD~\cite{kang2023explainable} & EMSE  & Codex,CodeGen,GPT-3.5 & Java,Python & N.A. \\
    2025  & RepairLLaMA~\cite{silva2023repairllama} & TSE   & CodeLLaMA & Java  & \href{https://anonymous.4open.science/r/repairllama-BC13}{URL} \\
    \bottomrule
    \end{tabular}%
    }
  \label{tab:bug_semantic_tech}%
\end{table}%

\begin{table}[htbp]
\footnotesize
  \centering
  \caption{\revise{A summary of representative LLM-based empirical studies on semantic bugs}}
  \resizebox{\textwidth}{!}{
    \begin{tabular}{lllp{4.5cm}p{2cm}l}
    \toprule
    {Year} & Technique & {Venue} & Model & Language & URL \\
    \midrule
    2021  & Mashhadi~\cite{mashhadi2021applying} & MSR   & CodeBERT & Java  & \href{https://github.com/EhsanMashhadi/MSR2021-ProgramRepair}{URL} \\
    2022  & Kolak~et~al.~\cite{kolak2022patch} & ICLR  & Codex, PolyCoder & Java, Python & N.A. \\
    2023  & Huang~et~al.~\cite{huang2023empirical} & ASE   & CodeBERT, GraphCodeBERT, PLBART, CodeT5, UniXcoder & Java, C, C++, JS & \href{https://github.com/LLMC-APR/STUDY}{URL} \\
    2023  & Wu~et~al.~\cite{wu2023exploring} & arxiv & GPT-4, GPT-3.5 & Java, C, Python & N.A. \\
    2023  & Zirak~et~al.~\cite{zirak2023improving} & TOSEM & CodeBERT, TFix & Java  & \href{https://github.com/arminzirak/TFix}{URL} \\
    2023  & Gao~et~al.~\cite{gao2023makes} & ASE   & Codex, GPT-3.5, text-davinci-003 & Java  & \href{https://github.com/shuzhenggao/ICL4code}{URL} \\
    2023  & Jiang~et~al.~\cite{jiang2023impact} & ICSE  & PLBART, CodeT5, InCoder, CodeGen & Java  & \href{https://github.com/lin-tan/clm}{URL} \\
    2023  & Tian~et~al.~\cite{tian2023chatgpt} & arxiv & GPT-3.5 & Python & \href{https://github.com/HaoyeTianCoder/ChatGPT-Study}{URL} \\
    2023  & Xia~et~al.~\cite{xia2023automated} & ICSE  & GPT-Neo, GPT-J, GPT-NeoX, Codex, CodeT5, InCoder & Java, C, Python & \href{https://zenodo.org/records/7592886}{URL} \\
    2024  & RTT~\cite{ruiz2024novel} & TOSEM & PLBART, InCoder, GPT-4, TransCoder, SantaCoder, StarCoder, CodeT5, GPT-3.5 & Java  & \href{https://zenodo.org/records/10500594}{URL} \\
    2024  & MGDebugger~\cite{shi2024code} & arxiv & DeepSeekCoder-V2, CodeQwen1.5, GPT-4 & Python & \href{https://github.com/YerbaPage/MGDebugger}{URL} \\
    2024  & Prenner~et~al.~\cite{prenner2023out} & ICSE  & CodeT5 & Java, Python & \href{https://github.com/giganticode/out\_of\_context\_paper\_data}{URL} \\
    2025  & Cao~et~al.~\cite{cao2023study} & AUSE  & GPT-3.5 & Python & N.A. \\
    2025  & Kong~et~al.~\cite{kong2025demystifying} & FSE/ESEC & GPT-3.5, CodeLLaMA & Java  & \href{https://sites.google.com/view/memprompt}{URL} \\
    2025  & Luo~et~al.~\cite{luo2024fine} & TOSEM & CodeLLaMA, DeepseekCoder, WizardCoder, Mistral, CodeQwen1.5 & Java  & \href{https://github.com/stringing/Federated-LLM-Based-APR}{URL} \\
    2025  & DSrepair~\cite{machavcek2025impact} & ICSME & CodeT5, CodeGen, StarCoder, DeepseekCoder, BLOOM, CodeLlama-2 & Java  & \href{https://doi.org/10.5281/zenodo.16359186}{URL} \\
    2025  & MANIPLE~\cite{parasaram2024fact} & ICSE  & LLaMA3, GPT-3.5 & Python & \href{https://github.com/PyRepair/maniple}{URL} \\
    2023  & Defects4J-Nl2fix~\cite{fakhoury2023towards} & ICSE  & Codex, GPT-3.5 & Java  & N.A. \\
    \bottomrule
    \end{tabular}%
    }
  \label{tab:bug_semantic_empirical}%
\end{table}%

Semantic bugs refer to logical errors in a syntactically correct program when the code does not do what the programmer intends, even though it may compile and run without syntax errors. 
\delete{Considering that the majority of LLM-based APR studies are concentrated in the semantic bug field, we summarize these representative studies based on three ways of using LLMs.}
\revise{
Considering that the majority of LLM-based APR studies are concentrated in the semantic bug field, we organize this rich body of work into two complementary parts to provide a clear and structured overview.
To better capture their methodological diversity, Sections~\ref{sec:semantic_fine_tuning}–\ref{sec:semantic_zero_shot} summarize technical studies that propose new repair methods under fine-tuning, few-shot prompting, or zero-shot prompting paradigms, whereas Section~\ref{sec:semantic_empirical} reviews empirical studies that focus on evaluating, benchmarking, and analyzing existing LLMs for automated program repair.
}

\subsubsection{Repair with Fine-tuning}
\label{sec:semantic_fine_tuning}
As early as 2021, Drain~et al.~\cite{drain2021generating} introduce DeepDebug, which learns to fix bugs in Java methods mined from real-world GitHub repositories through fine-tuning BART on the BFP~\cite{tufano2019empirical} dataset.
In 2022, Yuan~et al.~\cite{yuan2022circle} propose CIRCLE, an LLM-based multi-lingual APR framework by fine-tuning a pre-trained T5 with continual learning.
Particularly, CIRCLE utilizes a difficulty-based rehearsal strategy to achieve lifelong learning without access to all historical data and elastic regularization to resolve catastrophic forgetting.
CIRCLE is the first approach to repair multiple programming language bugs with a single repair model in the continual learning setting.
Hao~et al.~\cite{hao2023enhancing,hao2024exploring} propose APRFiT by fine-tuning CodeT5 and GraphCodeBERT with a curricular learning framework.
In 2023, RAP-Gen~\cite{wang2023rap} further improves the repair performance of fine-tuning LLMs via a retrieval-augmented generation framework.
RAP-Gen first utilizes a hybrid patch retriever to search for a relevant fix pattern from an external codebase, and utilizes Codet5 as the foundation model to synthesize candidate patches with the augmented inputs, i.e., the retrieved fix pattern and original buggy code.
\revise{Recently, Silva~\cite{silva2023repairllama} escalate the parameters of fine-tuned LLMs to the billion level, and propose RepairLLaMA based on CodeLlama-7B.
RepairLLaMA utilizes a parameter-efficient fine-tuning technique called LoRA to train repair adapters with appropriate code representations.
}

\subsubsection{Repair with Few-shot \delete{Learning}\revise{Prompting}}
\label{sec:semantic_few_shot}
In few-shot learning, LLMs take a prompt as input , which contains natural language instructions, a few examples of task demonstration, and a query, and generate an output. 
In 2023, Nashid~et al.~\cite{nashid2023retrieval} propose a CEDAR, retrieval-based prompt selection approach for several code generation tasks, including program repair.
CEDAR first automatically retrieves bug-fixing demonstrations relevant to the buggy code based on embedding or frequency similarity, and creates prompts to query Codex to generate correct patches in a few-shot learning manner.
Meanwhile, Do~et al.~\cite{do2023using} conduct a preliminary study with ChatGPT and  Galpaca-30b (a variant of LLaMA) by using a few-shot example generation pipeline.

\subsubsection{Repair with Zero-shot \delete{Learning}\revise{Prompting}}
\label{sec:semantic_zero_shot}
Unlike the above fine-tuning-based APR work, which relies on high-quality bug-fixing code pairs, some approaches are proposed to directly leverage off-the-shelf LLMs without any training.
Such approaches attempt to generate patches in a zero-shot setting and can be categorized into two types, namely cloze-style APR and conversation-style APR.

\textbf{Cloze-style APR} leverages the pre-training objective (e.g., model language modeling) of LLMs to predict correct code tokens with the help of repair patterns.
In 2022, Xia~et al.~\cite{xia2022less} introduce AlphaRepair, a cloze-style APR tool to directly query LLMs to generate patches in a zero-shot learning setting.
AlphaRepair first replaces the buggy line with a mask line and queries CodeBERT to fill the mask line to generate candidate patches.
Furthermore, FitRepair~\cite{xia2023plastic} is a more advanced cloze-style APR approach based on CodeT5 and the plastic surgery hypothesis\cite{barr2014plastic,barr2015automated}.
FitRepair utilizes knowledge-intensified fine-tuning and repair-oriented fine-tuning to help CodeT5 learn the buggy project-specific knowledge and the cloze-style task knowledge, respectively.
FitRepair then retrieves relevant identifiers with static analysis, which are fed into fine-tuned CodeT5 to generate candidate patches.
Similarly, Repilot~\cite{wei2023copiloting} improves the cloze-style APR AlphaRepair with a completion engine.
Repilot builds an interaction between LLMs and a completion engine to generate more valid patches by first pruning away infeasible tokens suggested by LLMs and then completing the token based on the suggestions provided by the completion engine.
GAMMA~\cite{zhang2023gamma} further explores the potential of using LLMs to generate patches in a zero-shot learning scenario with a list of well-summarized fix templates.
Particularly, GAMMA attempts to address the donor code retrieval issue of traditional template-based APR (e.g., TBar~\cite{liu2019tbar}) and regards patch generation as a fill-in-the-blank task by querying LLMs to predict the correct code for masked tokens in a pre-defined fix pattern.
Unlike the cloze-style APR, Ribeiro~et al.~\cite{ribeiro2022framing} frame the APR problem as a code completion task and apply CodeGPT to fix bugs from ManySStuBs4J~\cite{karampatsis2020often}.

\textbf{Conversation-style APR} leverages the powerful natural language and programming language understanding capabilities of LLMs to generate patches iteratively using test failure information.
In 2023, Xia~et al.~\cite{xia2024automated} propose ChatRepair, the first conversation-driven APR approach that interleaves patch generation with instant feedback to perform APR in a dialogue manner.
ChatRepair constructs prompts with test failure information and queries ChatGPT to generate correct patches from previously incorrect and plausible patches.
However, ChatRepair mainly relies on negative feedback (i.e., failure information derived from failing tests) to guide the conversations, which may not always offer specific and adequate prompts for an effective repair.
Thus,  Kong~et al.~\cite{kong2024contrastrepair} introduce ContrastRepair, which includes positive feedback from passing tests to supplement the negative feedback.
Given a buggy program and a failing test case, ContrastRepair generates a similar passing test case by making minimal modifications to the failing test case. 
ContrastRepair then constructs a contrastive pair to LLMs, allowing them to better pinpoint the root cause of the bug and generate accurate patches.
In the above conversation-style APR scenarios, LLMs take a prompt containing some tokens about buggy code as the input and infer the following tokens about patches as the output.
During the conversations, all tokens in the input prompt and output answer consume the computational and financial costs, such as \$0.06 per 1k input tokens and \$0.03 per 1k generated tokens for GPT-4.
To reduce the computational cost of ChatRepair, Hidvegi~et al.~\cite{hidvegi2024cigar} propose CigaR, a token-efficiency LLM-based APR approach that concentrates on token cost minimization of ChatGPT.
CigaR designs three prompts to help ChatGPT minimize the overall token cost with previous responses, including (1) an initiation prompt to initialize the repair process, (2) an improvement prompt to refine partial patches, avoiding discarding potentially valuable patches, and (3) a multiplication prompt builds upon the already generated plausible patches to synthesize more plausible patches with diversity maximization.
Unlike previous work relying on a fixed prompt template, RepairAgent~\cite{bouzenia2024repairagent} treats ChatGPT as an agent capable of autonomously planning and executing actions to generate patches by using dynamic prompts and a state machine to select suitable tools.

\subsubsection{Empirical Study}
\label{sec:semantic_empirical}
In addition to the above novel techniques, researchers have conducted numerous empirical studies to explore the capabilities of LLMs in fixing semantic bugs.
As early as 2021, Mashhadi~et al.~\cite{mashhadi2021applying} preliminarily evaluate the performance of fine-tuning CodeBERT for fixing Java bugs from ManySStuBs4J.
Lajko~et al.~\cite{lajko2022towards} empirically fine-tune GPT-2 to automatically generate candidate patches for JavaScript bugs.
Prenner~et al.~\cite{prenner2022can} investigate CodeX in fixing  QuixBugs with a zero-shot setting and Sobania~et al.~\cite{sobania2023analysis} utilizes a more powerful LLM ChatGPT with prompt engineering to generate patches for QuixBugs.
In 2023, Horvath~et al.~\cite{horvath2023extensive} explore the impact of model architectures and program representations, involving two popular programming languages, i.e., Java and JavaScript, three different code representations, i.e., raw text, command sequences, and ASTs, and four LLMs, i.e., T5, CodeT5, RoBERTa, and GPTNeo. 
Meanwhile, Zhao~et al.~\cite{zhao2023right} conduct a comprehensive investigation into the effective utilization of LLMs for repairing code review bugs, involving two LLMs (i.e., ChatGPT and GPT-4) in a zero-shot learning manner, and six LLMs (InCoder, CodeT5+, CodeFuse, LLaMA, CodeGen-2, CodeLLaMA) in a fine-tuning manner.
Recently, inspired by the phenomenon that grammatical errors in natural language can be fixed by round-trip translation, i.e., translating sentences to another intermediate language and then back, Ruiz~et al.~\cite{ruiz2024novel} investigate to what extent the RTT pipeline can fix bugs in code in a zero-shot fashion.
The comprehensive study involves eight LLMs: PLBART, CodeT5, TransCoder, SantaCoder, InCoder, StarCoderBase, GPT-3.5, and GPT-4, and four APR benchmarks: Defects4J-v1.2, Defects4J-v2.0, QuixBugs, and HumanEval-Java.

Furthermore, more comprehensive empirical studies are published in top-tier SE conferences.
Xia~et al.~\cite{xia2023automated} conduct an comprehensive study to explore the performance of LLMs in program repair, involving nine LLMs from two categories (i.e., infilling and generative models), and five datasets across three programming languages.
They explore three ways to use LLMs for patch generation: complete function generation, correct code infilling, and single-line generation. 
Meanwhile, Jiang~et al.~\cite{jiang2023impact} empirically explore the fixing capabilities of ten variants from four LLMs under zero-shot and fine-tuning settings, involving four Java benchmarks.
They also construct a new benchmark, HumanEval-Java, which none of the LLMs has seen during training to address the data leakage issue.
Huang~et al.~\cite{huang2023empirical} conduct an empirical study on fixing capabilities of LLMs in the fine-tuning paradigm, involving five LLMs, three programming languages, and three repair scenarios.

\subsection{Programming Problems}
This type of bug refers to incorrect solutions to programming problems from competition platforms, such as LeetCode.
In 2022, Zhang~et al.~\cite{zhang2022repairing} propose MMAPR to repair both syntactic and semantic mistakes from introductory Python programming through Codex and few-shot learning.
Fan~et al.~\cite{fan2023automated} systematically evaluate whether repair techniques can fix the incorrect solutions produced by LLMs in LeetCode contests.
Similarly, Liu~et al.~\cite{liu2024refining} analyze the correctness of ChatGPT-generated code in Java and Python for 2,033 programming tasks from LeetCode and query ChatGPT to mitigate incorrect code with different prompts.
Recently, Zhang~et al.~\cite{zhang2023critical} empirically evaluate the performance of ChatGPT in fixing incorrect submissions from programming problems.
Similar studies are conducted by Haque~et al.~\cite{haque2023fixeval} and Tian~et al.~\cite{tian2024debugbench}.
Ishizue~et al.~\cite{ishizue2024improved} combine two LLMs (i.e., ChatGPT and GPT-4) with Refactory~\cite{hu2019re} to fix programming assignment problems.
They leverage LLMs to repair programs with syntax errors and those that Refactory fails to repair. 
Recently, Zhao~et al.~\cite{zhao2024peer} explore how LLMs, including ChatGPT, StarCoder, and CodeLlama, perform in repairing student assignments from higher-level programming courses.
They construct a student assignment dataset named Defects4DS, which contains 682 submissions from 4 programming assignments, and introduce a repair framework named PaR, which generates patches by selecting peer solutions to create prompts.

\subsection{Security Vulnerabilities}

\begin{table}[t]
\footnotesize
  \centering
  \caption{\revise{A summary of representative LLM-based studies on security vulnerabilities}}
    \begin{tabular}{lllllll}
    \toprule
    {Year} & Technique & {Venue} & {Paper Type} & Model & {Language} & Link \\
    \midrule
    2022  & VulRepair~\cite{fu2022vulrepair} & FSE/ESEC & Approach & CodeT5 & C     & \href{https://github.com/awsm-research/VulRepair}{URL} \\
    2023  & VQM~\cite{fu2023vision} & TOSEM & Approach & CodeT5 & C,C++ & \href{https://github.com/awsm-research/VQM}{URL} \\
    2023  & Huang~et~al.~\cite{huang2023empirical} & ASE   & Empirical & \makecell[l]{CodeBERT, GraphCodeBERT, \\PLBART, CodeT5, UniXcoder} & Java,C,C++,JS & \href{https://github.com/LLMC-APR/STUDY}{URL} \\
    2023  & Pearce~et~al.~\cite{pearce2023examining} & S\&P  & Empirical & Codex, PolyCoder, Jurassic-1 & C,C++ & \href{https://drive.google.com/drive/folders/1xJ-z2Wvvg7JSaxfTQdxayXFEmoF3y0ET?usp=sharing}{URL} \\
    2023  & ZeroLeak~\cite{tol2023zeroleak} & ESORICS & Empirical & GPT-4, PaLM, Llama-2, GPT-3.5 & C,JS  & N.A. \\
    2023  & Wu~et~al.~\cite{wu2023effective} & ISSTA & Empirical & \makecell[l]{CodeT5, PLBART, InCoder, \\CodeGen, Codex} & Java  & \href{https://github.com/lin-tan/llm-vul}{URL} \\
    2023  & Wu~et~al.~\cite{wu2023exploring} & arxiv & Empirical & GPT-4, GPT-3.5 & Java,C,Python & N.A. \\
    2023  & Zhang~et~al.~\cite{zhang2023pre} & TDSC  & Empirical & \makecell[l]{CodeT5, CodeBERT, UniXcoder, \\GraphCodeBERT, CodeGPT} & C     & \href{https://github.com/iSEngLab/LLM4VulFix}{URL} \\
    2024  & VRpilot~\cite{kulsum2024case} & AIware & Approach & GPT-3.5 & C,Java & \href{https://drive.google.com/drive/folders/1yrYUJS1r2cu7G6D3ezASsfugrNqgwc3v}{URL} \\
    2024  & Le~et~al.~\cite{le2024study} & WWW   & Empirical & GPT-4, Bard & JS    & \href{https://doi.org/10.5281/zenodo.10783763}{URL} \\
    2024  & Mulpor~\cite{lin2024one} & ISSTA & Approach & CodeT5 & Java  & \href{https://zenodo.org/records/12660892}{URL} \\
    2024  & Liu~et~al.~\cite{liu2023chatgpt} & Security & Empirical & GPT-3.5, GPT-4 & C,C++ & \href{https://anonymous.4open.science/r/DefectManagementEvaluation-0411}{URL} \\
    2024  & Nong~et~al.~\cite{nong2024chain} & arxiv & Approach & Llama-2, Falcon, GPT-3.5 & C,C++ & N.A. \\
    2024  & VulAdvisor~\cite{zhang2024vuladvisor} & ASE   & Approach & CodeT5, Codex, GPT-3.5 & C,C++ & \href{https://github.com/zhangj111/VulAdvisor}{URL} \\
    2024  & VulMaster~\cite{zhou2024large} & ICSE  & Approach & CodeT5 & C,C++ & \href{https://github.com/soarsmu/VulMaster\_}{URL} \\
    2025  & ESBMC-AI~\cite{charalambous2023new} & AST   & Approach & GPT-3.5 & C,C++ & N.A. \\
    2025  & VulnTrace~\cite{gao2025teaching} & FSE/ESEC & Approach & \makecell[l]{UniXcoder, GraphCodeBERT, \\CodeBERT, RoBERTa} & C,C++ & \href{https://github.com/amiaog/Teaching-AI-the-Why-and-How-of-Software-Vulnerability-Fixes}{URL} \\
    2025  & NTR~\cite{huang2025template} & ICSE  & Approach & CodeT5, StarCoder, CodeLLaMA & Java  & \href{https://sites.google.com/view/neuraltemplaterepair}{URL} \\
    2025  & APPATCH~\cite{nong2025appatch} & Security & Approach & \makecell[l]{GPT-4, Gemini-1.5,\\ Claude-3.5, LLaMA-3.1} & C,C++ & \href{https://zenodo.org/records/14741018}{URL} \\
    2025  & Steenhoek~et~al.~\cite{steenhoek2025closing} & ICSE  & Human Study & GPT-4 & Java  & \href{https://doi.org/10.6084/m9.figshare.26367139}{URL} \\
    2025  & PATCHAGENT~\cite{yu2025patchagent} & Security & Approach & GPT-4, GPT-4o, Claude-3 & C,C++ & \href{https://osf.io/8k2ac}{URL} \\
    \bottomrule
    \end{tabular}%
  \label{tab:bug_security}%
\end{table}%

A software vulnerability refers to a flaw, bug, or weakness in the design, implementation, operation, or management of software.
When exploited, vulnerabilities can lead to unauthorized access, manipulation, or disruption of the software's intended functionality. 
In 2022, Fu~et al.~\cite{fu2022vulrepair} propose VulRepair, an LLM-based automated program repair approach by fine-tuning CodeT5 with vulnerability datasets.
Furthermore, in 2023, considering that a vulnerable function only contains a few core elements that need to be fixed, Fu~et al.~\cite{fu2023vision} propose VQM, an LLM-based automated vulnerability repair approach that leverages VIT-based approaches for object detection to help CodeT5 focus more on vulnerable code areas during patch generation.
Inspired by the relationship between semantic bugs and security vulnerabilities, Zhang~et al.~\cite{zhang2023pre} propose an enhanced LLM-based vulnerability framework based on transfer learning, demonstrating superior performance against VulRepair.
In 2024, Zhou~et al.~\cite{zhou2024large} propose VulMaster, a CodeT5-based approach to address several limitations of VulRepair, i.e., (1) handling lengthy vulnerable code by the Fusion-in-Decoder framework; (2) capturing the structure of the vulnerable code by AST; and (3) understanding the expert vulnerability knowledge by the CWE system.
De~et al.~\cite{de2024enhanced} leverage Quantized Low-Rank Adaptation (QLoRA)~\cite{li2024loftq} to fine-tune two advanced 7-billion-parameter LLMs, i.e., CodeLlama~\cite{roziere2023code} and Mistral~\cite{mistral}, to fix C vulnerabilities.

\textbf{Empirical Study}.
Pearce~et al.~\cite{pearce2023examining} evaluate the performance of LLMs in repair vulnerabilities in a zero-shot manner, involving five commercially available black-box and two open-source LLMs.
They particularly investigate the potential of designing prompts that query LLMs to patch vulnerable code.
Furthermore, Zhang~et al.~\cite{zhang2023pre} conduct an empirical study investigating the performance of fine-tuning LLMs for vulnerability repair, involving five LLMs from three categories, two C vulnerability datasets and more than 100 trained LLMs, which is the largest set to date.
They also explore the potential of ChatGPT in repairing security vulnerabilities in a zero-shot manner.
Similarly, Wu~et al.~\cite{wu2023effective} explore the fixing capabilities of five LLMs in a zero-shot manner and four LLMs in a fine-tuning manner on two real-world Java vulnerability datasets.
Recently, Le~et al.~\cite{le2024study} conduct a preliminary study of ChatGPT and Bard in detecting and fixing security vulnerabilities in JavaScript programs.

\subsection{Repository-level Issues}
\begin{table}[t]
\footnotesize
  \centering
  \caption{\revise{A Summary of representative LLM-based studies on repository-level issues.}}
    \begin{tabular}{lllllll}
    \toprule
    Year & Tool name & Venue & Type  & Model & Language & Link \\
    \midrule
    2024  & MASAI~\cite{arora2024masai} & NeurIPS & Approach & GPT-4o & Python & N.A. \\
    2024  & CodeR~\cite{chen2024coder} & arxiv & Approach & GPT-4 & Python & \href{https://github.com/NL2Code/CodeR}{URL} \\
    2024  & SWE-bench~\cite{jimenez2024swebench} & ICLR  & Benchmark & CodeLLaMA & Python & \href{https://github.com/princeton-nlp/SWE-bench}{URL} \\
    2024  & SWE-agent~\cite{yang2024swe} & NeurIPS & Approach & GPT-4,Claude-3 & Python & \href{https://github.com/SWE-agent/SWE-agent}{URL} \\
    2024  & AutoCodeRover~\cite{zhang2024autocoderover} & ISSTA & Approach & GPT-4 & Python & \href{https://autocoderover.dev/}{URL} \\
    2024  & MarsCode~\cite{liu2024marscode} & arxiv & Approach & GPT-4,Claude-2 & Python & N.A. \\
    2024  & DRCodePilot~\cite{zhao2024enhancing} & ASE   & Approach & GPT-4 & Java  & \href{https://figshare.com/s/82ed8e86e88d3268b4c1}{URL} \\
    2025  & Ehsani~et~al.~\cite{ehsani2025towards} & MSR   & Approach & GPT-3.5 & Python & \href{https://anonymous.4open.science/r/prompt-knowledge-gap-BE45}{URL} \\
    2025  & OmniGIRL~\cite{guo2025omnigirl} & ISSTA & Benchmark & \makecell[l]{GPT-4o,Claude-3.5,\\Deepseek-v2.5} & \makecell[l]{Python,JS,\\TS,Java} & \href{https://github.com/DeepSoftwareAnalytics/OmniGIRL}{URL} \\
    2025  & SWE-GPT~\cite{ma2025swe} & ISSTA & Approach & Qwen-2.5,Qwen2.5-Coder & Python & \href{https://github.com/LingmaTongyi/Lingma-SWE-GPT}{URL} \\
    2025  & SWT-Bench~\cite{mundler2024swt} & NeurIPS & Benchmark & \makecell[l]{GPT-4,GPT-4o-mini,\\Claude-3,Claude-3.5,Mistral} & Python & \href{https://github.com/logic-star-ai/SWT-Bench}{URL} \\
    2025  & Moatless~\cite{antoniades2024swe} & ICLR  & Approach & \makecell[l]{GPT-4o,GPT-4o-mini,Qwen-2.5,\\Deepseek-v2.5,LLaMA 3.1} & Python & \href{https://github.com/aorwall/moatless-tree-search}{URL} \\
    2025  & RepoGraph~\cite{ouyang2024repograph} & ICLR  & Approach & GPT-4o,GPT-4 & Python & \href{https://github.com/ozyyshr/RepoGraph}{URL} \\
    2025  & SWE-RL~\cite{wei2025swe} & NeurIPS & Approach & Llama-3.3 & Python & \href{https://github.com/facebookresearch/swe-rl}{URL} \\
    2025  & Agentless~\cite{xia2025demystifying} & FSE/ESEC & Approach & GPT-4o & Python & \href{https://github.com/OpenAutoCoder/Agentless}{URL} \\
    2025  & MAGIS~\cite{tao2024magis} & NeurIPS & Approach & GPT-4 & Python & N.A \\
    2025  & SWE-bench M~\cite{yang2024swe} & ICLR  & Benchmark & GPT-4o-mini,Claude-3.5 & JS,TS & \href{https://swebench.com/multimodal}{URL} \\
    2025  & OpenHands~\cite{wang2024openhands} & ICLR  & Approach & GPT-4o,GPT-4o-mini,Claude-3.5 & Python & \href{https://github.com/All-Hands-AI/OpenHands}{URL} \\
    \bottomrule
    \end{tabular}%
  \label{tab:bug_issue}%
\end{table}%

Repository-level issues refer to real-world issues (such as bug reports or feature requests) in large-scale repositories.
Recent LLM-based APR research increasingly targets repository-level issues, whose resolution requires long-context understanding over a large codebase, accurate fault localization from informal reports, dependency, and code edits across multiple modules, and robust validation of the resulting pull request.
A prominent benchmark for this scenario is SWE-bench~\cite{jimenez2024swebench}, which comprises 2294 real GitHub issues with paired ground-truth pull requests from 12 popular Python repositories. 
In addition, several other datasets, such as SWT-bench~\cite{mundler2024swt}, OmniGIRL~\cite{guo2025omnigirl}, and SWE-bench M~\cite{yang2024swem}, are used to evaluate LLMs under varying APR settings.
To address repository-level repair challenges, researchers have explored two main categories of approaches: procedural~\cite{xia2025demystifying,ouyang2024repograph} and agentic~\cite{zhang2024autocoderover,antoniades2024swe,chen2024coder,wang2024openhands}.
In particular, similar to few-shot or zero-shot prompting approaches, procedural approaches rely on expert-designed workflows where LLMs run at a fixed pipeline, and every choice is hard-coded.
Agentic approaches query LLMs to autonomously plan and execute each choice by equipping LLMs with various external tools

\subsection{Static Warnings}
Static warnings refer to automated alerts generated by static tools that examine code to identify potential errors, inefficiencies, or security vulnerabilities without executing the program.
In 2021, Berabi~et al.~\cite{berabi2021tfix} introduce TFix, which is considered a pioneer in incorporating LLMs into APR.
Inspired by T5, TFix formulates the program repair problem as a text-to-text task on code sequence, i.e., given a coding error as text, predicting a new text as the patch.
TFix is particularly fine-tuned from T5 with a high-quality dataset consisting of ~100k bug-fixing pairs across 52 error types detected by a static analyzer, ESLint.
Similar to RAP-Gen~\cite{wang2023rap}, InferFix~\cite{jin2023inferfix} employs a retrieval-augmented generation manner to fine-tune LLMs for program repair, i.e., (1) utilizing a retriever to search for semantically equivalent bugs and corresponding fixes from an external codebase, and (2) fine-tuning LLMs on supervised bug-fixing datasets with prompts augmented via adding retrieved similar fixes.
However, RAP-Gen targets semantic bugs while InferFix repairs static errors detected by the static analyzer Infer; and RAP-Gen utilizes CodeT5 while InferFix has a significantly larger model Codex with 12 billion parameters.
Recently, Wadhwa~et al.~\cite{wadhwa2023frustrated} present CORE to resolve code quality issues flagged by static analysis tools.
CORE utilizes \delete{ChatGPT}\revise{GPT-3.5} as the proposer LLM to generate candidate code revisions and GPT-4 as the ranker to rank the candidate revisions before presenting them to the developer.
Alrashedy~et al.~\cite{alrashedy2023can} introduce FDSP to resolve repair security vulnerabilities detected by a static code analysis tool, Bandit~\cite{bandit}, in a self-debugging manner.

In addition to the above technical papers, Kim~et al.~\cite{kim2022empirical} empirically investigate the performance of TFix in fixing errors from industrial Samsung Kotlin projects detected by a static analysis tool SonarQube.
Mohajer~et al.~\cite{mohajer2023skipanalyzer} conduct a more comprehensive study of LLMs in the static code analysis domain, and propose SkipAnalyzer, an LLM-based powered tool to perform three related tasks: detecting bugs, filtering false positive warnings, and patching the detected bugs.

\subsection{Syntax Errors}

Syntax errors refer to parsing mistakes that occur when code does not follow the rules or grammar of the programming language, such as invalid statements and expressions.
These errors are detected by the compiler or interpreter during the parsing phase before the program is executed. 
Ahmed~et al.~\cite{ahmed2022synshine} propose SYNSHINE to fix syntax errors by pre-training and fine-tuning RoBERTa with compiler diagnostics.
RustAssistant~\cite{deligiannis2023fixing} utilizes GPT-3.5 and GPT-4 to fix Rust compilation errors with prompt engineering.
RING~\cite{joshi2023repair} is a multilingual syntax error repair approach powered by Codex and few-shot learning and is proven promising in six different languages.
PCR-Chain~\cite{huang2023chain} attempts to resolve fully-qualified names and fix last-mile syntax errors based on ChatGPT and Chain-of-Thought Prompting.

\subsection{Hardware Bugs}
APR is well-explored in the software domain, while the research is less mature for hardware description languages, such as Verilog.
Ahmad~et al.~\cite{ahmad2024hardware} empirically explore the potential of LLMs to repair security-relevant bugs in hardware designs.
They leverage prompt engineering to query Codex, ChatGPT, GPT-4, and CodeGen for 15 security-related bugs from 10 CWEs.
Inspired by the retrieval-augmented generation, Tsai~et al.~\cite{tsai2023rtlfixer} introduce RTLFixer, a framework to fix Verilog syntax errors with LLMs and iterative prompting. 
RTLFixer formulates an input prompt to query GPT-4 to generate the correct code.
RTLFixer then utilizes error logs from the compiler and human guidance from a retrieval database as feedback to perform an interactive debugging loop until all errors are resolved.
Yao~et al.~\cite{yao2024hdldebugger}\delete{ from Huawei} focus on the domain of chip design and propose HDLdebugger, an LLM-assisted hardware description language debugging framework.
This framework encompasses data generation through reverse engineering, a search engine to enhance the retrieval-augmented generation, and a fine-tuning approach to train retrieval-augmented LLMs.
Similarly, Fu~et al.~\cite{fu2023llm4sechw} focus on the hardware design iteration process and introduce LLM4SECHW, an LLM-based hardware debugging framework.
LLM4SECHW constructs a hardware debugging-oriented dataset from open-source hardware projects and fine-tunes a suite of hardware domain-specific LLMs capable of automatically reading hardware designs and fixing bugs.

\subsection{Test Repair}
\revise{Test repair focuses on fixing broken or outdated test cases to ensure they remain valid and meaningful throughout software evolution, whereas traditional APR studies modify production code to pass existing tests.
Recent research has increasingly explored the potential of LLMs to automate test repair across diverse scenarios, including outdated tests, flaky tests, non-idempotent outcomes, and UI tests.
For example, to address outdated test cases, Yaraghi et al.~\cite{saboor2024automated} introduce TARGET, an LLM-based framework that treats test repair as a language translation task.
They construct TARBENCH, a comprehensive benchmark containing 45,373 broken test repairs across 59 open-source projects, and fine-tune off-the-shelf LLMs (i.e., PLBART, CodeGen, and CodeT5+) to generate correct test code with essential code context.
Fatima~et~al.~\cite{fatima2024flakyfix} introduce FlakyFix, a framework that leverages LLMs to predict fix categories for flaky tests and synthesize corresponding repairs
Ke~et~al.~\cite{ke2025niodebugger} propose NIODebugger, an agentic LLM-based framework for repairing non-idempotent-outcome tests that exhibit inconsistent results across repeated executions.
NIODebugger adopts a three-stage agentic loop: (1) analyzing the root cause of test non-determinism (e.g., environment or concurrency issues), (2) generating repair hypotheses and corresponding patches, and (3) verifying them through re-execution and iterative feedback.
Focusing on test script repair, Xu~et~al.~\cite{xu2023guiding} conduct the first feasibility study that integrates traditional heuristics with ChatGPT to repair Web UI tests (e.g., Selenium tests).
They first use traditional repair approaches to generate a preliminary list of candidate-matched elements and then use ChatGPT to perform global matching to select the best-matched element.
}

\subsection{Type Errors}
Type errors occur when an operation or function is applied to an object of an inappropriate type.
\delete{This indicates that the operation expects a certain data type but receives another, thus leading to an exception.}
\revise{This indicates that the operation expects a certain data type but receives another, thus leading to unhandled exceptions that cause program crashes. }
These errors are prevalent in dynamically typed languages such as Python and JavaScript, which do not allow operations that are inherently impossible with the given data types.
In 2024, Chow~et al.~\cite{chow2024pyty} construct PyTyDefects, a dataset consisting of 2,766 type error-fix pairs from 176 GitHub repositories.
They also introduce PyTy, a T5-based repair technique that fine-tunes the off-the-shelf LLM TFix with PyTyDefects for fixing type errors in Python. 
Unlike the fine-tuning-based approach PyTy, TypeFix~\cite{peng2024domain} is a prompt-based repair approach for Python type errors in a zero-shot manner.
Similar to GAMMA, TypeFix queries LLMs to generate patches by filling the masks in code prompts using fix templates. 
However, TypeFix distinguishes itself by automatically mining these templates through a hierarchical clustering algorithm, rather than relying on predefined ones in GAMMA.
\delete{Furthermore, Ribeiro et al.~\cite{ribeiro2023gpt} present Mentat, a type error repair technique for OCaml programs powered by GPT-3.}
\revise{Although type errors are more prevalent in dynamically typed languages, they can also arise in statically typed languages (e.g., OCaml) due to incorrect type inference, mismatched annotations, and violations of declared type constraints during compilation.
For example, Ribeiro et al.~\cite{ribeiro2023gpt} present Mentat, a type error repair technique for OCaml programs powered by GPT-3.}
Mentat first analyzes the source code to generate contextually relevant prompts, then exploits GPT-3’s advanced language understanding and generation capabilities to produce potential patches.

\subsection{GUI Bugs}
\revise{
GUI (Graphical User Interface) bugs refer to defects in the visual or interactive layer of software systems, including layout misalignment, incorrect rendering of UI components, inconsistent visual states, and broken interactions under different environments.
Compared with traditional APR, which focuses on source code-level bugs, GUI bugs involve multimodal contexts, including source code, visual layouts, and user interactions, which demand perception of visual cues and reasoning across the image–code domain.
Building upon SWE-bench~\cite{jimenez2023swe}, Yang~et~al.~\cite{yang2024swem} introduce SWE-bench~M, an extension targeting visual and GUI-related issues in JavaScript projects.
SWE-bench~M comprises 617 visual tasks from 17 open-source repositories, each associated with real-world issues containing screenshots or rendered UI components.
Huang~et~al.~\cite{huang2025seeing} propose GUIRepair,  a cross-modal reasoning framework that incorporates visual context into GUI bug fixing for SWE-bench~M.
GUIRepair consists of two core modules: (1) Image2Code, which localizes the source code regions responsible for visual symptoms based on UI screenshots, and (2) Code2Image, which re-renders candidate patches to verify whether the visual defects are eliminated.
Unlike GUIRepair, which focuses on general GUI functionality and rendering issues, Yuan~et~al.~\cite{yuan2024designrepair} present DesignRepair, a dual-stream framework for repairing UI design-quality issues.
DesignRepair leverages both a low-level component representation and a high-level design-guideline knowledge base, integrating retrieval-augmented LLMs to correct design inconsistencies such as misalignment, color misuse, and accessibility violations.
}

\subsection{Code Review}
\revise{Code review aims to improve code quality and maintainability by reviewing submitted changes and providing actionable comments.
Guo~et~al.~\cite{guo2023exploring} conduct the first empirical study investigating the potential of ChatGPT for automated code refinement based on existing review comments.
They leverage prompt engineering to compare ChatGPT with CodeReviewer~\cite{li2022automating} using two datasets: the established CodeReview~\cite{li2022automating} dataset and a newly curated CodeReview-New dataset.
To further improve ChatGPT's performance, they design several enhancement strategies, including leveraging more advanced models and optimized prompting.
Furthermore, Wang~et~al.~\cite{wang2024divide} propose CodeReviser, a two-phase ``localization-and-revision'' framework that follows the divide-and-conquer principle for automated code revision.
Instead of asking a single model to both find and fix issues, CodeReviser first localizes the code entities to modify via a generative localizer and then revises them using a fine-tuned CodeT5-based reviser.}

\subsection{Performance Bugs}
\delete{Performance bugs refer to inefficient code snippets that do not interfere with functionality but lead to unnecessary time and resource consumption.}
\revise{Performance bugs refer to non-functional defects that degrade efficiency while preserving functional correctness~\cite{garg2022deepdev}.
Such bugs, although not causing system failure, still represent undesired behavior because they violate performance expectations and can negatively impact user experience.}
Garg~et al.~\cite{garg2022deepdev} propose DeepDev-PERF, the first LLM-based approach to repair performance bugs for C\# applications by fine-tuning the BART model.
Building upon DeepDev-PERF, RAPGen~\cite{garg2023rapgen} further enhances the repair process through prompt engineering, which demonstrates superior efficiency over traditional intensive fine-tuning methods. 
RAPGen retrieves relevant instructions from a knowledge base, constructs an input prompt tailored to the specific bug, and then uses Codex to generate the optimized code. 
This approach is similar to CEDAR~\cite{nashid2023retrieval} yet uniquely leverages prompt engineering to achieve more effective and resource-efficient solutions.

\subsection{Smart Contracts}
Smart contracts are self-executing contracts on a blockchain network with the terms of the agreement directly written into code.
In 2023, Napoli~et al.~\cite{napoli2023evaluating} conduct a preliminary study to explore the capabilities of ChatGPT in fixing vulnerable smart contracts with 143 vulnerable Solidity codes.
In 2024, Zhang~et al.~\cite{zhang2024acfix} introduce ACFIX, a GPT-based approach to repair access control vulnerabilities in smart contracts based on static code slicing and Chain-of-Thought Prompting.
ACFIX mines common AC practices for major categories of code functionality.
These practices form a knowledge base that guides LLMs in repairing code with analogous functionalities.

\subsection{Misc Repair Types}
We summarize other repair scenarios that are covered in fewer than two papers as follows.

\textbf{API Misuse}.
Zhang~et al.~\cite{zhang2023evaluating} conduct an empirical study to explore the performance of learning-based APR techniques on API misuse repair, involving nine LLMs with millions of parameters under a fine-tuning setting, i.e., CodeBERT, GraphCodeBERT, CodeGPT, PolyCoder-160M, PolyCoder-0.4B, CodeTrans, PLBART, CodeT5 and UniXcoder.

\textbf{Crash Bugs}.
Crash bugs are considered critical issues as they might
cause unexpected program behaviors and termination.
Du~et al.~\cite{du2023resolving} conduct the first investigation into ChatGPT's capability to resolve real-world crash bugs, particularly code-related and environment-related ones.
They design different prompts and propose IntDiagSolver, an interaction approach that involves continuous interaction with LLMs.

\revise{\textbf{Error-Handling Bugs}.
Error-handling bugs arise when erroneous conditions are not correctly detected, propagated, or managed by the software, leading to silent failures, data corruption, system crashes, or security vulnerabilities.  
Liu et al.~\cite{liu2025error} conduct the first in-depth empirical analysis of propagated error-handling bugs in mature software systems.  
They also introduce EH-Fixer, a retrieval-augmented generation approach to repair error-handling bugs by constructing the error-propagation path, retrieving similar historical cases, and prompting LLMs to generate patches. }

\textbf{Software Formal Proof}.
Formal software verification aims to validate the correctness of software properties.
First~et al.~\cite{first2023baldur} introduce Baldur, an LLM-based approach to automate the generation and repair of whole formal proofs.
Baldur leverages two versions of Minerva~\cite{lewkowycz2022solving}, which is pre-trained on a mathematics corpus based on the PaLM~\cite{chowdhery2023palm}: one with eight billion parameters and another with 62 billion parameters.
Baldur first fine-tunes a generation model to synthesize whole proofs for theorems on a proof dataset, and then fine-tunes a repair model based on the proof assistant's error messages to repair incorrectly generated proofs.

\revise{\textbf{Formal Specification}
Such bugs often arise in specification languages (e.g., Alloy) where unintended logical constraints, missing relations, or erroneous propagation lead to incorrect or incomplete system models.
Alhanahnah et al.~\cite{alhanahnah2025empirical} present a first systematic empirical investigation into the capability of LLMs for repairing faulty Alloy specifications. 
They design a novel dual-agent pipeline comprising a repair agent and a prompt agent, iteratively refining prompt-generation and repair proposals.}

\delete{\textbf{Web UI}.
Xu~et al.~\cite{xu2023guiding} conduct the first feasibility study on integrating traditional Web UI test repair techniques with ChatGPT for enhancing Web UI test repair.
They first utilize traditional repair approaches to generate a preliminary list of candidate-matched elements and employ ChatGPT to execute a global matching to select the best-matched element further.}

\textbf{Motion Planning Algorithm}.
In the automated vehicles scenario, notion planners refer to algorithms that are responsible for computing a viable path for the vehicle to travel from an initial state to a designated goal region within a predefined time frame.
Lin~et al.~\cite{lin2024drplanner} introduce DrPlanner, the first automated framework that utilizes GPT-4 to diagnose and repair motion planners.
DrPlanner first designs a structured prompt to describe the planners using both natural and programming languages.
DrPlanner then queries GPT-4 to repair algorithms with continuous diagnostic feedback in a closed-loop manner.

\textbf{Translation Bugs}.
Code translation refers to the process of converting source code from one language into another without changing the original program's functionality or logic. 
Pan~et al.~\cite{pan2024lost} present a large-scale empirical study to investigate the ability of LLMs for code translation across five different languages, including C, C++, Go, Java, and Python.
They provide a bug taxonomy of unsuccessful translations across 15 categories and 5 groups, and design an iterative translation bug repair approach to fix such failures from LLMs, using a set of heuristics for prompt crafting.

\delete{\textbf{Github Issue}.
Unlike most prior work~\cite{chen2021evaluating}, which evaluates LLMs in fixing self-contained problems, such as programming problems, Jimenez~et al.~\cite{jimenez2024swebench} explore the potential of LLMs to resolve GitHub issues in a realistic software engineering setting.
Given an issue (such as a bug report or a feature request) submitted to popular GitHub Python repositories, they fine-tune CodeLlama-7B and CodeLlama-13B to generate a patch that passes the unit and system tests.
}

\delete{
\textbf{Code Review}.
Guo~et al.~\cite{guo2023exploring} conduct the first empirical study to explore the potential of ChatGPT in code review, specifically focusing on automated code refinement based on existing code reviews.
They leverage prompt engineering to compare ChatGPT with CodeReviewer~\cite{li2022automating} using two datasets: an established one named CodeReview~\cite{li2022automating} and a newly introduced one named CodeReview-New.
They also design several strategies to improve the performance of ChatGPT, such as using more advanced models.}

\subsection{Trend Analysis}
\revise{
Table \ref{tab:bug_types} summarizes the distribution of LLM-based APR studies across various bug categories from 2020 to 2025. 
We discuss the research trend of LLM-based APR across different bug types.

\textbf{Dominant Focus on Semantic Bug Repair.}
Semantic bug repair remains the core and most mature research direction in LLM-based APR.
From 2020 to 2025, the number of studies addressing semantic bugs increased dramatically, from 4 papers in 2021 to more than 20 papers annually since 2023.
This sharp growth reflects two parallel trends:
(1) model evolution, from fine-tuned models (e.g., CodeT5~\cite{yuan2022circle}) to general-purpose LLMs such as GPT-4 and Claude-3~\cite{hu2025aprmcts}; 
and (2) methodological diversification, shifting from cloze-style styles~\cite{zhang2023gamma} to conversational and agentic paradigms (e.g., ChatRepair~\cite{xia2024automated}, ContrastRepair~\cite{kong2024contrastrepair}, RepairAgent~\cite{bouzenia2024repairagent}).
These successful explorations in semantic bug repair have laid the foundation and inspired subsequent studies on other bug types.

\textbf{Rising Interest in Domain-Specific Bug Types}.
Several domain-oriented bug types have emerged since 2023, marking a broadening of APR's application scope beyond classical semantic repair.
For example, starting in 2023, programming problems form the second-largest research stream, with the number of related studies surging to 13 in 2024.
Security vulnerability repair has become a persistent and practically important theme since 2022, maintaining 6–7 papers per year.
Most studies fine-tune domain-specific code models (e.g., CodeT5, UniXcoder~\cite{zhang2023pre}) or use GPT-4 for zero-shot~\cite{ahmad2024hardware} repair on vulnerability datasets.
Such diversification demonstrates the transferability of LLM reasoning to heterogeneous domains and suggests that APR research is progressively converging with broader software applications.

\textbf{Emerging and Niche Bug Types}.
Until 2022, LLM-based APR research is limited to about five major bug types.
However, since 2023, the field has expanded rapidly, with researchers starting to explore a wide spectrum of specialized or cross-domain bugs, each addressed by only one or two studies so far.
For example, repository-level issue repair has represented a new frontier since 2024.
These works (e.g., SWE-agent~\cite{yang2024swe}, RepoGraph~\cite{ouyang2024repograph}, OpenHands~\cite{wang2024openhands}) extend APR to realistic GitHub repositories requiring long-context reasoning, multi-file edits, and tool orchestration.
The rapid expansion from 7 to 11 papers between 2024 and 2025 demonstrates the growing emphasis on scalability and autonomous agent behavior.

Overall, the evolution of LLM-based APR research exhibits a clear trajectory, from foundational studies centered on semantic bug repair to increasingly diversified efforts tackling domain-specific, large-scale, and cross-modal bug types.
This transition reflects not only the rapid advancement of model capabilities but also the community's growing ambition to address real-world software engineering challenges.
As LLMs continue to improve in reasoning, memory, and tool integration, future APR research is expected to move toward unified frameworks that can adaptively diagnose and repair heterogeneous bugs across different abstraction levels and programming environments.

}

\summary{3}{
(1) LLMs have been applied in a wide array of repair scenarios in the literature, involving \bugnum{} bug types.
(2) In some common scenarios dominated by traditional APR, such as semantic bugs, researchers continue to invest substantial efforts in investigating the application of LLMs.
(3) Thanks to LLMs' general knowledge learned from all possible Internet data, LLM-based APR has been extended to some rare scenarios that are previously unexplored, such as hardware bugs and Web UI.
}

\section{RQ4: What key factors contribute to the integration of LLMs for APR?}
\label{sec:rq4}

\subsection{What sources of datasets are used to evaluate LLM APR studies?}
\label{sec:benchmark}

\begin{table}[t]
  \centering
  \caption{Distribution of publications across datasets}
    \resizebox{0.98\linewidth}{!}{
    \begin{tabular}{lrrrrrrr|lrrrrrrr}
    \toprule
    \textbf{Model} & \textbf{2020} & \textbf{2021} & \textbf{2022} & \textbf{2023} & \textbf{2024} & \textbf{2025} & \multicolumn{1}{l|}{\textbf{Total}} & \textbf{Model} & \textbf{2020} & \textbf{2021} & \textbf{2022} & \textbf{2023} & \textbf{2024} & \textbf{2025} & \multicolumn{1}{l}{\textbf{Total}} \\
    \midrule
    Defects4J & 1     & 1     & 2     & 12    & 19    & 16    & 51    & Codeflaws & 0     & 0     & 0     & 0     & 2     & 1     & 3 \\
    QuixBugs & 0     & 1     & 5     & 9     & 10    & 4     & 29    & ManySStuBs4J & 0     & 1     & 1     & 0     & 1     & 0     & 3 \\
    HumanEval-Java & 0     & 0     & 0     & 1     & 3     & 8     & 12    & DebugBench & 0     & 0     & 0     & 0     & 1     & 2     & 3 \\
    BFP   & 0     & 1     & 1     & 8     & 1     & 0     & 11    & DeepDev-PERF & 0     & 0     & 1     & 0     & 0     & 1     & 2 \\
    SWE-bench lite & 0     & 0     & 0     & 0     & 4     & 6     & 10    & DB    & 0     & 0     & 0     & 2     & 0     & 0     & 2 \\
    HumanEval & 0     & 0     & 0     & 4     & 4     & 2     & 10    & SWE-bench Verified & 0     & 0     & 0     & 0     & 0     & 2     & 2 \\
    CVEfixes & 0     & 0     & 1     & 3     & 3     & 1     & 8     & Refactory & 0     & 0     & 0     & 1     & 1     & 0     & 2 \\
    TFix  & 0     & 1     & 0     & 5     & 0     & 0     & 6     & RWB   & 0     & 0     & 0     & 0     & 1     & 1     & 2 \\
    MBPP  & 0     & 0     & 0     & 1     & 5     & 0     & 6     & MITRE & 0     & 0     & 0     & 1     & 1     & 0     & 2 \\
    Big-Vul & 0     & 0     & 0     & 3     & 3     & 0     & 6     & BIFI  & 0     & 0     & 0     & 2     & 0     & 0     & 2 \\
    SWE-Bench & 0     & 0     & 0     & 0     & 3     & 2     & 5     & TypeBugs & 0     & 0     & 0     & 0     & 2     & 0     & 2 \\
    BugsInPy & 0     & 0     & 0     & 0     & 2     & 2     & 4     & HumanEvalFix & 0     & 0     & 0     & 1     & 1     & 0     & 2 \\
    Vul4J & 0     & 0     & 0     & 1     & 1     & 2     & 4     & VJBench & 0     & 0     & 0     & 1     & 1     & 0     & 2 \\
    N.A.  & 0     & 0     & 0     & 3     & 1     & 0     & 4     & Pearce et al. & 0     & 0     & 0     & 1     & 1     & 0     & 2 \\
    ManyBugs & 0     & 0     & 1     & 1     & 2     & 0     & 4     & BugAID & 0     & 0     & 1     & 0     & 1     & 0     & 2 \\
    APPS  & 0     & 0     & 0     & 2     & 2     & 0     & 4     & NC    & 0     & 0     & 0     & 2     & 0     & 0     & 2 \\
    ExtractFix & 0     & 0     & 0     & 0     & 2     & 1     & 3     & ConDefect & 0     & 0     & 0     & 0     & 1     & 1     & 2 \\
    Bears & 0     & 0     & 0     & 1     & 2     & 0     & 3     & CPatMiner & 0     & 0     & 1     & 1     & 0     & 0     & 2 \\
    \bottomrule
    \end{tabular}%
    }
  \label{tab:benchmark}%
\end{table}%

Benchmarking is crucial for evaluating the performance of APR techniques, fundamentally shaping the direction of development within the research community.
For example, Defects4J~\cite{just2014defects4j} has been the common practice over the last decade, facilitating a standard comparison scenario to understand the strengths and weaknesses of proposed approaches and guiding researchers towards addressing the most pressing challenges in the field.
We identify a total of \datanum{} benchmarks from all collected studies and present the benchmarks utilized more than once in Table~\ref{tab:benchmark}.

\revise{We find that Defects4J~\cite{just2014defects4j} (51 papers) is the most popular benchmark, followed by QuixBugs~\cite{lin2017quixbugs} (29 papers), HumanEval-Java~\cite{jiang2023impact} (12 papers), both of which are existing benchmarks from previous APR studies.
This phenomenon is reasonable, as existing benchmarks are usually well-constructed and have been inspected and utilized by the community, making them the preferred choice for evaluating new APR techniques.
For example, a mass of LLM-based APR techniques are evaluated with existing popular benchmarks, such as CIRCLE~\cite{yuan2022circle}, AlphaRepair~\cite{xia2022less}, and GAMMA~\cite{zhang2023gamma}.

Besides, we notice some newly constructed benchmarks in LLM-based APR, such as HumanEval-Java~\cite{jiang2023impact} (12 papers) and Swe-bench lite (10 papers).
We summarize these new benchmarks into three categories.
The first category dataset is tailored for rare repair scenarios.
As discussed in Section~\ref{sec:rq3_repair_scenarios}, LLMs have been used in a variety of scenarios, some of which have not been considered by previous work, leading to a gap in relevant benchmarks.
As a result, with the advent of LLM-based APR techniques, researchers have also developed corresponding new datasets, such as TFix~\cite{berabi2021tfix} for static warnings, DeepDev-PERF~\cite{garg2022deepdev} for performance bugs, Du~et al.~\cite{du2023resolving} for crash bugs, and Zhang~et. al~\cite{zhang2023evaluating} for API misuse.
The second category dataset attempts to address the limitations of previous benchmarks.
For example, considering BFP~\cite{tufano2019empirical} lacking test suites, FixEval~\cite{haque2023fixeval} offers a collection of unit tests for a large scale of competitive programming problems and is evaluated with PLBART and CodeT5.
The third category of datasets is designed to address the issues unique to LLMs, particularly the data leakage problem.
For example, Jiang~et al.~\cite{jiang2023impact} create a new evaluation benchmark, HumanEval-Java, that has not been seen by LLMs during pre-training.
Zhang~et al.~\cite{zhang2023critical} extensively explore the data leakage issue of ChatGPT in the APR domain and introduce EvalGPTFix, a new benchmark from competitive programming problems after the training cutoff point of ChatGPT.
DebugBench~\cite{tian2024debugbench} is a follow-up of EvalGPTFix with a larger scale and more diverse types of bugs.
DebugBench contains 4,253 buggy programs from the
LeetCode community, covering four major bug categories and 18 minor types in C++, Java, and Python.
Similarly, ConDefects~\cite{wu2023condefects} contains 1,254 Java faulty programs and 1,625 Python faulty programs from the online competition platform AtCoder.
These collected programs are produced between October 2021 and September 2023 to address the data leakage issue for LLM-based APR approaches.
Different from the aforementioned benchmarks derived from programming problems\cite{guo2024codeeditorbench}, Silva~et al.~\cite{silva2024gitbug} introduce GitBug-Java, a reproducible benchmark comprising 199 recent Java bugs. These bugs are extracted from the 2023 commit history of 55 notable open-source repositories to mitigate the risk of data leakage.
}

\subsubsection{Trend Analysis}
\label{sec:dataset_trend}
\revise{
From Table~\ref{tab:benchmark}, a clear temporal trend is evident in the evolution of benchmarks used to evaluate LLM-based APR techniques. 
Before 2022, studies mainly rely on traditional benchmarks such as Defects4J, BFP, and QuixBugs, which are originally designed for traditional APR approaches. 
These benchmarks primarily target Java or Python bugs and offer limited coverage of diverse programming languages and bug categories. 
With the emergence of LLM-based repair around 2022, researchers begin to explore more complex and large-scale datasets. 
The number of datasets increases substantially after 2023, reflecting both the expansion of APR scenarios and the community's growing emphasis on comprehensive and standardized evaluation.

In particular, Defects4J remains the most widely used benchmark across all years (51 studies), underscoring its established role as a standard testbed in the APR community. 
However, since 2023, there is a clear shift toward more diversified benchmarks, such as HumanEval-Java (12 studies), indicating a move toward benchmarks that better align with the scale and reasoning capabilities of modern LLMs. 
These newer benchmarks are often multi-language, task-diverse, and intentionally designed to mitigate issues such as data leakage (e.g., HumanEval-Java and DebugBench). 
Furthermore, the introduction of SWE-bench in 2024–2025 reflects a methodological shift toward verifiable, reproducible, and large-scale benchmarks tailored to LLM evaluation.

Overall, the trend indicates a transition from traditional, manually curated datasets to large-scale and repository-aware benchmarks. 
This evolution marks a maturation of the field, from proof-of-concept experiments on small datasets to robust and reproducible evaluations grounded in realistic software repair scenarios. 
It also highlights a growing consensus within the community to establish standardized evaluation protocols for LLM-based APR, paving the way for fairer and more interpretable comparisons among future approaches.}

\subsection{What input forms are software bugs transformed into when utilizing LLMs?}
\begin{table}[t]
\footnotesize
  \centering
  \caption{\revise{Distribution of publications across input forms}}
    \begin{tabular}{lllp{10cm}}
    \toprule
    Input & Year & Count & Citations \\
    \midrule
    \multicolumn{1}{l}{\multirow{5}[2]{*}{Raw}} & 2021  & 3     & \cite{jiang2021cure} \cite{drain2021generating} \cite{mashhadi2021applying} \\
    \multicolumn{1}{l}{} & 2022  & 6     & \cite{ahmed2022synshine} \cite{fu2022vulrepair} \cite{li2022dear} \cite{garg2022deepdev} \cite{lajko2022towards} \cite{hu2022fix} \\
    \multicolumn{1}{l}{} & 2023  & 8     & \cite{fu2023vision} \cite{huang2023empirical} \cite{hao2023enhancing} \cite{zhang2023pre} \cite{zirak2023improving} \cite{horvath2023extensive} \cite{paul2023enhancing} \cite{zhang2023evaluating} \\
    \multicolumn{1}{l}{} & 2024  & 4     & \cite{saboor2024automated} \cite{gharibi2024t5apr} \cite{zhao2024repair} \cite{prenner2023out} \\
    \multicolumn{1}{l}{} & 2025  & 4     & \cite{wong2024investigating} \cite{gao2025teaching} \cite{luo2024fine} \cite{machavcek2025impact} \\
    \midrule
    \multicolumn{1}{l}{\multirow{5}[2]{*}{Prompt}} & 2021  & 1     & \cite{berabi2021tfix} \\
    \multicolumn{1}{l}{} & 2022  & 5     & \cite{yuan2022circle} \cite{kolak2022patch} \cite{prenner2022can} \cite{sobania2023analysis} \cite{kim2022empirical} \\
    \multicolumn{1}{l}{} & 2023  & 25    & \cite{alrashedy2023can} \cite{du2023resolving} \cite{fan2023automated} \cite{fu2023llm4sechw} \cite{huang2023chain} \cite{pearce2023examining} \cite{tol2023zeroleak} \cite{jin2023inferfix} \cite{joshi2023repair} \cite{wu2023exploring} \cite{wang2023rap} \cite{zhao2023right} \cite{zirak2023improving} \cite{liu2024refining} \cite{nashid2023retrieval} \cite{mohajer2023skipanalyzer} \cite{napoli2023evaluating} \cite{gao2023makes} \cite{ribeiro2023gpt} \cite{koutcheme2023training} \cite{paul2023enhancing} \cite{liventsev2023fully} \cite{tian2023chatgpt} \cite{xia2023automated} \cite{zhang2023critical} \\
    \multicolumn{1}{l}{} & 2024  & 33    & \cite{ahmad2024hardware} \cite{chow2024pyty} \cite{fatima2024flakyfix} \cite{guo2023exploring} \cite{yang2024swe} \cite{hao2024retyper} \cite{liu2024marscode} \cite{li2024exploring} \cite{kulsum2024case} \cite{liu2024craftrtl} \cite{liu2024fastfixer} \cite{le2024study} \cite{liu2023chatgpt} \cite{omari2024investigating} \cite{pan2024lost} \cite{nong2024chain} \cite{peng2024domain} \cite{zhang2024vuladvisor} \cite{tsai2023rtlfixer} \cite{wadhwa2023frustrated} \cite{wadhwa2024core} \cite{wang2024divide} \cite{chen2024large} \cite{yao2024hdldebugger} \cite{hidvegi2024cigar} \cite{hossain2024deep} \cite{zhang2024pydex} \cite{zhao2024peer} \cite{zhou2024leveraging} \cite{ruiz2024novel} \cite{shi2024code} \cite{xiang2024far} \cite{xue2024exploring} \\
    \multicolumn{1}{l}{} & 2025  & 22    & \cite{dai2025less} \cite{garg2023rapgen} \cite{ehsani2025towards} \cite{ma2025swe} \cite{wei2025swe} \cite{xu2023guiding} \cite{xia2025demystifying} \cite{yuan2024designrepair} \cite{zhang2024acfix} \cite{nong2025appatch} \cite{cao2023study} \cite{feng2025integrating} \cite{hu2025aprmcts} \cite{kong2025demystifying} \cite{lin2025haprepair} \cite{ma2025alibaba} \cite{parasaram2024fact} \cite{xu2025aligning} \cite{yang2024multi} \cite{huang2025template} \cite{zhang2025repair} \cite{silva2023repairllama} \\
    \midrule
    \multicolumn{1}{l}{\multirow{3}[2]{*}{Mask}} & 2022  & 2     & \cite{xia2022less} \cite{ribeiro2022framing} \\
    \multicolumn{1}{l}{} & 2023  & 8     & \cite{first2023baldur} \cite{wu2023effective} \cite{xia2023plastic} \cite{zhang2023gamma} \cite{jiang2023impact} \cite{koutcheme2023automated} \cite{wei2023copiloting} \cite{xia2023automated} \\
    \multicolumn{1}{l}{} & 2024  & 4     & \cite{peng2024domain} \cite{wan2024automated} \cite{lin2024one} \cite{yang2024revisiting} \\
    \midrule
    \multicolumn{1}{l}{\multirow{4}[2]{*}{Conversation}} & 2022  & 1     & \cite{sobania2023analysis} \\
    \multicolumn{1}{l}{} & 2023  & 10    & \cite{madaan2023selfrefine} \cite{huang2023agentcoder} \cite{jiang2023selfevolve} \cite{moon2023coffee} \cite{olausson2024is} \cite{liu2023automated} \cite{zhang2023steam} \cite{wuisang2023evaluation} \cite{zhang2023critical} \cite{zhang2023self} \\
    \multicolumn{1}{l}{} & 2024  & 12    & \cite{alhanahnah2025empirical} \cite{chen2024teaching} \cite{ding2024cycle} \cite{hu2024leveraging} \cite{zhao2024enhancing} \cite{lin2024drplanner} \cite{tang2024code} \cite{yang2024cref} \cite{zheng2024opencodeinterpreter} \cite{zhong2024ldb} \cite{xia2024automated} \cite{yin2024thinkrepair} \\
    \multicolumn{1}{l}{} & 2025  & 10    & \cite{brancas2025combining} \cite{deligiannis2023fixing} \cite{liu2025error} \cite{orvalho2025counterexample} \cite{charalambous2023new} \cite{bouzenia2024repairagent} \cite{ehsani2025bug} \cite{kong2024contrastrepair} \cite{ruiz2025art} \cite{kang2023explainable} \\
    \midrule
    \multicolumn{1}{l}{\multirow{2}[2]{*}{Repository}} & 2024  & 3     & \cite{arora2024masai} \cite{chen2024coder} \cite{zhang2024autocoderover} \\
    \multicolumn{1}{l}{} & 2025  & 11    & \cite{huang2025seeing} \cite{guo2025omnigirl} \cite{mundler2024swt} \cite{ke2025niodebugger} \cite{antoniades2024swe} \cite{ouyang2024repograph} \cite{tao2024magis} \cite{wang2024openhands} \cite{yu2025patchagent} \cite{lee2024unified} \cite{ye2025adverintent} \\
    \midrule
    \multicolumn{1}{l}{\multirow{4}[2]{*}{Conversation}} & 2022  & 1     & \cite{li2022dear} \\
    \multicolumn{1}{l}{} & 2023  & 2     & \cite{horvath2023extensive} \cite{zhang2023neural} \\
    \multicolumn{1}{l}{} & 2024  & 2     & \cite{zhang2024autocoderover} \cite{zhou2024large} \\
    \multicolumn{1}{l}{} & 2025  & 2     & \cite{li2025hybrid} \cite{ouyang2025knowledge} \\
    \midrule
    Multimodal & 2025  & 1     & \cite{huang2025seeing} \\
    \bottomrule
    \end{tabular}%
  \label{tab:input}%
\end{table}%

\delete{Thanks to the powerful natural language understanding capabilities of LLMs, the inputs of LLM-based APR contain rich information, thus more complex than that of traditional APR techniques~\cite{zhang2023survey}.
We summarize various input forms into five categories according to their data types.
As illustrated in \delete{Fig}\revise{Table}.~\ref{fig:input}, we find 52\% of collected papers leverage prompt engineering to feed LLMs with bug-fixing information, and 18\% utilize a conversational-style representation to provide dynamic information.
We also find only 18\% of LLM-based APR studies adopt raw bug-fixing inputs in a manner similar to traditional DL/ML models~\cite{zhang2023survey}.
We will discuss the five input representations utilized by LLMs as follows.}
\revise{

Thanks to the powerful natural language understanding and code reasoning capabilities of LLMs, the inputs used in LLM-based APR are much richer and more diverse than those in traditional APR techniques~\cite{zhang2023survey}.
We categorize these input representations into six major types according to their data modalities and structural forms.
As shown in Table \ref{tab:input}, prompt-based inputs remain the dominant form, adopted by 47.78\% (86) of all collected studies, where LLMs are guided through carefully engineered textual instructions or templates describing the bug-fixing task.
Meanwhile, chat-based inputs (18.33\%) have gained popularity with the emergence of conversational LLMs, enabling iterative and context-aware repair dialogues.
Only 13.89\% of studies still rely on raw code inputs similar to those used in traditional DL/ML repair models.
In addition, a small but growing proportion of works explore mask-based (7\%) and repository-level (5\%), indicating a trend toward richer contextualization and multi-source reasoning in future APR systems.
We will discuss the five input representations utilized by LLMs as follows.
}

\ding{182}\textbf{Raw Bug-fixing Input}.
Similar to most traditional learning-based APR, this type of input regards APR as an NMT task, which translates a sentence from one source language (i.e., buggy code) to another target language (i.e., fixed code).
Such representation directly feeds LLMs with the buggy code snippet and has been typically employed to train LLMs with supervised learning in semantic bugs~\cite{drain2021deepdebug,mashhadi2021applying,zirak2023improving} security veulnerabilities~\cite{zhang2023pre,fu2022vulrepair}, and static warnings~\cite{kim2022empirical}.
For example, Zhang~et al.~\cite{zhang2023pre} investigate the performance of three bug-fixing representations (i.e., context, abstraction, and tokenization) to fine-tune five LLMs for vulnerability repair.

\ding{183}\textbf{Prompt Input}.
This type of input incorporates more information into the buggy code.
The prompt concatenates different input components with a prefixed prompt, thus effectively bridging the gap between pre-trained tasks and the APR downstream task.
For example, CIRCLE~\cite{yuan2022circle} utilizes a manually-designed prompt template to convert buggy code and corresponding context into a unified fill-in-the-blank format.
Particularly, they utilize ``Buggy line:'' and ``Context:'' to denote the buggy and contextual code, and they utilize ``The fixed code is:'' to query a T5-based model to generate candidate patches according to the previous input.
Besides, TFix~\cite{berabi2021tfix}, Zirak~et al.~\cite{zirak2023improving} and Kim~et al.~\cite{kim2022empirical} represent all valuable information about the bug as a single piece of text, including bug type, bug message, bug line, and bug context.
Furthermore, InferFix~\cite{jin2023inferfix} and RAP-Gen~\cite{wang2023rap} construct prompts by retrieving relevant repair examples from an external codebase.

\ding{184}\textbf{Mask Input}.
This type of input masks the buggy code and queries LLMs to fill the masks with the correct code tokens.
Unlike the above input forms, the mask input reformulates the APR problem as a cloze-style task and directly leverages LLMs' pre-training objectives in a zero-shot setting.
AlphaRepair~\cite{xia2022less} is considered the first work to demonstrate the potential of mask inputs, and researchers have proposed various follow-ups to better perform mask prediction for patch generation, such as GAMMA~\cite{zhang2023gamma} with well-summarized repair patterns, FitRepair~\cite{xia2023plastic} with the plastic surgery hypothesis, Repilot~\cite{wei2023copiloting} with a completion engine, as well as empirical studies~\cite{xia2023automated,jiang2023impact}.

\ding{185}\textbf{Conversation-Style Input}
This type of input further extends the prompt input with feedback-driven chats like humans.
Conversation-style representation contains more complex information, such as dynamic execution results, while iteratively improving generated patches through multiple rounds of dialogue.
For example, Sobania~et al.~\cite{sobania2023analysis} conduct an early exploration into the feasibility of leveraging ChatGPT's conversational capabilities for program repair, motivating some follow-ups~\cite{zhang2023critical,xia2024automated}.

\ding{186}\textbf{Structure-Aware Input}.
This type of input represents source code as syntactic structures, such as Abstract Syntax Trees (ASTs).
For example, Horvath~et al.~\cite{horvath2023extensive} utilizes RoBERTa and GPTNeo to encode ASTs for program repair.
Besides, VulMaster~\cite{zhou2024large}  utilizes the AST as part of its input to capture the structural aspects of the vulnerable code.

\ding{187}\textbf{Repository-level Input}.
This type of input extends beyond single-file or localized code fragments to encompass the entire software repository.
Such inputs enable LLMs to access multi-file dependencies and repository-level relationships, thereby capturing the holistic semantics required for realistic bug fixing.
Unlike earlier APR paradigms that rely on static context windows, repository-level inputs are dynamically retrieved, updated, and reasoned about by autonomous LLM agents capable of tool invocation and code search.
For example, approaches such as SWE-agent~\cite{yang2024swe}, AutoCodeRover~\cite{zhang2024autocoderover}, and RepoGraph~\cite{ouyang2024repograph} leverage agents equipped with retrieval and execution tools to navigate the repository, localize faults, edit multiple files, and validate candidate patches.

\subsubsection{Trend Analysis}
\label{sec:input_trend}

\revise{
Table~\ref{tab:input} presents a clear evolution in how software bugs are represented as inputs for LLM-based APR. 
First, in the early years (2021–2022), most studies use raw inputs similar to those of traditional learning–based APR models. 
At this stage, researchers mainly adapt models such as CodeBERT and CodeT5 to program repair tasks, where the input consists primarily of buggy code and its immediate context. 
There are also some studies that adopt mask-based inputs to simulate the pre-training objectives of LLMs, allowing models to directly predict the correct code tokens without additional fine-tuning. 
Second, starting from 2023, prompt-based inputs rapidly become dominant, accounting for nearly half of all studies (47.78\%). 
This trend coincides with the widespread availability of instruction-tuned models such as GPT-3.5 and with the increasing use of prompt engineering to align model behavior with repair objectives. 
Prompt-based inputs integrate natural language instructions, examples, and bug context, marking a significant shift from pure code transformation toward task-oriented reasoning.
At the same time, conversation-style inputs emerge as an important paradigm from 2023 onward, extending prompt-based formats with dynamic, feedback-driven interactions between the LLM and its environment (e.g., test execution results). 
The continuous increase in conversation-style studies through 2025 reflects a growing interest in interactive and human-in-the-loop repair processes.
Very recently, researchers have attempted to explore repository-level and multimodal inputs (first appearing in 2024–2025). 
These approaches move beyond single-file contexts to capture multi-file dependencies, project histories, and even visual information, supporting real-world repair scenarios that require long-context reasoning and cross-file understanding.

Overall, the evolution of input forms reflects a broader shift in the field, from static, isolated code snippets to dynamic, context-aware, and semantically rich representations. 
This trend reflects how LLM-based APR increasingly emphasizes reasoning, interactivity, and holistic program understanding, rather than treating bug fixing as a simple sequence-to-sequence translation problem in learning-based APR~\cite{zhang2023survey}.
}

\subsection{How are LLMs used to support patch correctness?}
\label{sec:apca}

Test-overfitting has recently been the focus of APR research due to the mainstream test-driven generate-and-validate repair workflow in the community.
Particularly, after candidate patches are generated, the developer-written test suites are utilized as the oracle to validate the correctness of patches, and the passing patches are returned to developers.
However, as an incomplete specification, test suites only describe a part of the program's behavioral space, resulting in plausible yet overfitting patches without fixing bugs.
Thus, developers need to spend enormous effort filtering out overfitting patches manually, even resulting in a negative debugging performance~\cite{tao2014automatically,zhang2022program1}.
Researchers have proposed various automated \textbf{patch correctness} assessment (APCA) techniques to identify whether a plausible patch is overfitting or not, so as to improve the quality of returned patches.
For example, PATCH-SIM~\cite{xiong2018identifying} assesses the correctness of patches by calculating the similarity of dynamic execution traces.
PATCH-SIM is acknowledged as a foundational work in the APCA field, providing crucial guidance for the development of follow-up works, particularly learning-based or LLM-based ones~\cite{zhang2023survey}.

\begin{table*}[htbp]
  \centering
  \footnotesize
  \caption{A summary of existing APR studies using LLMs to predict patch correctness.}
    \begin{tabular}{cccl}
    \toprule
    \textbf{Year} & \textbf{Study} & \textbf{LLMs} & \textbf{Repository} \\
    \midrule
    2020  & Tian~~et~al.~\cite{tian2020evaluating} & BERT  & \url{https://github.com/TruX-DTF/DL4PatchCorrectness} \\
    2022  & Quatrain~\cite{tian2022change} & BERT  & \url{https://github.com/Trustworthy-Software/Quatrain} \\
    2023  & Tian~~et~al.~\cite{tian2023best} & BERT  & \url{https://github.com/HaoyeTianCoder/Panther} \\
    2023  &  Invalidator~\cite{le2023invalidator} & CodeBERT & \url{https://github.com/thanhlecongg/Invalidator} \\
    2023  & PatchZero~\cite{zhou2024leveraging} & StarCoder & N.A. \\
    2024  & APPT~\cite{zhang2024appt} & BERT, CodeBERT, GraphCodeBERT & \url{https://github.com/iSEngLab/APPT} \\
    \bottomrule
    \end{tabular}%
  \label{tab:rq3_apca}%
\end{table*}%

Table~\ref{tab:rq3_apca} presents existing APCA studies involving LLMs.
We summarize them into three stages.

\ding{182}\textbf{LLMs as Feature Extractor.}
In 2020, Tian~et al.~\cite{tian2020evaluating,tian2023best} empirically explore the performance of code embeddings via representation learning models in reasoning about patch correctness.
Following the similarity-based pipeline from PATCH-SIM, they first calculate the similarities of patched and buggy code snippets based on code embeddings, and then predict patch correctness with a binary classifier.
They consider four embedding models, including re-trained (i.e., Doc2vec, code2vec and CC2vec) and pre-trained models (i.e., BERT), which is the first APCA study empowered with LLMs.
Recently, Le~et al.~\cite{le2023invalidator} propose Invalidator, to assess the correctness of patches via semantic and syntactic reasoning.
Similar to Tian~et al.~\cite{tian2020evaluating}, they utilize CodeBERT to extract code features and train a classifier for prediction.
Unlike the above studies calculating similarities of patches, Tian~et al.~\cite{tian2022change} formulate APCA as a question-answering (QA) problem and propose Quatrain.
Quatrain first utilizes CodeBERT to encode bug reports and patch descriptions and trains a QA model for prediction.

\ding{183}\textbf{Fine-tuning LLM-based APCA.}
In 2024, Zhang~et al.~\cite{zhang2024appt} propose APPT, equipped with BERT as the encoder stack, followed by an LSTM stack and a deep learning classifier.
Unlike previous studies~\cite{tian2020evaluating,tian2023best} limiting BERT to extract features without benefiting from training, APPT further fine-tunes LLMs in conjunction with other components as a whole pipeline to fully adapt it specifically for reasoning about patch correctness.
APPT is implemented with BERT by default and is also proven generalizable to other advanced LLMs, such as CodeBERT and GraphCodeBERT.

\ding{184}\textbf{Zero-shot LLM-based APCA.}
In 2023, Zhou~et al.~\cite{zhou2024leveraging} propose PatchZero to explore the feasibility of LLMs in predicting patch correctness with a zero-shot setting.
PatchZero directly queries LLMs to generate the next token about patch correctness (i.e., a token either ``correct'' or ``overfitting'') based on previous tokens, which is similar to LLMs' original pre-training objective.

\subsection{How are LLMs utilized to facilitate both code generation and program repair?}

Compared with traditional APR approaches~\cite{zhang2023survey} that usually employ heuristics or neural networks to generate a multitude of patches in one go, LLMs can iteratively refine the generated patches based on the outcomes of dynamic execution.
As mentioned in Section~\ref{sec:three_ways}, the iterative patch generation capabilities brought by LLMs have facilitated the emergence of conversational-based repair techniques~\cite{pearce2023examining,peng2024domain,deligiannis2023fixing,zhang2023critical,prenner2022can,sobania2023analysis,napoli2023evaluating}.
In addition to conversational-based repair, we have identified some efforts that improve code generation performance by repairing LLM-generated code with feedback, referred to as self-repair.
Different from conversational-based repair, which utilizes feedback for patch generation, self-repair approaches leverage LLMs to identify code errors implemented by themselves via investigating execution results and explaining the generated code.
Such self-repair approaches integrate code generation and program repair, which can be considered a further step toward automated programming.

\begin{table}[htbp]
  \centering
  \footnotesize
  \caption{A summary of existing studies that use APR to boost code generation.}
  \resizebox{\linewidth}{!}{
    \begin{tabular}{llll}
    \toprule
    \textbf{Year} & \textbf{Study} & \textbf{LLMs} & \textbf{Repository} \\
    \midrule
    2023  & AgentCoder~\cite{huang2023agentcoder} & GPT-3.5,GPT-4,Claude,PaLM & N.A. \\
    2023  & Self-Edit~\cite{zhang2023self} & InCoder,CodeGen,Codex & \url{https://github.com/zkcpku/Self-Edit} \\
    2024  & Self-Debugging~\cite{chen2024teaching} & \delete{ChatGPT}\revise{GPT-3.5},GPT-4,Codex,StarCoder & N.A. \\
    2024  & Zheng~et al.~\cite{zheng2024opencodeinterpreter} & CodeLLaMA,DeepseekCoder & \url{https://github.com/OpenCodeInterpreter/OpenCodeInterpreter} \\
    2024  & CYCLE~\cite{ding2024cycle} & CodeGen,StarCoder & \url{https://github.com/ARiSE-Lab/CYCLE\_OOPSLA\_24} \\
    2024  & LDB~\cite{zhong2024ldb} & StarCoder,CodeLLaMA,GPT-3.5 & \url{https://github.com/FloridSleeves/LLMDebugger} \\
    2024  & Olausson~et al.~\cite{olausson2024is} & GPT-3.5,GPT-4,CodeLLaMA & \url{https://github.com/theoxo/self-repair} \\
    2024  & Hu~et al.~\cite{hu2024leveraging} & GPT-4 & N.A. \\
    \bottomrule
    \end{tabular}%
    }
  \label{tab:self}%
\end{table}%

Table~\ref{tab:self} presents some LLM-based studies that leverage program repair to boost code generation.
Self-Edit~\cite{zhang2023self} represents the first attempt to adopt a neural code editor that takes both the generated code and error messages as inputs to improve the code quality on the competitive programming task.
Self-Edit is evaluated with both fine-tuned models (i.e., PyCodeGPT, GPT-Neo, CodeGen, GPT-Neo, InCoder, GPT-J) and prompt-based LLMs (i.e., InCoder, CodeGen, Codex).
Chen~et al.~\cite{chen2024teaching} from DeepMind propose Self-Debugging to teach LLMs to debug their own predicted code via few-shot prompting, including Codex, ChatGPT, GPT-4 and StarCoder.
There also exist some similar self-repair studies, including OpenCodeInterpreter~\cite{zheng2024opencodeinterpreter}, Cycle\cite{ding2024cycle}, LDB~\cite{zhong2024ldb}, SelfEvolve~\cite{jiang2023selfevolve}, Self-Refine~\cite{madaan2023selfrefine}, Hu~et al.~\cite{hu2024leveraging}, AgentCoder~\cite{huang2023agentcoder}.
Recently, Olausson~et al.~\cite{olausson2024is} conduct an empirical study to investigate the ability of CodeLlama, GPT-3.5, and GPT-4 to perform self-repair in code generation. 
They find that self-repair is not a panacea for code generation challenges, as existing LLMs often fail to provide reliable, accurate, and valuable feedback on why the code is incorrect.

\subsection{How often do the collected LLM-based APR papers provide publicly available artifacts?}

Open science plays a crucial role in advancing scientific progress through principles of transparency, reproducibility, and applicability. 
Given the benefits, the SE community has been actively promoting open science principles and encouraging all researchers to share their artifacts, thereby bolstering the reliability of research findings.
In this section, we investigate the extent to which the analyzed papers make their artifacts publicly accessible.

We find that 80 studies provide the replication packages in their papers, accounting for 62.99\% (80/\papernum{}) of all collected studies.
Among 78 studies that propose novel LLM-based APR approaches, which is the largest contribution type in Table~\ref{tab:contribution}, we find that 53.85\% (42/78) of them fail to make their artifacts publicly available.
This makes it difficult for researchers to validate experimental findings, conduct quantitative comparisons with existing studies, and build follow-ups instead of reinventing the wheel.
Considering that some papers have not been published, we then focus on top-tier SE venues, i.e., ICSE, ASE, FSE, ISSTA, TSE and TOSEM, and identify 86.84\% of papers (33/38) make related artifacts publicly open, indicating a high-quality commitment to reproducibility in high-quality papers.
Besides, some studies only provide datasets or trained models without source code or essential instructions.
Overall, open science is a critical challenge in advancing LLM-based APR research development because many factors, such as datasets, data pre-processing methods, source code, hyper-parameters, and documents, may lead to the reproducibility of studies.
Therefore, we hope that researchers in the LLM-based APR community can provide high-quality open-source artifacts for convenient reproduction.
\revise{For transparency and reproducibility, we provide the detailed artifact availability statistics and corresponding open-source links for all collected papers in our replication package repository.}

\summary{4}{
(1) We summarize \datanum{} different datasets that are utilized to benchmark LLMs in fixing bugs.
(2) Defects4J, QuixBugs, BFP, CVEfixes, and Big-Vul are most frequently adopted in the LLM-based APR.
(3) We categorize the input forms within all collected papers into five groups: raw bug-fixing input, prompt input, mask input, conversation-style input, and structure-aware input.
(4) Prompt input is the most frequently used form in applying LLMs to program repair, indicating that designing effective prompts is particularly important for leveraging LLMs' natural language processing capabilities.
(5) We summarize some studies that leverage LLMs to predict patch correctness.
(6) 62.99\% of all collected papers have made their artifacts open source, and the ratio increases to 86.84\% for top-tier SE publications. 
}

\section{\deletenew{Challenges \& Opportunities of LLM-based APR}\revisenew{Research Agenda for LLM-based APR}}
\label{sec:challenges}

\begin{figure}[t]
\centering
    \includegraphics[width=0.9\linewidth]{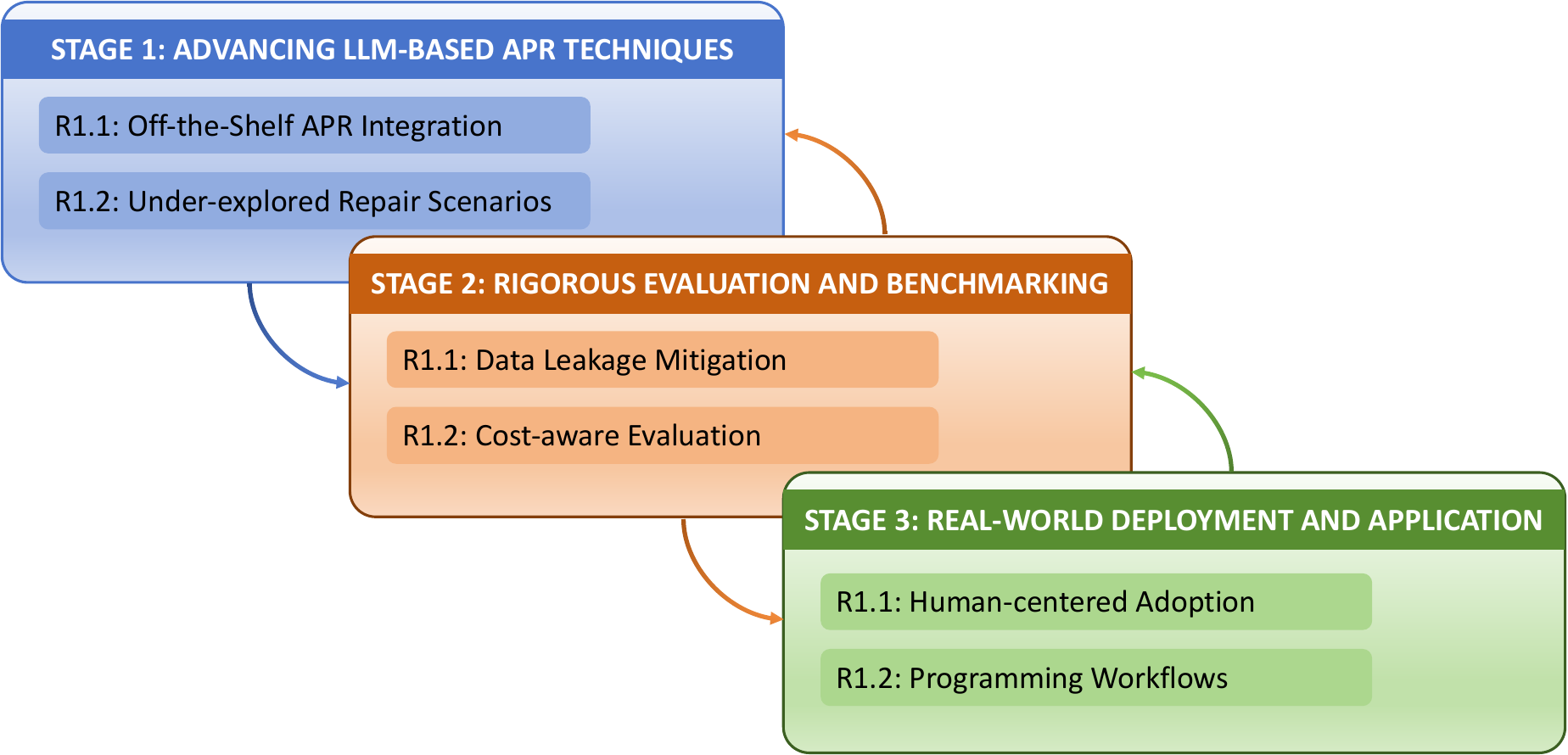}
    \caption{\revisenew{A three-stage future research roadmap for LLM-based APR.}}
    \label{fig:future_roadmap}
\end{figure}

\deletenew{Our study reveals the following crucial challenges and practical guidelines for future LLM-based APR work.}
\revisenew{
Based on our systematic analysis, this section presents a structured research agenda that outlines concrete and actionable directions for future studies on LLM-based APR.
As illustrated in Figure~\ref{fig:future_roadmap}, we organize future research directions into three progressive stages: 
(1) advancing LLM-based APR techniques, 
(2) rigorous evaluation and benchmarking, and 
(3) real-world deployment and adoption.
}

\subsection{\revisenew{Stage 1: Advancing LLM-based APR Techniques}}

\subsubsection{Off-the-Shelf APR Integration}
\revisenew{
\textbf{Current state.}
As mentioned in Section~\ref{sec:rq3_repair_scenarios}, researchers typically utilize LLMs as core backbones to design novel repair approaches in an end-to-end setting, such as sequence-to-sequence learning~\cite{yuan2022circle,berabi2021tfix}.
Parallelly, the community has seen some explorations treating LLMs as components integrated into existing repair workflow.
These studies attempt to boost the capabilities of off-the-shelf APR approaches instead of proposing new techniques.
For example, CURE~\cite{jiang2021cure} combines GPT and CoCoNut~\cite{lutellier2020coconut} to capture code syntax for the APR task.
Built on top of DLFix~\cite{li2020dlfix}, DEAR~\cite{li2022dear} attempts to fix multi-hunk, multi-statement bugs by fine-tuning BERT to learn fixing-together relationships among statements, i.e., identifying if two statements are needed to be fixed together or not.
Recently, GAMMA~\cite{zhang2023gamma} integrates LLMs into the traditional template-based APR TBar by querying LLMs to generate masked code tokens instead of retrieving donor code from local files.
These efforts demonstrate the potential of integrating LLMs with off-the-shelf APR techniques, yet there is currently a lack of more in-depth work in this area.

\textbf{Opportunities.}
In the future, researchers could attempt to combine LLMs with more traditional APR techniques.
For example, it is promising to utilize LLMs to help SMT solvers generate patches for constraint-based APR, or feed search algorithms with LLM-generated potential patches to build search space for heuristic-based APR.
Besides, domain-specific repair techniques can benefit from the powerful code-understanding capabilities of LLMs, thus extending to a broader range of repair scenarios.
For example, we can design fix templates for specific scenarios, such as static warnings, and then utilize the general knowledge contained in LLMs to generate correct patches.
}

\subsubsection{Under-explored Repair Scenarios}

\revisenew{
\textbf{Current state.}
As summarized in Section~\ref{sec:rq3_repair_scenarios}, we observe that most existing LLM-based APR studies are concentrated on a limited number of bug types, particularly semantic bugs.
However, there exist some rare repair scenarios that benefit less from LLMs, such as hardware bugs (one paper) and concurrency bugs (zero paper).
\deletenew{The key challenge lies in insufficient training data from which LLMs can learn.}
\revisenew{A key challenge lies in the scarcity of labeled training data and the increased complexity associated with these scenarios.}

\textbf{Opportunities.}
We suggest that future work concentrate on three possible directions to broaden the scope of LLM applications for more repair scenarios, such as software requirement~\cite{xie2023impact} and fuzzing~\cite{zhang2023program}.
\begin{itemize}
    \item Transfer learning is an effective training approach for rare scenarios.
    We can first fine-tune LLMs with abundant data in a source scenario and then utilize a small amount of data to transfer the acquired knowledge to a target scenario.
    The source and target scenarios should have similar data distributions.
    For example, Zhang~et al.~\cite{zhang2023pre} demonstrate that transferring learning from bug-fixing can improve the vulnerability repair performance of five LLMs by 9.40\% on average.

    \item It is promising to utilize the cloze-style APR with repair patterns to generate patches for rare scenarios.
    Unlike fine-tuning, which requires substantial labeled data, cloze-style APR effectively leverages expert domain knowledge to guide LLMs in performing APR based on their existing pre-trained knowledge without additional retraining.
    For example, researchers design repair patterns and exploit LLMs' pre-training objective (i.e., predicting masked tokens) to directly generate correct patches by filling masked tokens under these patterns.
    This makes the cloze-style APR particularly effective for scenarios with limited training data or domain-specific bug types.
    Besides, for scenarios not previously encountered by LLMs, researchers can employ unlabeled data in the target scenario to learn the data distribution of the project or language under test in an unsupervised learning setting.
    
    \item In-context learning and prompt engineering are feasible solutions for directly querying billion-parameter LLMs to generate correct code in a target scenario, given that these models are immensely large and encompass virtually all data available on the Internet.
\end{itemize}
}

\subsection{\revisenew{Stage 2: Rigorous Evaluation and Benchmarking}}

\subsubsection{Data Leakage Mitigation}
\revisenew{
\textbf{Current state.}
As highlighted in Section~\ref{sec:benchmark}, Zhang et al.~\cite{zhang2023critical} identify that existing repair benchmarks have been inadvertently included in the pre-training data of popular LLMs, such as ChatGPT, through web scraping and other methods. 
For example, ChatGPT is able to enumerate all projects within Defects4J~\cite{just2014defects4j}, one of the most popular APR benchmarks. 
Researchers~\cite{zhang2023gamma} can ascertain the exposure for open-source LLMs by inspecting pre-training data against benchmarks.
It is significantly more challenging with more powerful black-box LLMs due to a lack of training details.
However, as preliminary explorations, they are mainly limited in the quantity of involved bugs and the variety of bug types.
For example, all of them are created from programming problems with only small-scale or medium-scale buggy solutions, without delving into large-scale, real-world projects that encompass complex API calls, such as Defects4J~\cite{just2014defects4j}.
More importantly, data leakage issues in other repair scenarios, such as security vulnerabilities and API misuse, continue to be overlooked.
There may be overlaps among datasets across different repair scenarios, such as Defects4J, which also serves as a source for a subset of the API misuse dataset~\cite{zhang2023evaluating}.
Overall, the risk of data leakage introduces bias into the benchmarking of existing work, necessitating urgent efforts from researchers to mitigate it.

\textbf{Opportunities.}
We recommend that future work be conducted in the following two directions.
First, it is crucial to construct a large-scale benchmark free from data leakage that contains real-world projects so as to evaluate the actual fix capabilities of LLMs in a more practical debugging scenario.
Commercial closed-source software, real-time updated programming websites, or manually written programs may serve as potential data sources.
Second, considering the variety of bug types that LLMs have been applied to, researchers need to consider and attempt to address the data leakage risk when conducting related studies.
}

\subsubsection{Cost-aware Evaluation}
\revisenew{
\textbf{Current state.}
As mentioned in Section~\ref{sec:rq2_llms}, the literature tends to employ the growing size of LLMs to achieve better performance, such as from T5-250M in CIRCLE to Codex-12B in InferFix.
This trend is reasonable as Xia~et al.~\cite{xia2023automated} demonstrate that larger models typically generate more correct patches for software bugs.
However, APR, being a task highly relevant to humans, the millions, billions and even trillions of parameters pose significant challenges in the development workflow.
First, training LLMs with billions (or even more) of parameters is a highly time-consuming and resource-intensive process. 
The GPU resources required for such training are often prohibitively expensive, making them inaccessible for many researchers in both academic and industrial settings.
For example, in Section~\ref{sec:three_ways}, we observe that most fine-tuning-LLM-based APR studies utilize CodeT5/T5 or similar-sized models, except for InferFix from Microsoft, which fine-tunes Codex. 
However, InferFix is proposed by the world-leading technology company Microsoft and trained with industrial-grade hardware, i.e., 64 32-GB-V100-GPUs.
Second, despite the increased likelihood of generating correct patches with larger models, the patch generation time cost also increases.
For example, Jiang~et al.~\cite{jiang2023impact} demonstrate PLBART takes 0.70–0.89 seconds on average to generate a correct patch, CodeGen, although fixing more bugs than PLBART with more parameters, requires 3.64–13.88 seconds on average to generate a correct patch.
Third, the increase in patch generation time further compresses the time available for patch validation, as developers need to spend more time waiting for the model to complete its inference.
For example, Shi~et al.~\cite{shi2022compressing} demonstrate that the vast storage and runtime memory consumption, coupled with high inference latency, make these LLMs prohibitive for integration in modern IDEs, especially on resource-constrained or real-time terminal devices.

\textbf{Opportunities.}
In the future, to address the first challenge, it is promising to explore the potential of parameter-efficient fine-tuning approaches on APR, such as prefix-tuning and low-rank adaptation~\cite{ding2023parameter}.
\revise{In addition, recent sparsity-aware optimization frameworks such as SparseGPT~\cite{frantar2023sparsegpt}, SparseML~\cite{sparsemlNeuralMagic}, and DeepSparse~\cite{neuralmagic2024deepsparse} demonstrate promising directions for improving inference efficiency of LLMs, which could also benefit LLM-based APR in future work.}
To address the second challenge, researchers can optimize the size of LLMs without significantly compromising their performance, such as model pruning, quantization, and knowledge distillation~\cite{shi2022compressing}.
To address the third challenge, we recommend boosting patch validation with advanced strategies, such as mutation testing~\cite{demillo2006hints,xiao2024accelerating}, or utilizing LLMs to rank candidate patches before validation.
}

\subsection{\revisenew{Stage 3: Real-world Deployment}}

\subsubsection{Human-centered Adoption}
\revisenew{
\textbf{Current state.}
As summarized in Section~\ref{sec:rq3_repair_scenarios}, with the introduction of LLMs, APR has achieved groundbreaking progress in terms of the number of correctly fixed bugs in popular benchmarks, exemplified by the advancements on Defects4J-1.2 from 64 bugs by CIRCLE, to 74 bugs by AlphaRepair, and 82 bugs by GAMMA.
The progress prompts us to consider whether LLMs have indeed facilitated improvements in real-world debugging and how LLM-based APR has broader implications for developers' daily activities.
Previous research~\cite{zhang2022program,winter2022let} highlights the potential pitfalls of developing APR tools without adequate feedback from developers, which could compromise their effectiveness in real-world deployment.
However, there remains a significant gap in our understanding of how software engineers tackle software problems in practical settings, including their use of dedicated debugging tools and their expertise in debugging techniques.

\textbf{Opportunities.}
Researchers should conduct human studies to gain deeper insights into the maturity and reliability of LLM-based APR tools in terms of human factors.
Possible directions are to investigate whether LLMs can assist developers in reducing the debugging cost, such as fixing more bugs, accelerating the bug-fixing process, and handling more complex bugs. 
Besides, it would be valuable to investigate developers' perceptions and interactions with LLMs based on their practical experiences and established debugging practices.
}

\subsubsection{Programming Workflows}
\revisenew{
\textbf{Current state.}
As mentioned in Section~\ref{sec:background}, existing APR approaches operate under a ``near-correct assumption''~\cite{offutt1994empirical,demillo2006hints}, which posits that experienced programmers are capable of writing programs that are almost correct, requiring only minimal modifications to rectify bugs and ensure compliance with all test cases.
This assumption has long served as the foundation for APR research development. 
However, the evolution of LLMs and their application in programming suggests a future where APR can transcend its traditional boundaries, moving towards a more holistic approach in conjunction with fully autonomous programming.
In this context, APR can be reimagined not only just as a tool for correcting minor coding errors, but as an integral component of a self-correcting, self-improving system that iteratively enhances the quality of automatically generated code.
Fortunately, we have observed initial explorations into combining repair with programming, yet these efforts are far from sufficient.
For example, Fan~et al.~\cite{fan2023automated} employ LLMs to fix buggy solutions generated by themselves, and recent studies~\cite{zhang2023critical,xia2024automated} iteratively refine auto-generated code with dynamic execution.

\textbf{Opportunities.}
In the future, the opportunities presented by integrating APR with fully autonomous programming are vast~\cite{lyu2024automatic}.
First, it is flexible to implement collaborative Human-AI Programming tools~\cite{liang2021interactive}, where the initial code is written by developers and continuously optimized and repaired by LLMs. 
Particularly for complex problem-solving, LLMs suggest innovative solutions that had not been considered by human programmers.
This partnership could accelerate development cycles, reduce the burden of debugging on human developers, and lead to more creative and effective solutions.
Second, the general knowledge inherent in LLMs endows them with the capability to support multiple downstream tasks, which opens up the possibility of bridging the gap between code generation, testing, debugging, and fixing.
For example, fault localization is a precondition of patch generation, while patch validation can reflect the accuracy of fault localization, making these takes interconnected.
Thus, it is promising to explore the capabilities of LLMs in these connected tasks using real-time feedback within a unified framework.
}

\section{\revise{Threats to Validity}}

\revise{In this section, we discuss several potential threats to validity and the corresponding mitigation strategies below.

\subsection{Construct Validity}
Construct validity concerns whether our RQ formulation and categorization processes accurately reflect the intended objectives. 
In designing RQs, as the first literature review in the LLM-based APR field,  we followed prior LLM4SE~\cite{wang2024software} and APR~\cite{zhang2023survey} survey studies, focusing on three main dimensions (i.e., APR in Section~\ref{sec:rq1}, LLMs in Section~\ref{sec:rq2} and integration in Section~\ref{sec:rq3_repair_scenarios}) to systematically classify the collected papers. 
However, one threat arises from the possible ambiguity of paper categorization, as some studies could reasonably belong to multiple categories (e.g., Mashhadi~et al.~\cite{mashhadi2021applying} involving both fine-tuning and empirical evaluation). 
To mitigate this, we assign each paper to the category that best represented its , following consistent classification rules.

\subsection{Internal Validity}
Internal validity threats are related to potential bias in the process of paper selection, data extraction, and quality assessment. 
To reduce such risks, we followed a multi-stage procedure combining both manual and automated search. 
First, we performed manual searches in top venues to identify seed papers, which are then used to design the search keywords for automatic retrieval from digital libraries. 
Second, we applied inclusion and exclusion criteria consistently across all sources and conducted a forward and backward snowballing process to ensure completeness. 
Finally, each included paper underwent a quality assessment using a ten-question checklist. 
Two authors independently performed selection and evaluation, and any disagreement was resolved through discussion.

\subsection{External Validity}
External validity refers to the generalizability of our findings beyond the collected dataset. 
Although we carefully collected studies published between 2020 and September 2025 from major digital libraries and high-quality preprints, some relevant works may have been missed due to limitations in indexing or keyword variations. 
To mitigate this, we employed iterative search refinement, snowballing, and inclusion of diverse publication venues across SE, AI, and PL communities. 
Nevertheless, we acknowledge that the rapidly evolving nature of LLM-based APR research means that future studies may extend or refine our findings.
}

\section{Conclusion}
\label{sec:conclusion}

Automated Program Repair (APR) tackles the long-standing challenge of fixing software bugs automatically, thus facilitating software testing, validation, and debugging practices.
Very recently, Large Language Models (LLMs) have brought significant changes to the APR domain, already yielding impressive progress and further demonstrating a promising future in follow-up research.
In this paper, we provide a systematic literature review of existing LLM-based APR techniques from LLMs, APR and their integration perspectives
We summarize popular LLMs, typical utilization strategies, and repair scenarios.
We also discuss some crucial factors, such as input forms and self-debug metrics, within the LLM-based APR community. 
Finally, we outline several challenges, such as data leakage issues, and suggest potential directions for future research.
\revise{In the }

\revise{
Overall, LLM-based APR has rapidly emerged over the past few years, reflecting growing interest across diverse research communities (e.g., software engineering, artificial intelligence, and programming languages) and suggesting that many promising directions remain largely unexplored.
We conclude by discussing three complementary perspectives that not only appear particularly promising for future research but also shaped the way we structured this survey.
From the repair perspective, future work should explore broader repair scenarios and further evaluate how LLMs can effectively assist developers in real-world repair environments.
From the integration perspective, future research could design more advanced strategies to address emerging challenges (e.g., enabling long-term memory and repository-level reasoning), and establish comprehensive benchmark suites to systematically evaluate diverse repair capabilities across different settings.
}

\section*{Acknowledgments}
This research was supported in part by the Australian Research Council Discovery Projects under Grant No. DP200102491 and DP230101790, the National Natural Science Foundation of China (61932012, 62141215, 62372228), Natural Science Foundation of Jiangsu Province (BK20251458), and Fundamental Research Funds for the Central Universities (AE89991/463).

\bibliographystyle{ACM-Reference-Format}
\bibliography{reference}

\end{document}